# Degradation Analysis of Perovskite Solar Cells via Short-Circuit Impedance Spectroscopy: A case study on NiO$_x$ passivation

Osbel Almora,[1,*] Pilar López-Varo,[2,*] Renán Escalante,[3] John Mohanraj,[4] Luis F Marsal,[1] Selina Olthof,[4] Juan A. Anta[3,*]

[1] Departament d'Enginyeria Electrònica Elèctrica i Automàtica, Universitat Rovira i Virgili, 43007 Tarragona, Spain
[2] Institut Photovoltaïque d'Ile-de-France (IPVF), 91120 Palaiseau, France
[3] Department of Physical, Chemical and Natural Systems, Universidad Pablo de Olavide, Sevilla 41013, Spain
[4] Department of Chemistry, University of Cologne, Greinstrasse 4−6, Cologne 50939, Germany

* osbel.almora@urv.cat, pilar.lopez-varo@ipvf.fr, jaantmon@upo.es

## Abstract

Perovskite solar cells (PSCs) continue to be the "front runner" technology among emerging photovoltaic devices in terms of power conversion efficiency and application versatility. However, not only the stability but also the understanding of their ionic-electronic transport mechanisms continues to be challenging. In this work, the case study of NiO$_x$-based inverted PSCs and the effect of different interface passivating treatments on the device performance are approached. Our experiments include impedance spectroscopy (IS) measurements in short-circuit under different illumination intensities and operational stability tests under constant illumination intensity. It is found that certain surface treatments lead to more stable performance. However, protic anion donors can induce, both, an initial performance decrease and a subsequent reactivity during light exposure which apparently improves the cells performance. Our drift-diffusion simulations suggest that the modification of the interface with the hole transport material may have decrease the conductivity, as well as the ion and electron mobilities at the perovskite and the NiO$_x$, respectively. Importantly, capacitance and resistance are shown to peak maximum and minimum values, respectively, around specific ranges of mobile ion concentration. Our results introduce a general route for characterization of degradation paths in PSCs via IS in short-circuit.

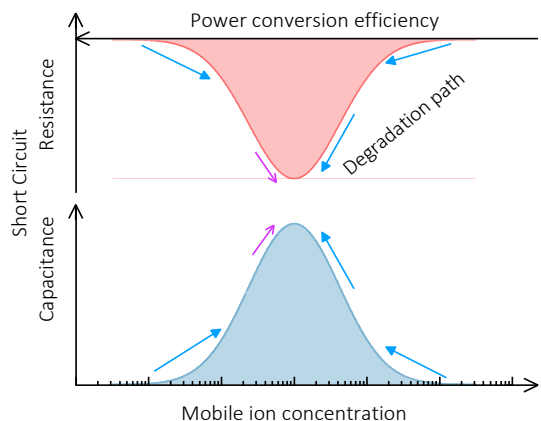

TOC figure:

# 1. Introduction

The optimal optoelectronic properties of metal-halide perovskites have gained major attention of the semiconductor device research community during the last decade resulting in unprecedented progress in several fields, such as photovoltaics,[1] light emitting diodes,[2] lasers[3] and ionizing radiation detectors.[4] Particularly, single junction perovskite solar cells (PSCs) have been reported with record power conversion efficiency (*PCE*) values >25% with relatively low-cost solution-based fabrications methods. These devices are based on the superposition of sequenced thin film layers where the perovskite absorber is sandwiched between the electron and hole selective layers. In addition, the compatibility with different substrates have produced a broad range of versatile applications, for instance in transparent/semitransparent and flexible photovoltaics. However, the understanding of the working mechanisms of these devices is still limited. Particularly, the long-term operational stability issues in PSCs, which remain a key limiting factor for upscaling and industrial deployment, are still in early phases of elucidation.[5]

The long-term degradation of thin film devices, such as PSCs, is a complex process that depends on several mechanisms, parameters and constituting elements (see Figure S1 in Section S1, supporting information). Most typically, instability originates from chemical reactions creating oxidation and/or unintended products between the pristine materials and reactant leftovers of each layer,[6, 7] the interfaces [8] and the air (e.g. humidity) in operational conditions.[9, 10] Moreover, reactivity can be catalyzed and/or triggered by mechanical,[9] thermal[9-11] and bias[11-14] stresses as well as photon interaction.[5, 9, 15] In addition, besides reactivity, the materials crystal structure of each layer and the interfaces can be modified, creating undesired defects that reduce photon absorption and enhance charge carrier recombination.[16] Among these non-reactive sources of defects one can find temperature stress and the migration of various species inside the layer due to either diffusion, illumination or electrical stressing.[13, 14, 17, 18] Notably, not only species from different layers present within the device may migrate increasing the leakage currents of the cell,[19] but also the ion migration of intrinsic halide vacancies and other charged defects[20, 21] have been demonstrated in halide perovskites.[22] Therefore, assessing the individual contribution of each mechanism and element in the device is challenging, which motivates the design of experiments where one can neglect some of the degradation agents. In this context, the use of advanced characterization techniques (beyond the routine measurement of current density-voltage $(J-V)$ curves) and support by numerical device simulation is particularly useful.

The device structure of PSCs includes a sequence of thin film layers where the intrinsic absorber perovskite is sandwiched between the electron and hole transport layers, ETL and HTL, respectively. The conductivity of the ETL and the HTL are n- and p-type, respectively, which define p-i-n or n-i-p structures, depending on which of the transport materials comes first in the direction of the incident light path. Particularly, p-i-n structures, also known as inverted PSCs, have reported efficiencies over 25%[23, 24] and are attractive due to their low-temperature fabrication methods,[25] and compatibility with existing industrial techniques.[26] An example of this structure is schemed in **Figure 1**a, where non-stoichiometric nickel oxide (NiO$_x$) is chosen as inorganic HTL and the organic semiconductor films comprised of C$_{60}$ and bathocuproine (BCP) serve as ETL and hole blocking layers, respectively. The use of organic ETLs in inverted PSCs has been suggested because of the facile low-temperature synthesis with orthogonal solvents and purification methods resulting in optimal performance with relatively low instability.[27] In addition, the fact that the p-i-n structure places the ETL behind the absorber perovskite in the direction of the light path reduces the radiative and temperature stresses on the organic HTL. On the other hand, the use of NiO$_x$ is proposed considering the already demonstrated *PCE* values over 24%.[28, 29]

The direct contact between as-prepared NiO$_x$ and perovskite has proven to be problematic for solar cell efficiency and stability, as for example summarized in the recent review by Cai et al.[30]. To circumvent this issue, the NiO$_x$/perovskite interface has been optimized via several materials and fabrication methods. For instance, the energy level alignment between NiO$_x$ and the perovskite has been optimized with the introduction of inorganic extrinsic doping such as the case of Wang et al.[31] who sited Ag via a sol-gel method and Yi et al.[32] who fabricated nanopatterned Zn:NiO$_x$ with an advantageous 1D nanoscale architecture and synergistic substitutional Zn doping. Alternatively, Hu et al.[33] used an organic doping by including 4-tert-butylpyridine (tBP) as additive in the NiO$_x$ precursor solution, whereas Kang et al.[34] used 4-iodophenylboronic acids to modify the NiO$_x$/perovskite layer interface, believed to be affected by the intrinsic defects (Ni vacancies) in the NiO$_x$ film and the I vacancies at the buried interface of the perovskite. Similarly, high performance devices were reported by Wang et al.,[29] who utilized a multi-fluorine organic molecule 6FPPY, to manage the buried interface of NiOx-based p-i-n PSC. Notably, Pu et al.[28] introduced a poly[bis(4phenyl)(2,4,6-trimethylphenyl)amine] (PTAA) interlayer between the NiO$_x$ and the perovskite, resulting in high efficiency and stability. Furthermore, Shen et al.[35] proposed that the hydrophilic chain of the amphipathic molecule Triton X100 can coordinate as a Lewis additive with the $Ni^{3+}$ on the $NiO_x$ surface, passivating interfacial defects and hindering the detrimental reactions at the $NiO_x$/perovskite interface.

From the device operation point of view, the degradation of a solar cell is commonly associated with the decrease of the $PCE$, as measured from the current density-voltage ($J-V$) curve under standard 1 sun illumination incident power density ($P_{in}$). The $PCE$ value can be expressed in terms of the complementary performance parameters as

$$PCE = \frac{V_{oc} J_{sc} FF}{P_{in}} \quad (1)$$

where $V_{oc}$ is the open-circuit, $J_{sc}$ is the short-circuit current, and $FF$ the fill factor. Accordingly, a decrease in $PCE$ can be found in terms of the decrease of one or up to three performance parameters (see Figure S2) in overlapping. In practice, $FF$ and $V_{oc}$ modifications are more likely to be connected since these parameters relate to the near-flat-band condition where charge carrier recombination rate approaches that of the photogeneration. Note that under illumination intensities close to and above 1 sun standard, high-performance devices are typically expected to achieve a maximum power point (MPP) voltage $V_{mpp} < V_{oc}$ close to the built-in voltage ($V_{bi}$). In contrast, the lower the illumination intensity, the farther the values of $V_{mpp}$ and $V_{oc}$ from $V_{bi}$.

Around flat-band (V=$V_{bi}$), the energy diagram is expected as shown by the simulation in **Figure 1**b for the device structure in **Figure 1**a. This is a regime where the diffusion current is similar to, if not larger than, drift current due to the low electric field (see **Figure 1**d), resulting in large gradients in the charge carrier density profiles (see **Figure 1**e). Diffusion-related long-term relaxations may overlap with degradation processes, hindering dedicated investigations regarding device stability. Moreover, the illumination intensity is not expected to produce a big change in the energy diagram around flat-band (dashed lines close to solid lines in **Figure 1**) because the slower electrical response hinders the separation between injected and photogenerated excess charge carriers, which recombine mostly radiatively. In addition, the dual electronic-ionic conductivity of metal halide perovskites not only hinders the evaluation of the MPP, which defines the $FF$, but also complicates the transport by adding field screening by ions to a regime where diffusion transport would otherwise be more dominant.[4, 36] The field screening effect can be seen in **Figure 1**d towards the interfaces of the perovskite and the mobile ions could distribute along the entire perovskite bulk, as shown in **Figure 1**e. Accordingly, it may be convenient to explore another biasing regime for studying stability.

At short-circuit ($V$=0 V), the electric field in the perovskite bulk is significantly higher and sensitive to the illumination intensity (see **Figure 1**d). Under illumination, this is a regime where the drift current and nonradiative recombination towards the interfaces are predominant due to the band leaning (see **Figure 1**c) and the lower concentrations of charge carriers located within the perovskite near the transport selective layers (see **Figure 1**f). In addition, the short-circuit and reverse bias regimes are commonly used for the characterization of the shunt resistance, which is a parameter typically affected during long-term operation. Interestingly, PSCs have been found to exhibit linear photo-shunt resistance (see Table S1)[37] and photo-capacitive[38] increase at short-circuit, which are properties characterized via impedance spectroscopy (IS). Notably, the use of IS analyses at short-circuit is preferrable due to the higher linearity of the signal response, compared to near flat-band situation. Moreover, unlike the IS studies in quasi-open-circuit condition,[39, 40] little attention have been paid in the literature to the IS spectra in short-circuit for PSCs. Table S2 in the supporting information summarizes the comparison between short-circuit and near flat-band regimes, from which one can assess the suitability of each condition for designing experiments.

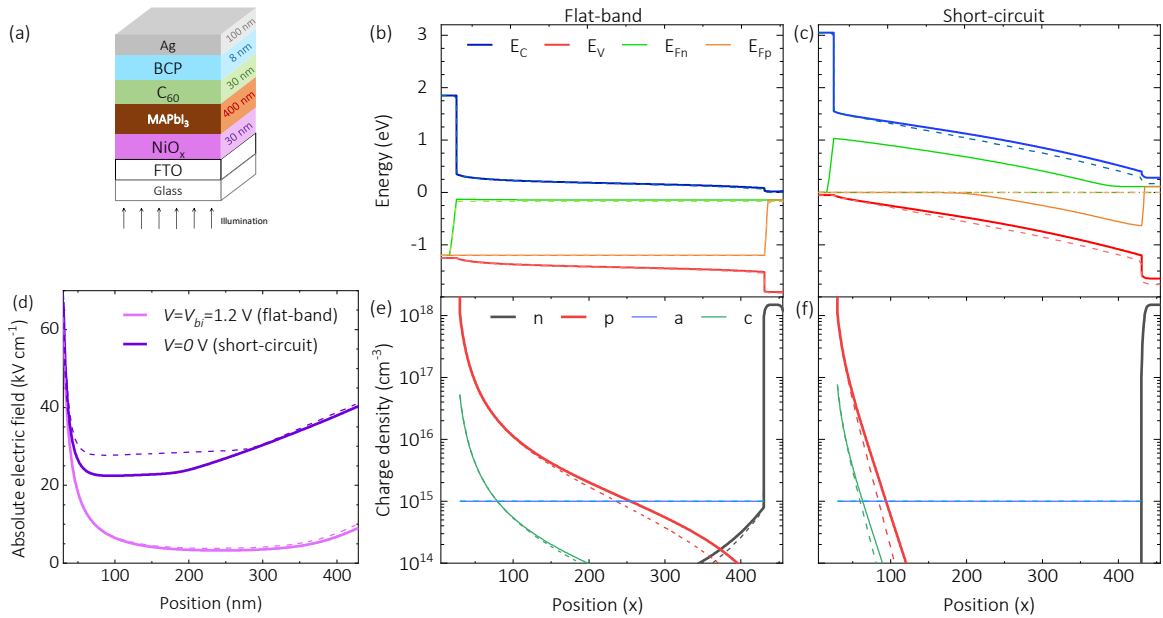

**Figure 1.** Simulated steady-state electrical response of the reference PSCs with the structure in (a) in terms of energy diagrams (b, c), electric field (d) and charge density profiles (e, f). The depicted conditions include flat-band (b, e), short-circuit (e, f), 1 sun illumination (solid lines) and dark (dashed lines). In (b, c), $E_C$ and $E_V$ are the conduction band minimum and valence band maximum energy levels, respectively; and $E_{Fn}$ and $E_{Fp}$ are the quasi-Fermi levels for electrons and holes, respectively. In (d), $V_{bi}$ is the built-in voltage. In (e, f), the densities are $n$, $p$, $a$, and $c$ for electrons, holes, mobile anions and mobile cations, respectively. Setfos-Fluxim[41] was used for these simulations using the parameters enclosed

in Table S9.

In this work, the stability of passivated-$NiO_x$ PSCs is analyzed with focus on the HTL-perovskite interface and the evolution of electrical properties. In a separate publication these passivating surface treatments of NiOx have been studied with respect to film formation, interface composition, as well as solar cell initial performance and stability. [Ref] In contrast, the focus here is set on the modeling and understanding of fundamental transport roperties. Experimentally, the current evolution over time under high power condition was sensed for several passivation methods targeting the interface between the $NiO_x$ and $MAPbI_3$ absorber layer. Moreover, impedance spectra were measured over time and under different illumination intensities at short-circuit in each case. The results were contrasted with equivalent circuit modeling and numerical simulations on Setfos Fluxim[41] and Driftfusion[42] transport equation solvers. Our results correlate different modification in the electron and mobile ion concentrations and mobilities with the degradation observed in the samples.

## 2. Methods

The fabrication details for the studied devices with structure $FTO/NiO_x/MAPbI_3/C_{60}/BCP/Ag$ (see **Figure 1**a) can be found in our simultaneous article [ref]. An adapted version is presented in Section S2 (supporting information) along with illustrative initial *PCE* characterization (see Figure S4 and Table S3). For the passivation, the selected materials distribute into three categories attending to the potential conductivity effect, as discussed in more detail in reference [ref]. Firstly, $PbI_2$ as cation donor which is a metallic salt with divalent Pb cation that may potentially occupy $Ni^{2+}$ vacant sites, and corresponding anions to interact with high valent (> 2+) Ni centers. Secondly, 1-Iodobutane ($C_4H_9I$, also labeled here as iodobutane) and 1-Phenylethylamine ($C_8H_{11}N$, also labeled here as amine) neutral bases, which are Lewis bases, i.e. neutral organic molecules with a capacity to donate electron pairs to the Ni atoms with higher formal charge (>2+). Thirdly, HI and MAI protic anion donors, i.e. Bronsted acids, which are organic molecules that could donate protons (or accept electron pairs) to O atoms with increased formal charge due to $Ni^{2+}$ vacancies (O-defect sites) and anions to interact with $Ni^{>2+}$.

The IS data were measured with an Autolab PGSTAT302N potentiostat including a FRA32M unit and a kit Autolab Optical Bench from MetroOhm. The samples were illuminated with a white LED (CREE XM-L3 U4 on star PCB, XMLDWT-U40E1) at different steady-state illumination intensities, then the short-circuit current was stabilized before applying 15 mV of alternating current (AC) mode perturbation. The details for the IS characterization under different illumination intensities are in Section S3. The stability assessment was carried out under 0.2 sun equivalent illumination intensity, as measured with a calibrated reference cell 91150-112/PVM 164 from Newport. The samples were under continuous $N_2$ flux, at room temperature, and were kept in operation at a forward bias close MPP under illumination with systematic switching to short-circuit for IS measurements (see Figure S16). Section S4 provides further details on the operational stability test procedure and current and IS data.

The equivalent circuit modeling was carried out with ZView from Scribners. Details of each equivalent circuits and fitting parameters can be found in Sections S3 and S4. SETFOS Fluxim[41] software and MATLABS's Driftfusion[42] code were used for the numerical simulations of the time-dependent solutions of the transport equations with electronic and mobile ion charge carriers. The corresponding details can be found in Sections S5 and S6, in the supporting information.

## 3. Results and discussion

The different NiOx surface treatments employed in the various samples in this study resulted in two main behaviours, as seen in the solar cell characteristics presented in Figure S4. On the one hand, the $NiO_x$ surface passivation with 1-Iodobutane, 1-Phenylethylamine and $PbI_2$ improved the performance mostly due to a slight increase of the photocurrent. The resultant PCE values ranged around 13% for the un-passivated reference sample ($NiO_x$ Ref.) and nearly 15% for the passivated samples, in line with state-of-the-art pure $MAPbI_3$ $NiO_x$-based PSCs.[43, 44] This suggests that the treatment with these passivation agents could reduce electrical losses due to charge carrier recombination at the interface. Nevertheless, these passivation processes could also modify the morphology and optoelectronic properties of the perovskite layers resulting in further effects such as the reduction of optical losses due to interference and the decrease of bulk recombination in the perovskite. On the other hand, the MAI and HI passivated samples showed a significant PCE reduction to less than 6% due to a remarkable decrease of the fill factor and the photocurrent. The absence of rectifying behavior in these samples indicate major modifications of the charge carrier density profiles, large parasitic resistive effects and large recombination rates.

For the understanding of these two different behaviors, complementary experiments were performed with a focus on the short-circuit condition, the different illumination intensities and the IS analysis. This is presented in Section 3.1 where the equivalent circuit modeling is used to estimate resistive, capacitive and characteristic response times. Subsequently, the operational stability and the evolution of the IS spectra over time under illumination is presented in Section 3.2.

### 3.1. **Different illumination intensities in short-circuit.**

The impedance spectra under different illumination intensities were measured in short-circuit condition for representative samples of the studied device set with different passivation treatment on the NiO$_x$ interface with the perovskite, as described in Section S3 of the supporting information and summarized in **Figure 2**. The characteristic two-arcs spectra in the impedance Nyquist plots are present in the reference and optimized samples (amine, iodobutane, PbI$_2$), as indicated in **Figure 2**a. Typically, the high- and low-frequency semicircles relate to electronic and ionic-electronic resistance (*R*)-capacitance (*C*) contributions, respectively. In the capacitance Bode plot representation, two plateaus can be identified (see **Figure 2**b), where the nearly constant values toward high-frequency (Hf, >10 kHz) are related with the geometrical dielectric capacitance[39] and the capacitance step-like increase toward low-frequency (Lf, <10 Hz) has been associated with electrode polarization and mobile ion accumulation at the interfaces.[40] However the HI and MAI samples not only demonstrate smaller resistances in SC, but also show three arcs in the impedance Nyquist representation (see **Figure 2**a) and a gradual capacitance increase in the capacitance Bode plot (see **Figure 2**b). Notably, the higher the illumination intensity, the clearer the definition of the three arcs in the Nyquist plot. Then, a transition from a "two (larger)" to a "three (smaller)" arcs situation complicates the "high-frequency electronic" versus "low-frequency ionic" interpretation of the IS spectra.

The equivalent circuit models used for parameterization of the IS spectra are shown in **Figure 2**c,d for the two- and three-$RC$ constants, respectively. Particularly, for the reference and optimized samples two resistors and two capacitors were fitted for the high- and low-frequency features of the spectra as $R_{Hf}$, $R_{Lf}$, $C_{Hf}$ and $C_{Lf}$, respectively. In contrast, for the MAI and HI treated samples an additional set of medium-frequency resistor ($R_{Mf}$) and capacitor ($C_{Mf}$) were considered. The resultant values for the resistances as a function of the short-circuit current under different illumination intensities are summarized in **Figure 2**e. First, in general, all the samples follow a similar trend where a saturation is observed towards dark condition and a decrease $R \propto J_{sc}^{-1}$ follows as the illumination intensity increases. This is a common trend in photovoltaic solar cells (see Table S1) due to the photoconductivity increase upon charge carrier generation. Second, the low-frequency resistance is always higher than that of the high-frequency resistance. This may suggest that in short-circuit condition the ion migration and/or ion-related charge carrier transport is significantly hindered in comparison to the faster "pure" electronic response. Third, the higher the short-circuit resistance the higher the device operational performance. This behavior can be directly related with the shunt resistance ($R_{sh}$) of the samples which mostly affects the fill factor and, ultimately, the photovoltage. Interestingly, **Figure 2**e highlights one of the advantages of the IS over the $J-V$ curve for the estimation of the $R_{sh}$. Not only hysteresis issues are avoided but also smaller resistance contributions are identified, which would not be measured in series connected added resistances with direct current (DC) mode methods. In addition, the fact that low-frequency resistances saturate in dark suggests that ion migration is occurring in the grain boundaries and surface defects, in addition to the bulk perovskite.[45]

The low-frequency capacitance linear increase as a function of short-circuit current $C_{Lf} \propto J_{sc}$ for different illumination intensities is illustrated in **Figure 2**f, whereas the constant high-frequency can be found in Section S7. This increase of low-frequency capacitance in short circuit is a characteristic feature of PSCs[38] due to mobile ion accumulation toward the interfaces (see **Figure 1**f). Interestingly, the reference and surface-treated (amine, iodobutane, PbI$_2$) samples show a threshold illumination value for $J_{sc} \approx 10^{-5}$ A·cm$^{-2}$ which indicates a transition between a nearly constant behavior and the linear trend. In contrast, the HI and MAI samples show a gradual increase of $C_{Lf}$ without an apparent baseline capacitance in the explored range of illumination intensity. Furthermore, by coupling the capacitors and the corresponding discharge resistors, the characteristic short-circuit charge carrier response times $\tau = RC$ can be accessed, as presented in **Figure 2**g for the high-frequency times. The low-frequency, and slower, response times can be found in Section S3 showing nearly linear or slightly increasing trends with the augment of illumination intensity. On the other hand, the fastest high-frequency response times evolve from saturated maximum values up to linear decreasing trends as the illumination intensity is augmented. This can be interpreted as a transition of main charge carrier recombination mechanism between non-radiative and radiative, when dark and under illumination, respectively. Notably, the higher the response time the higher the operational performance of the samples in terms of PCE.

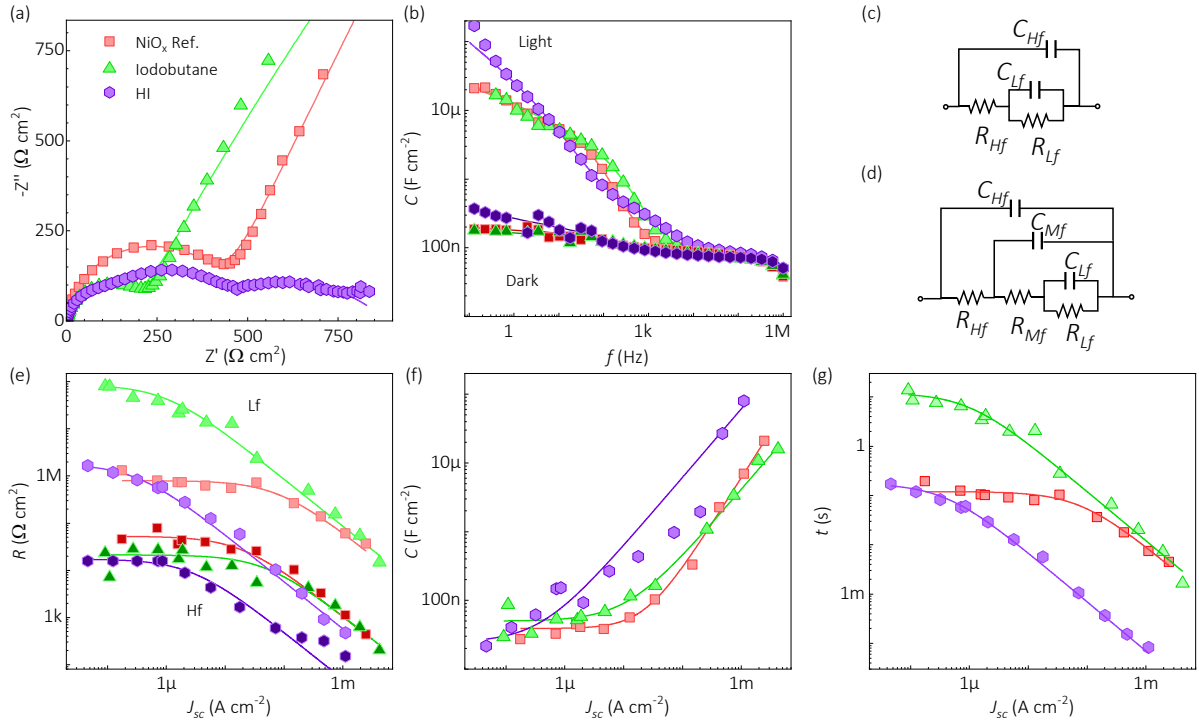

**Figure 2.** Experimental selected IS spectra under different illumination intensities in short-circuit condition ($V$=0V) for illustrative samples with different passivation procedures, as indicated, and equivalent circuit parameterization. In (a) there are impedance Nyquist plots under 0.2 sun equivalent LED light and (b) shows the capacitance Bode plots in perspective with the dark spectra (darker dots). Solid lines in (a) and (b) are the fitting to the equivalent circuits in (c) and (d), in each case. The corresponding (e) resistance, (f) capacitance (low-frequency) and (g) a characteristic response time (high frequency) are plotted in the lower panel with darker and lighter dot fill colours for high and low frequency related components, respectively. The lines in (e-g) are empirical fittings introduced in Section S3.

The experimental evidence shown in **Figure 2** and Section S3 with IS spectra over a range of illumination intensities highlights the paramount importance of the HTL/perovskite interface for the device operation. Nevertheless, several aspects require further understanding. For instance, the role of 1-iodobutane, 1-phenylethylamine and PbI$_2$ treatments appears to be passivating surface defects and improving the charge extraction by hindering recombination. However, it is unclear how the MAI and HI treatments deteriorate the rectifying barrier, reducing the overall resistance and showing the characteristic "three arcs" in the impedance Nyquist plots. Furthermore, the fact that a large low-frequency impedance arc evolves into two smaller arcs in the Nyquist plot could be related with ion migration and the modification of the nature or concentration of the mobile ions and electrons in the perovskite [ref].

### 3.2. Constant illumination over operational time

The operational stability of the samples was studied over time under constant illumination intensity, at room temperature, and with nitrogen flux circulation. The stability tests procedure is schemed and described in detail in Section S4. The sample was exposed to constant 0.2 sun white LED equivalent illumination intensity, next a voltage $V_{hpp}$ close to the MPP was applied and the current was recorded for measuring the output power ($P_{out}$). Subsequently, the biasing was switched to short-circuit and IS spectra were acquired in each iteration of the loop. This process was continued until the samples broke, the duration ranged from 20 h to 50 h. The overall $P_{out}$ performance of the devices is depicted in **Figure 3**a, where two main trends are observed. For the reference and surface-optimized (amine, iodobutane, PbI$_2$) samples, a rather stable $P_{out}$ with a slight increase is detected. On the other hand, the HI- and MAI-treated cells experienced a large increase of $P_{out}$ with an apparent extrapolated saturation around 300% of the initial value. Nevertheless, despite the final tripled $P_{out}$ values for the HI and MAI samples, the performance of the reference and optimized cells was still 30% to 50% better, as summarized in **Figure 3**b and Figure **S17**. When comparing final versus initial $P_{out}$ values in this plot, the closer the sample is to the "x=y" diagonal (dashed line in **Figure 3**b), the higher the stability. Alternatively, considering the different initial performance and test duration of each sample, one can also use the concept of effective or overall degradation rate[1]

$$DR = \frac{P_{out,final} - P_{out,initial}}{t_{STD}} \quad (2)$$

where $P_{out,initial}$ and $P_{out,final}$ are the initial and final output powers for the stability test of duration $t_{STD}$. In the definition of equation (2), the smaller the value of *DR* the better. Moreover, despite the typical decrease of performance would entail *DR*<0, the report of *DR*>0 during the first 200 hours of operation under constant illumination is well-known in the literature,[1] and typically precedes a subsequent performance decrease period with *DR*<0. In **Figure 3**b (right axis) the studied samples show *DR*>0 with the highest and lowest values for the HI and the PbI$_2$. samples, respectively. Interestingly, the reference sample resulted in a DR value close to that of the PbI$_2$-treated device. This latter behavior mismatches that of previous MPP tracking experiments [ref] where the reference device was more instable. Nevertheless, in that study [ref] there were not intermittent short-circuit periods for current stabilization and IS measurements. Therefore, one can hypothesize that the short-circuit and near-flat-band hinders and favors the degradation of the untreated sample, respectively; and the opposite can be true for the PbI$_2$-treated samples.

The IS spectra in short-circuit condition were obtained along the above-described operational stability tests and the results are depicted in **Figure 3**c-e (see also Section S4). Analogously to the previous discussion, two main trends were observed. For the reference and optimized samples, a relatively stable evolution of the IS spectra was found. **Figure 3**c zooms the high-frequency region in the impedance Nyquist plots for the reference sample, which slightly decrease the high-frequency resistance over time during the degradation test. This highlights the accuracy of the IS experiments, even though the main contribution to resistance and device performance comes from the larger low-frequency range of the spectra. In **Figure 3**d (see also Figure S18), the complete Nyquist plots for the $PbI_2$-treated sample illustrate the apparent constancy of the device over time of degradation test. At the same time, the spectra are also shown similarly unchanged in the impedance and capacitance Bode plots in figures S19 and S20, respectively. This steadiness of the reference and optimized samples in short-circuit suggests the stability of the devices shunt resistance which agrees with the slight increase of the output power shown in **Figure 3**a. Furthermore, it is implied that any significant improvement of the device operation performance should be mostly related with the modification of optical absorption and/or the improvement of the transport properties which affects the near-flat-band regime.

The HI and MAI cells show an unstable behavior in the IS spectra over time under constant illumination, as in **Figure 3**e-f (see also Figure S18-20). Not only the impedance is smaller in these samples with respect to the reference and optimized ones, but also a clear increase is observed. The longer the operation time under constant illumination, the higher the impedance of the sample which correlates with the performance improvement in **Figure 3**a. Interestingly, the three-similar-arcs feature in the Nyquist plot evolves into the more typical two-arcs shape where the high frequency arc is significantly smaller than that of the low-frequency range. Once more, the qualitative change of the low frequency part of the IS spectra suggests that the nature and/or quantity of the mobile ion properties may be modified by the action of the current flow of current under continuous illumination.

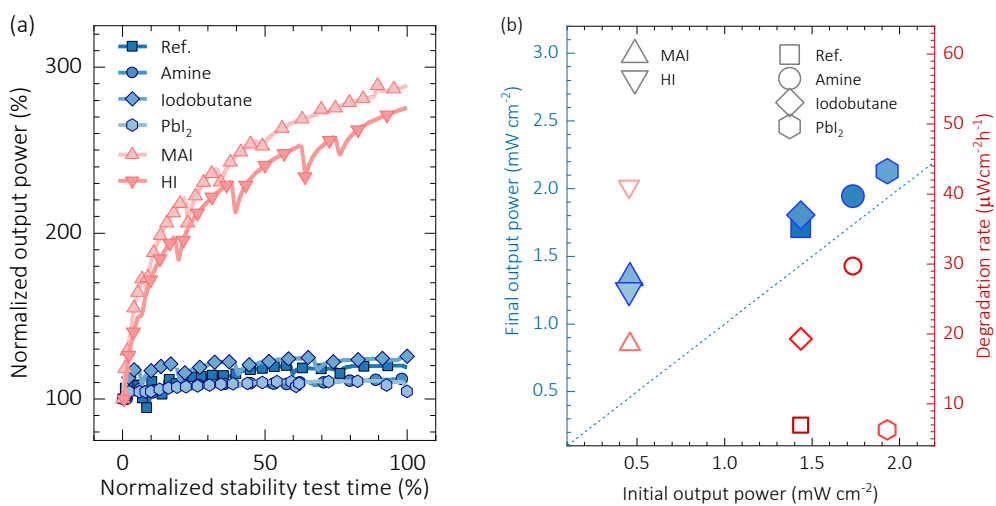

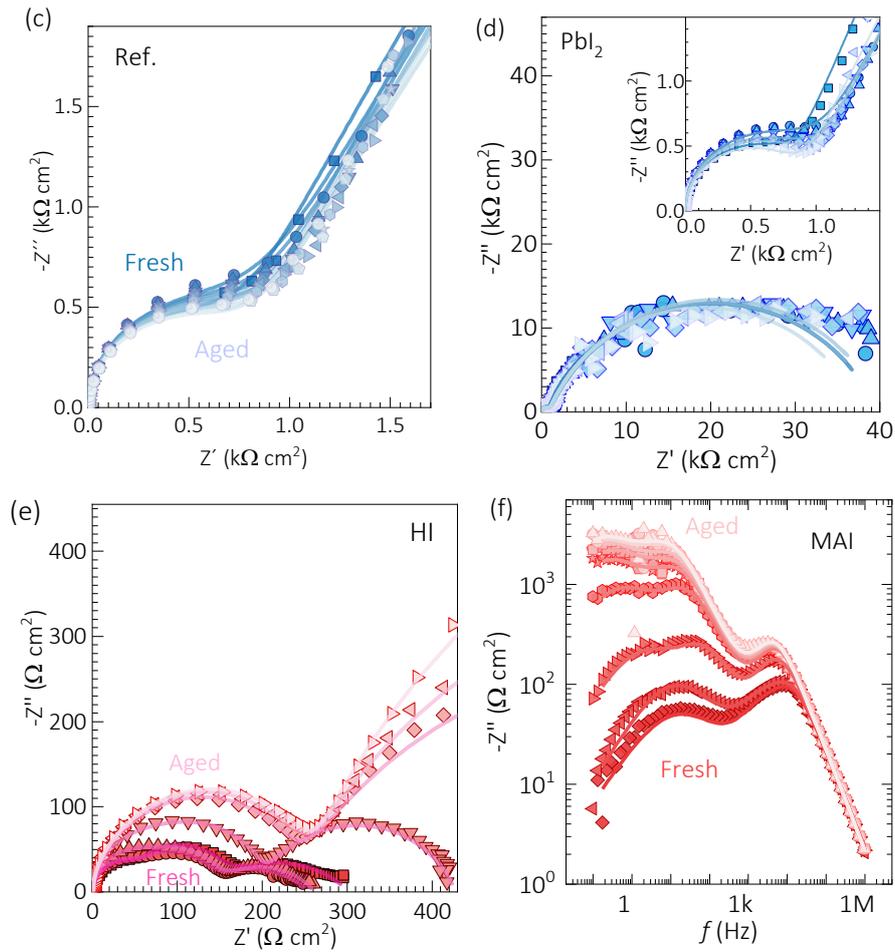

**Figure 3.** Experimental stability test under 0.2 sun white LED equivalent illumination intensity. The normalized time evolution of output power in (a) is summarized in (b) with the final values (left axis, filled symbols) and degradation rates (right axis, open symbols) as a function of the initial output power for each sample, as indicated. Illustrative impedance spectra in short-circuit under constant illumination over operation time are in Nyquist representation in (c-e) and Bode plots (f) for different samples, as indicated. Dots and lines are the experimental data and simulated spectra from fitting to equivalent circuit models in Figure S5, respectively. The dashed line in (b) indicates the x=y coordinates. In (c-f), the lighter the colour the longer the operational time in the IS spectra; and the inset in (d) magnifies the high frequency part of the spectra.

The instability of the MAI and HI sample indicates the presence of intrinsic reactivity, which could be understood in terms of the creation of a non-stoichiometric excess of MA, I or H that modifies the local composition of the perovskite in contact with the $NiO_x$ interface. It may appear counter-intuitive that these excess reagents contribute to performance improvement rather than further degradation. However, one must bear in mind that intrinsic reactivity transforms the material composition and properties regardless of whether it contributes to or hinders the final device operational performance. In addition, the chemical "softness" of the perovskite under biasing is herein highlighted since only 0.2 sun white LED equivalent illumination intensity with room temperatures (<40°C) have been sufficient to catalyze the modification of the samples, when compared with previous studies [ref] on unbiased samples comprising MAPbI$_3$ on treated NiO$_x$ films.

Evidence that the use of non-stoichiometric formulations using volatile precursors of PSCs modifies the operational stability of the PSC[46] which create an excess of mobile ions already exists in the literature. For instance, Lammar et al.[18] found that the excess of a protic precursor FAI, linked to a larger concentration of mobile ionic defects, also leads to an initial increase of the PCE under illumination, due to ionic redistribution. HI and MAI are not only more acidic but also smaller than the rest of the studied additives and, consequently, more likely to infiltrate in the perovskite lattice and create ionic defects [ref].

### 3.3. **Drift-diffusion simulations**

The experiments described in sections 3.1 and 3.2 illustrate the effects of using different treatments to the surface of the NiO$_x$ HTL before subsequent deposition of the absorber perovskite layer. Two main trends are found. On the one hand, 1-iodobutane, 1-phenylethylamine and PbI$_2$ treatments result in similar, if not better, device performance and stability when compared with the reference untreated samples. On the other hand, MAI- and HI-treated samples not only show very low fill factors in the *J-V* curves and anomalous three-arcs low impedance IS spectra in the Nyquist plots, but also behave unstable under constant illumination in operation over time. These observations rise several questions on the nature of the material modifications and the corresponding implications for the transport properties. Accordingly, this section presents a series of numerical simulations which are firstly validated by qualitatively reproducing the experimental behaviors and subsequently explore the possible origins and consequences of the observed phenomena. The simulation parameters and detailed description of the simulated spectra can be found in Section S5 and S6 for the simulations made with SETFOS-Fluxim software[41] and MATLAB's Driftfusion code.[42] The use of two simulators not only provide further validation to the analyses but also allows to assess the impact of the different boundary conditions to solve similar problems.

The initial current-voltage curves under standard 1 sun illumination were qualitatively fitted to match the DC electrical response before simulation of the IS spectra. This is shown in **Figure 4**a,b (see also Sections S5.1 and S6.1) within a range of mobile ion concentration between $10^{15}$ and $10^{19}$ cm$^{-3}$, which includes the minimum defect concentration measurable with IS methods and the effective density of states at the bands.[47] Notably, in order to reproduce the $J-V$ response from the HI- and MAI-treated samples, three main parameters were modified with respect to those used to simulate the reference and the optimized samples. First, a change of absorptivity was considered by reducing the charge carrier generation rate from $G_{ref}$=3.3×10$^{21}$ cm$^{-3}$ to $G_{HI}$ = 2.2×10$^{21}$ cm$^{-3}$, mostly accounting for the $J_{sc}$ difference between the reference and the HI samples, respectively. Second, an increase of series resistance from 2.5 up to 25 Ω·cm$^2$ was taken for the HI sample that directly diminishes the $FF$ from around 70%, for the reference and optimized cells, to less than 40% for the MAI and HI devices. Third, the mobility of electron and hole charge carriers was reduced at the NiO$_x$ HTL in the HI sample simulations to a tenth of those used in the reference cell. The modification of this parameter can be correlated with the modification of the HTL during the surface treatment, resulting in the decrease of the $FF$ in the $J-V$ characteristic. Moreover, the modification of the electron and hole mobilities in the perovskite layer has also been found to alter the shape of the $J-V$ curve accounting for some of the above-described lessening in the $J_{sc}$ and the $FF$.

The increase of mobile ion concentration ($N_{ion}$) decreases the $J_{sc}$ of the simulated $J-V$ curves in **Figure 4**a,b, and the $FF$ also appears to be inversely proportional to $N_{ion}$. The reason for this can be associated to the electric field screening due to ion space charge regions toward the electrode. The higher the $N_{ion}$ the smaller the effective built-in field of the junction, which diminishes the drift current producing photocurrent around the short-circuit condition (see Figure S24).[18] On the other hand, the closer to the flat-band condition, which for these samples approaches the open-circuit voltage ($V_{oc}$), the smaller the effect of $N_{ion}$ resulting in very similar currents for the same bias. Around flat-band condition, the diffusion current takes over as the main transport mechanism, which is no longer that dependent on the electric field shielding effect of the mobile ions. Notably, the curves of **Figure 4**a,b can be considered quasi-steady-state solutions since little hysteresis effect was obtained in the simulations (see Section S6.1) and the broad range of $N_{ion}$ values that reproduce similar $J-V$ shapes suggest that further evidence should be considered.

The simulated IS spectra in short-circuit under constant 0.2 sun illumination intensity are presented in **Figure 4**c,d for the reference and HI samples, respectively. A common feature for both device types is that the impedance does not behave smoothly with the increase of $N_{ion}$. The highest impedance arcs in the Nyquist plot representation are obtained for the lowest and highest values of $N_{ion}$ whereas a minimum impedance is found for intermediate values. For $N_{ion}$ <10$^{15}$ cm$^{-3}$ the electric field screening is not sufficient to shield the built-in field and the conduction is hindered. On the other hand, for $N_{ion}$ >10$^{18}$ cm$^{-3}$ the excess mobile ions behave like a doping concentration that no longer shield but maximize the built-in field. In an intermediate range $10^{15}<N_{ion}<10^{18}$ cm$^{-3}$ the charge density profile of mobile ions is modified, which favors diffusion current and thus conductivity. This increase of diffusion current enhances the associated low frequency capacitance, which behaves opposite to the impedance: minimum capacitance values are found for the lowest and the highest $N_{ion}$ values (see Section S6).

For the IS spectra of the reference sample in **Figure 4**c, our simulations suggest mobile ion concentrations in the range $10^{17}<N_{ion}<10^{18}$ cm$^{-3}$. This is concluded from the proportion between the low-frequency and high-frequency arcs of the Nyquist plots. **Figure 2**d illustrates the experimental behavior of the IS spectra after equivalent circuit parameterization resulting in $R_{Lf}/R_{Hf} > 10$. Therefore, the simulations with a low-frequency semicircle smaller than that of the high frequency range in the Nyquist plot would not relate to the studied samples. Moreover, the IS spectra of the HI sample in **Figure 4**d is complicated by the appearance of the characteristic three-arcs feature in the Nyquist plot representation. Particularly, the closest qualitative resemblance between simulations and the experiments (see **Figure 2**e) was obtained for $N_{ion} \sim 10^{16}$ cm$^{-3}$. This may suggest that the absolute concentration of mobile ions had decreased in the HI and MAI samples, with respect to the reference sample. However, when considering the stability data in **Figure 3** and the IS spectra in **Figure 4**d, one can correlate the reactivity over time of operation under constant illumination with the increase of concentration of mobile ions in the perovskite layer. More importantly, an increase of the mobile ion mobility is inferred from the following simulations, which may account as a major contribution.

A narrow range of mobile ion concentrations has been found key to reproduce the experimental trends. However, the multi-dependance of the simulated $J-V$ curves and IS spectra on several parameters should be considered in detail. For instance, in sections S5.2 and S6.2, our simulations of IS spectra as a function of illumination intensity nicely reproduce the $R_{Lf} \propto P_{in}^{-1}$ and $C_{Lf} \propto P_{in}$ trends which collapse until $R_{sh}$ limits the transport in the experiments (see **Figure 2** and Section S3). Yet, spurious results with $R_{Lf}/R_{Hf}<1$ can be found when simulating IS spectra with low concentration of mobile ions and shunt resistance. Purposely, the effect of shunt resistance is analyzed in sections S5.3 and S6.3 where an exclusively resistive effect is identified. Shunt resistance do not modify the capacitance of the sample but instead alters the $FF$ of the $J-V$ curves and the $R_{Lf}/R_{Hf}$ ratio of the IS spectra for intermediate ranges of ion concentrations. Interestingly, the mobile ion concentration and shunt resistance modify $R_{Lf}$ and $R_{Hf}$ simultaneously, whereas the ion mobility ($\mu_{ion}$) of the perovskite only changes the low-frequency part of the spectra (see section S5.4).[48]

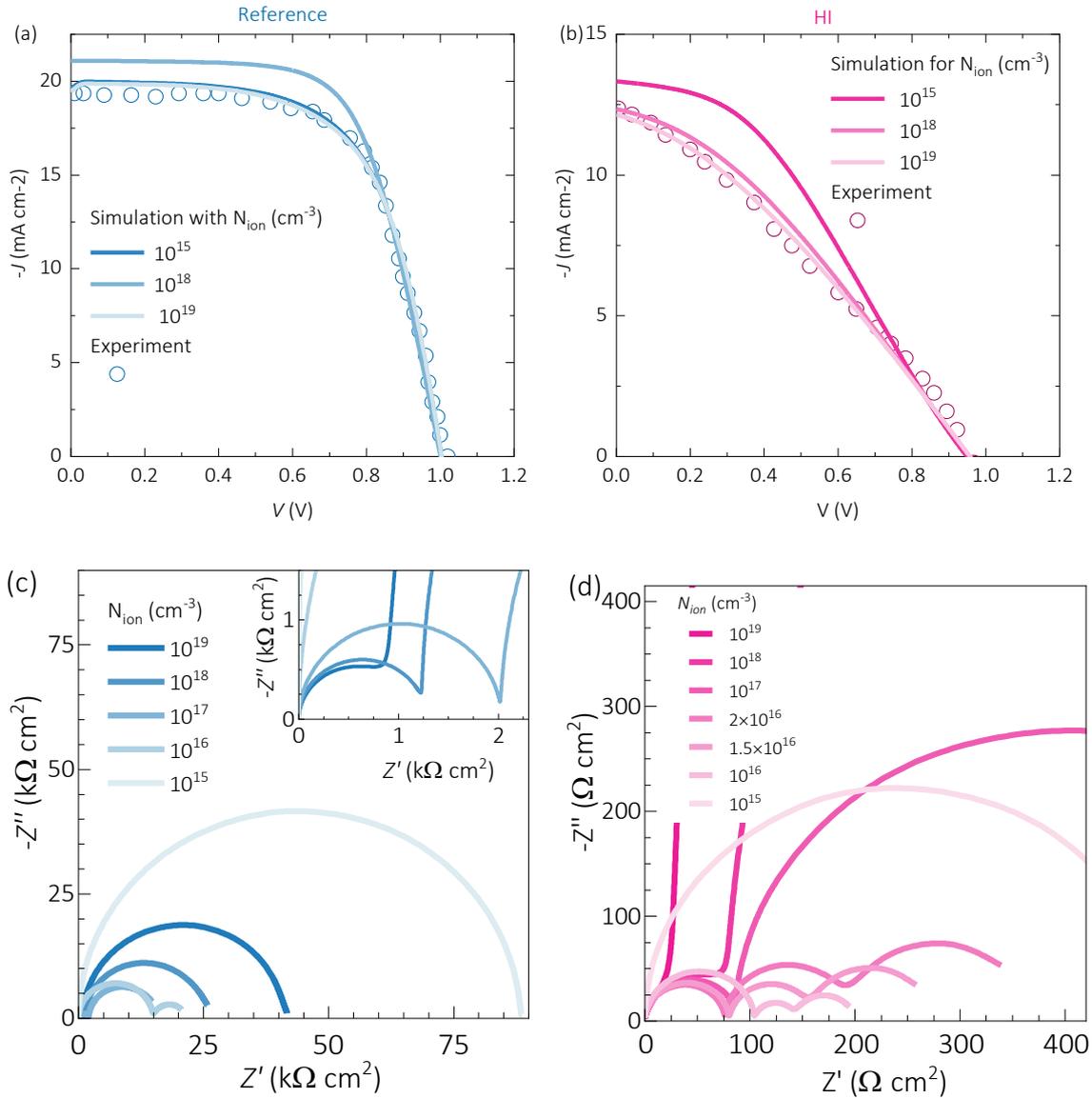

**Figure 4.** Simulated (a,b) current-voltage under 1 sun illumination intensity and (c, d) impedance spectra in short-circuit under 0.2 sun illumination intensity for (a,c) the reference and (b,d) the HI devices considering different mobile ion concentrations, as indicated. The simulation were done with MATLAB's Driftfusion[42] code with details as described in Section S6.

Charge carrier recombination is also closely linked with the mobile ion concentration and the electrical response of the samples. However, care must be taken with the model employed for the simulations. For instance, the *Setfos*-software by Fluxim utilizes the recombination velocity at the interface between the NiO$_x$ HTL and the perovskite. With this model, Section S5.5 shows an impedance decrease with the augment of the recombination velocity with a more significant modification of the low-frequency part of the spectrum. Since a large contribution to the current throughout the device in short-circuit is due to recombination, the higher the surface recombination velocity ($v_s$), the higher the leakage current and the smaller the impedance. Alternatively, the MATLAB's code *Driftfusion* simulates the interfaces with narrow interlayers where a surface recombination lifetime $\tau_s$ is set, as in section S6.5. Our simulations with this approach show analogous reduction of the impedance with the decrease of the $\tau_s$, but a clear dependency on $R_{sh}$ and $N_{ion}$ is also identified. For instance, the capacitance always peaks a maximum for intermediate values of $N_{ion}$, regardless of $R_{sh}$ or $\tau_s$. Similarly, when $R_{sh}$ is sufficiently small that controls the transport in SC, the total impedance only varies the $R_{Lf}/R_{Hf}$ ratio and behaves steady regardless of $N_{ion}$ or $\tau_s$. Moreover, for high $R_{sh}$ and all the calculated interface $\tau_s$ values, the impedance reaches a minimum for intermediate values of $N_{ion}$ which often includes spurious $R_{Lf}/R_{Hf}>1$ ratios. Furthermore, decreasing bulk recombination lifetime ($\tau_b$) in the perovskite layer similarly reduces the impedance, as presented in Section S6.6. However, unlike $\tau_s$ whose decrease slightly declines $J_{sc}$ and $FF$, the reduction of $\tau_b$ implies a large lessening of $J_{sc}$, $FF$ and even the $V_{oc}$. From all these simulations, capacitance and resistance are shown to peak maximum and minimum values, respectively, around specific threshold mobile ion concentration as schemed in **Figure 5**c. In this figure, we also highlight the inverse relation between the $PCE$ and $N_{ion}$: the higher the mobile ion density, the smaller the device efficiency. Accordingly, and assuming constant ion mobility, degradation routes with decrease or increase of device performance can be outlined in terms of the resistance and capacitance in short-circuit condition, (see arrows in **Figure 5**c).

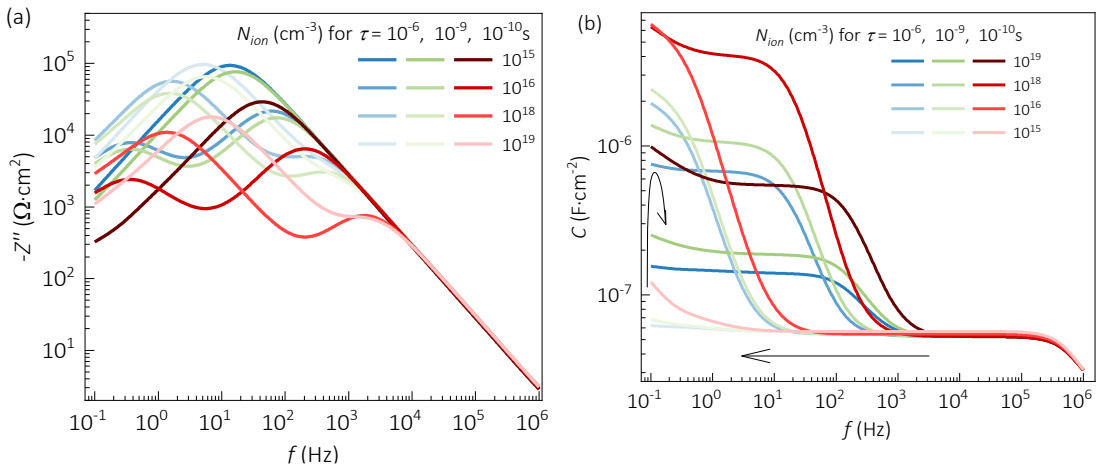

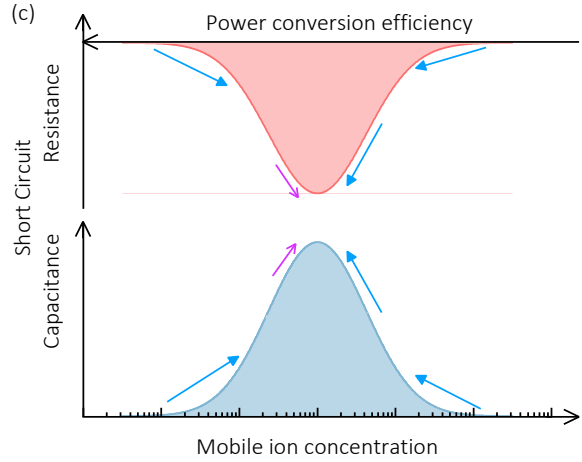

**Figure 5.** Simulated impedance spectra in short-circuit for different concentrations of mobile ions ($N_{ion}$= $10^{15}$, $10^{16}$, $10^{18}$, $10^{19}$ cm$^{-3}$) and trap recombination lifetime values at the interface ($\tau = 10^{-6}$, $10^{-9}$, $10^{-10}$ s) considering a shunt resistance of $R_{sh}$ = 1 MΩ in imaginary impedance (a) and (b) capacitance Bode plots. In (c) there is a schematic on these trends for the reference and optimized devices including degradation paths with PCE increase (purple open arrows) and decrease (blue filled arrows), assuming constant ion mobility.

The effects of the electron and hole mobilities ($\mu_{e,h}$) in the perovskite were also simulated, as displayed in Section S6.7, and no simple trend is observed. Instead, a multi-dependent relationship is found whereby the concentration of mobile ions increases or decreases the impedance and the capacitance in different ranges. Remarkably, the perovskite electronic mobility has a high impact on the $J_{sc}$ and $FF$ of the $J-V$ curve, although mostly negligible on the $V_{oc}$. A similar behavior is also found for the hole mobility ($\mu_h$) at the HTL in terms of the variated relation of the impedance and capacitance dependence on the mobility and the concentration of mobile ions. In contrast, the change of mobility of the NiO$_x$ layer (see Section S6.8) is mostly affecting the $FF$, whereas the $J_{sc}$ and the $V_{oc}$ are mostly independent. Last but not least, the effect of the perovskite dielectric constant ($\varepsilon$) was also investigated in section S6.9. It is shown that the increase of the dielectric permittivity of the absorber layer reduces the impedance and increases the capacitance in the high-frequency range where the geometrical contribution is dominant.

The multiple relations between simulation parameters and the behaviors of the IS spectra in short-circuit condition and the $J-V$ curves under illumination are summarized in **Table 1** with a comprehensible "arrow-based" scheme. It is highlighted, that while some parameters are straightforwardly related, some others present complicated dependencies that hinder the interpretation of the experimental data in terms of more simplistic models.

**Table 1.** Summary of different effects on the impedance spectra in short-circuit and current-voltage characteristic due to the modification of simulation parameters. The up/down arrows indicate direct or inverse proportionality, respectively. Several arrows indicate intensity of dependence when in the same direction or complex multi-dependent relation when in different directions.

| Parameter | IS spectra in SC | | $J-V$ characteristic | | | Sections in SI |
|---|---|---|---|---|---|---|
| | Impedance | Capacitance | $J_{sc}$ | FF | $V_{oc}$ | |
| $N_{ion}$ | ↓↑ | ↑↓ | ↓ | ↓ | ↓ | S5.1, S6.1 |
| $\mu_{ion}$ | ↑ | - | | | | S5.4 |
| $P_{in}$ | ↓ | ↑ | ↑ | ↑ | ↑ | S5.2, S6.2 |
| $R_{sh}$ | ↑ | - | - | ↑ | ↑ | S5.3, S6.3 |
| $v_s$ | ↑↑↓ | ↑↑↓ | ↑ | ↑↑ | - | S5.5 |
| $\tau_s$ | ↓ | ↓ | ↓ | ↓ | - | S6.5 |
| $\tau_b$ | ↓ | ↓ | ↓ | ↓ | ↓ | S6.6 |
| $\mu_{e,h}$ (perovskite) | ↓↓↑ | ↑↑↓ | ↑↑ | ↑ | - | S6.7 |
| $\varepsilon$ (perovskite) | ↓ | ↑ | | | | S6.9 |
| $\mu_h$ (HTL) | ↑↑↓ | ↑↑↓ | - | ↓ | - | S6.8 |

## 4. Conclusions

This work summarizes an optoelectronic characterization of perovskite solar cells with different passivation treatments on the interface between the perovskite absorber material and the NiO$_x$ hole transport layer. From the initial assessment of the device performance and the subsequent operational stability test two main trends where identified. The untreated reference sample and those utilizing 1-iodobutane, 1-phenylethylamine and PbI2 as passivators resulted in optimal and rather stable performances. In contrast, the MAI- and HI-treaded samples not only resulted in low fill factor and photocurrent but also evidences substantial instability with an apparent improvement of the output power up to three times the initial value during the operational stability test.

The use of impedance spectroscopy analysis in short-circuit condition under different illumination intensities and over time during operational stability test have been introduced, discussed and carried out as a resourceful procedure for understanding the electrical response of solar cells. Several practical and theoretical advantages of this approach have been discussed over the more common open-circuit/flat-band and/or MPP studies. The experimental spectra showed that the untreated and optimized devices behave in a more typical way where two main $RC$ constants produce two arcs in the impedance Nyquist representation. By using equivalent circuit modelling the resistances for the low and high frequency ranges of spectra can be identified as $R_{Lf} \gg R_{Hf}$. Characteristically, the MAI and HI treated samples evidenced up to three $RC$ constant with an extra middle frequency feature that apparently merged with the low frequency part of the spectra by reduction of the illumination intensity or exposure time during the operational stability test. Notably, the continuous switching between the operational forward bias and the short-circuit conditions was found to break the cells, which could be explored as a reliability test protocol in future works.

Our drift-diffusion simulations were able to qualitatively reproduce most of the experimental behaviors for the current-voltage curves and the impedance spectra. The explored simulation parameters suggest that the HI and MAI treatments on the NiO$_x$ HTL reduced the device absorptance, which could be associated with an interface morphology change that increases photon reflectivity and electrical series resistance. At the same time, the apparently damaged HTL experienced a decrease in the electronic mobility. Moreover, the ionic mobility and concentrations of the perovskite were also found to be modified with respect to the values in the reference device. This work not only presents a comprehensible analysis on our study case, but also explore a wide range of simulation parameters and its simultaneous effects on the J-V curve and the IS spectra. These simulations results can be used to further comprehend the working mechanism of the perovskite solar cells, their ionic properties and stability.

**Authors Information**

**Corresponding Authors**

Osbel Almora — Department of Electronic, Electric and Automatic Engineering, Universitat Rovira i Virgili, 43007 Tarragona, Spain; https://orcid.org/0000-0002-2523-0203 ; Email: osbel.almora@urv.cat
Pilar López-Varo — Institut Photovoltaïque d'Île-de-France (IPVF), 91120 Palaiseau, France; https://orcid.org/0000-0002-2170-1581 ; Email : pilar.lopez-varo@ipvf.fr
Juan A. Anta — Department of Physical, Chemical and Natural Systems, Universidad Pablo de Olavide, Sevilla 41013, Spain; https://orcid.org/0000-0002-8002-0313; Email: jaantmon@upo.es

**Authors**

Renán Escalante — Department of Physical, Chemical and Natural Systems, Universidad Pablo de Olavide, Sevilla 41013, Spain; https://orcid.org/0000-0002-5100-5448; Email: raescqui1@upo.es
Lluis F. Marsal — Department of Electronic, Electric and Automatic Engineering, Universitat Rovira i Virgili, 43007 Tarragona, Spain; https://orcid.org/0000-0002-5976-1408; Email: lluis.marsal@urv.cat

**Notes**

The authors declare no competing financial interests.

## Acknowledgment

S. Mohamed acknowledges the financial support from Programa Martí i Franquès. M. Ramírez-Como acknowledges the financial support from Diputació de Tarragona under Grant 2021CM14 and 2022PGR-DIPTA-URV04. This work was further supported by the Spanish Ministerio de Ciencia e Innovación (MICINN/FEDER) under Grants PDI2021-128342OB-I00 and RTI2018-094040-B-I00, by the Agency for Management of University and Research Grants (AGAUR) ref. 2017-SGR-1527, and from the Catalan Institution for Research and Advanced Studies (ICREA) under the ICREA Academia Award. O.A. thanks the National Research Agency (Agencia Estatal de Investigación) of Spain for the Juan de la Cierva 2021 grant (FJC2021-046887-I). P.LV. thanks the French Government in the frame of the program of investment for the future (Programme d'Investissement d'Avenir – ANR-IEED-002-01). J.M. and S.O. thank the Ministry of Economic Affairs Innovation, Digitalization and Energy of the State of North Rhine-Westphalia for funding under the grant SCALEUP (SOLAR-ERA.NET Cofund 2, id: 32).## References

[1] Osbel Almora, Carlos I. Cabrera, Sule Erten-Ela, Karen Forberich, Kenjiro Fukuda, Fei Guo, Jens Hauch, Anita W. Y. Ho-Baillie, T. Jesper Jacobsson, Rene A. J. Janssen, Thomas Kirchartz, Maria A. Loi, Xavier Mathew, David B. Mitzi, Mohammad K. Nazeeruddin, Ulrich W. Paetzold, Barry P. Rand, Uwe Rau, Takao Someya, Eva Unger, Lídice Vaillant-Roca, Christoph J. Brabec, Device Performance of Emerging Photovoltaic Materials (Version 4), *Adv. Energy Mater.* **2024**, 14, 2303173, https://doi.org/10.1002/aenm.202303173

[2] Giulia Pacchioni, Highly efficient perovskite LEDs, *Nat. Rev. Mater.* **2021**, 6, 108, https://doi.org/10.1038/s41578-021-00280-5

[3] Jiyoung Moon, Yash Mehta, Kenan Gundogdu, Franky So, Qing Gu, Metal-Halide Perovskite Lasers: Cavity Formation and Emission Characteristics, *Adv. Mater.* **2023**, n/a, 2211284, https://doi.org/10.1002/adma.202211284

[4] Agustín O. Álvarez, Ferdinand Lédée, Marisé García-Batlle, Pilar López-Varo, Eric Gros-Daillon, Javier Mayén Guillén, Jean-Marie Verilhac, Thibault Lemercier, Julien Zaccaro, Lluis F. Marsal, Germà Garcia-Belmonte, Osbel Almora, Ionic field screening in MAPbBr$_3$ crystals from remnant sensitivity in X-ray detection, *ACS Phys. Chem. Au* **2023**, 3, 386, https://doi.org/10.1021/acsphyschemau.3c00002

[5]     Jing Zhou, You Gao, Yongyan Pan, Fumeng Ren, Rui Chen, Xin Meng, Derun Sun, Jizhou He, Zonghao Liu, Wei Chen, Recent Advances in the Combined Elevated Temperature, Humidity, and Light Stability of Perovskite Solar Cells, *Solar RRL* **2022**, 6, 2200772, https://doi.org/10.1002/solr.202200772

[6]     Subrata Ghosh, Snehangshu Mishra, Trilok Singh, Antisolvents in Perovskite Solar Cells: Importance, Issues, and Alternatives, *Adv. Mater. Interfaces* **2020**, 7, 2000950, https://doi.org/10.1002/admi.202000950

[7]     Anatolii Belous, Sofiia Kobylianska, Oleg V'yunov, Pavlo Torchyniuk, Volodymyr Yukhymchuk, Oleksandr Hreshchuk, Effect of non-stoichiometry of initial reagents on morphological and structural properties of perovskites $CH_3NH_3PbI_3$, *Nanoscale Research Letters* **2019**, 14, 4, https://doi.org/10.1186/s11671-018-2841-6

[8]     You Gao, Fumeng Ren, Derun Sun, Sibo Li, Guanhaojie Zheng, Jianan Wang, Hasan Raza, Rui Chen, Haixin Wang, Sanwan Liu, Peng Yu, Xin Meng, Jizhou He, Jing Zhou, Xiaodong Hu, Zhengping Zhang, Longbin Qiu, Wei Chen, Zonghao Liu, Elimination of unstable residual lead iodide near the buried interface for the stability improvement of perovskite solar cells, *Energy Environ. Sci.* **2023**, 16, 2295, https://doi.org/10.1039/D3EE00293D

[9]     Kusuma Pinsuwan, Chirapa Boonthum, Thidarat Supasai, Somboon Sahasithiwat, Pisist Kumnorkaew, Pongsakorn Kanjanaboos, Solar perovskite thin films with enhanced mechanical, thermal, UV, and moisture stability via vacuum-assisted deposition, *J. Mater. Sci.* **2020**, 55, 3484, https://doi.org/10.1007/s10853-019-04199-9

[10]   F. Valipour, E. Yazdi, N. Torabi, B. F. Mirjalili, A. Behjat, Improvement of the stability of perovskite solar cells in terms of humidity/heat via compositional engineering, *J. Phys. D: Appl. Phys.* **2020**, 53, 285501, https://doi.org/10.1088/1361-6463/ab8511

[11]   Dongchen Lan, Martin A. Green, Combatting temperature and reverse-bias challenges facing perovskite solar cells, *Joule* **2022**, 6, 1782, https://doi.org/10.1016/j.joule.2022.06.014

[12]   Anoop K. M, Mark V. Khenkin, Francesco Di Giacomo, Yulia Galagan, Stav Rahmany, Lioz Etgar, Eugene A. Katz, Iris Visoly-Fisher, Bias-Dependent Stability of Perovskite Solar Cells Studied Using Natural and Concentrated Sunlight, *Solar RRL* **2020**, 4, 1900335, https://doi.org/10.1002/solr.201900335

[13]   Mark V. Khenkin, Anoop K. M, Eugene A. Katz, Iris Visoly-Fisher, Bias-dependent degradation of various solar cells: lessons for stability of perovskite photovoltaics, *Energy Environ. Sci.* **2019**, 12, 550, https://doi.org/10.1039/C8EE03475C

[14]   Kristijan Brecl, Marko Jošt, Matevž Bokalič, Jernej Ekar, Janez Kovač, Marko Topič, Are Perovskite Solar Cell Potential-Induced Degradation Proof?, *Solar RRL* **2022**, 6, 2100815, https://doi.org/10.1002/solr.202100815

[15]   M. I. El-Henawey, Istiaque M. Hossain, Liang Zhang, Behrang Bagheri, Ranjith Kottokkaran, Vikram L. Dalal, Influence of grain size on the photo-stability of perovskite solar cells, *J. Mater. Sci.: Mater. Electron.* **2021**, 32, 4067, https://doi.org/10.1007/s10854-020-05148-y

[16]   Arthur Julien, Jean-Baptiste Puel, Jean-François Guillemoles, Distinction of mechanisms causing experimental degradation of perovskite solar cells by simulating associated pathways, *Energy Environ. Sci.* **2023**, 16, 190, https://doi.org/10.1039/D2EE03377A

[17]   Hee Joon Jung, Daehan Kim, Sungkyu Kim, Joonsuk Park, Vinayak P. Dravid, Byungha Shin, Stability of Halide Perovskite Solar Cell Devices: In Situ Observation of Oxygen Diffusion under Biasing, *Advanced Materials* **2018**, 30, 1802769, https://doi.org/10.1002/adma.201802769

[18]   Stijn Lammar, Renán Escalante, Antonio J. Riquelme, Sandra Jenatsch, Beat Ruhstaller, Gerko Oskam, Tom Aernouts, Juan A. Anta, Impact of non-stoichiometry on ion migration and photovoltaic performance of formamidinium-based perovskite solar cells, *J. Mater. Chem. A* **2022**, 10, 18782, https://doi.org/10.1039/D2TA04840J

[19]   Yong Huang, Pilar Lopez-Varo, Bernard Geffroy, Heejae Lee, Jean-Eric Bourée, Arpit Mishra, Philippe Baranek, Alain Rolland, Laurent Pedesseau, Jean-Marc Jancu, Jacky Even, Jean-Baptiste Puel, Marie Gueunier-Farret, Detrimental effects of ion migration in the perovskite and hole transport layers on the efficiency of inverted perovskite solar cells, *J. Photon. Energy* **2020**, 10, 024502, https://doi.org/10.1117/1.JPE.10.024502

[20]   Xin Yan, Wenqiang Fan, Feiyu Cheng, Haochun Sun, Chenzhe Xu, Li Wang, Zhuo Kang, Yue Zhang, Ion migration in hybrid perovskites: Classification, identification, and manipulation, *Nano Today* **2022**, 44, 101503, https://doi.org/10.1016/j.nantod.2022.101503

[21]   Christopher Eames, Jarvist M. Frost, Piers R. F. Barnes, Brian C. O'Regan, Aron Walsh, M. Saiful Islam, Ionic Transport in Hybrid Lead Iodide Perovskite Solar Cells, *Nat. Commun.* **2015**, 6, 7497, https://doi.org/10.1038/ncomms8497

[22]   Kostiantyn Sakhatskyi, Rohit Abraham John, Antonio Guerrero, Sergey Tsarev, Sebastian Sabisch, Tisita Das, Gebhard J. Matt, Sergii Yakunin, Ihor Cherniukh, Martin Kotyrba, Yuliia Berezovska, Maryna I. Bodnarchuk, Sudip Chakraborty, Juan Bisquert, Maksym V. Kovalenko, Assessing the Drawbacks and Benefits of Ion Migration in Lead Halide Perovskites, *ACS Energy Lett.* **2022**,

7, 3401, https://doi.org/10.1021/acsenergylett.2c01663

[23]　Qi Jiang, Jinhui Tong, Yeming Xian, Ross A. Kerner, Sean P. Dunfield, Chuanxiao Xiao, Rebecca A. Scheidt, Darius Kuciauskas, Xiaoming Wang, Matthew P. Hautzinger, Robert Tirawat, Matthew C. Beard, David P. Fenning, Joseph J. Berry, Bryon W. Larson, Yanfa Yan, Kai Zhu, Surface reaction for efficient and stable inverted perovskite solar cells, *Nature* **2022**, https://doi.org/10.1038/s41586-022-05268-x

[24]　Sanwan Liu, Vasudevan P. Biju, Yabing Qi, Wei Chen, Zonghao Liu, Recent progress in the development of high-efficiency inverted perovskite solar cells, *NPG Asia Mater.* **2023**, 15, 27, https://doi.org/10.1038/s41427-023-00474-z

[25]　Xuesong Lin, Danyu Cui, Xinhui Luo, Caiyi Zhang, Qifeng Han, Yanbo Wang, Liyuan Han, Efficiency progress of inverted perovskite solar cells, *Energy Environ. Sci.* **2020**, 13, 3823, https://doi.org/10.1039/D0EE02017F

[26]　Bowei Li, Wei Zhang, Improving the stability of inverted perovskite solar cells towards commercialization, *Commun. Mater.* **2022**, 3, 65, https://doi.org/10.1038/s43246-022-00291-x

[27]　Ahmed Ali Said, Jian Xie, Qichun Zhang, Recent Progress in Organic Electron Transport Materials in Inverted Perovskite Solar Cells, *Small* **2019**, 15, 1900854, https://doi.org/10.1002/smll.201900854

[28]　Xingyu Pu, Junsong Zhao, Yongjiang Li, Yixin Zhang, Hok-Leung Loi, Tong Wang, Hui Chen, Xilai He, Jiabao Yang, Xiaoyan Ma, Xuanhua Li, Qi Cao, Stable $NiO_x$-based inverted perovskite solar cells achieved by passivation of multifunctional star polymer, *Nano Energy* **2023**, 112, 108506, https://doi.org/10.1016/j.nanoen.2023.108506

[29]　Haoxin Wang, Wei Zhang, Biyi Wang, Zheng Yan, Cheng Chen, Yong Hua, Tai Wu, Linqin Wang, Hui Xu, Ming Cheng, Modulating buried interface with multi-fluorine containing organic molecule toward efficient $NiO_x$-based inverted perovskite solar cell, *Nano Energy* **2023**, 111, 108363, https://doi.org/10.1016/j.nanoen.2023.108363

[30]　X. Cai, T. Hu, H. Hou, P. Zhu, R. Liu, J. Peng, W. Luo, H. Yu, A review for nickel oxide hole transport layer and its application in halide perovskite solar cells, *Mater. Today Sust.* **2023**, 23, 100438, https://doi.org/10.1016/j.mtsust.2023.100438

[31]　Haibin Wang, Zhiyin Qin, XinJian Li, Chun Zhao, Chao Liang, High-Performance Inverted Perovskite Solar Cells with Sol–Gel-Processed Sliver-Doped NiOX Hole Transporting Layer, *Energy Envirom. Mater.* **2023**, n/a,

e12666, https://doi.org/10.1002/eem2.12666

[32] Xin Yin, Lixin Song, Pingfan Du, Bingang Xu, Jie Xiong, Cation exchange strategy to construct nanopatterned Zn:NiO$_x$ electrode with highly conductive interface for efficient inverted perovskite solar cells, *Chem. Eng. J.* **2023**, 457, 141358, https://doi.org/10.1016/j.cej.2023.141358

[33] Taotao Hu, Hongming Hou, Jin Peng, Qiaofeng Wu, Jialong He, Hua Yu, Rui Liu, Tian Hou, Xiangqing Zhou, Meng Zhang, Xiaolong Zhang, Xinchun Yang, Yuanmiao Sun, Xuanhua Li, Yang Bai, 4-tert-butylpyridine induced Ni$^{3+}$/Ni$^{2+}$ ratio modulation in NiO$_x$ hole transport layer towards efficient and stable inverted perovskite solar cells, *Mater. Today Energy* **2023**, 32, 101245, https://doi.org/10.1016/j.mtener.2023.101245

[34] Xinxin Kang, Dourong Wang, Kun Sun, Xue Dong, Hui Wei, Baohua Wang, Lei Gu, Maoxin Li, Yaqi Bao, Jie Zhang, Renjun Guo, Zerui Li, Xiongzhuo Jiang, Peter Müller-Buschbaum, Lin Song, Unraveling the Modification Effect at NiO$_x$/perovskite interfaces for Efficient and Stable Inverted Perovskite Solar Cells, *J. Mater. Chem. A* **2023**, https://doi.org/10.1039/D3TA05069F

[35] Guibin Shen, Hongye Dong, Fan Yang, Xin Ren Ng, Xin Li, Fen Lin, Cheng Mu, Application of an amphipathic molecule at the NiO$_x$/perovskite interface for improving the efficiency and long-term stability of the inverted perovskite solar cells, *J. Energy Chem.* **2023**, 78, 454, https://doi.org/10.1016/j.jechem.2022.12.015

[36] Osbel Almora, Daniel Miravet, Ilario Gelmetti, Germà Garcia-Belmonte, Long-term Field Screening by Mobile Ions in Thick Metal Halide Perovskites: Understanding Saturation Currents, *Phys. Status Solidi RRL* **2022**, 16, 202200336, https://doi.org/10.1002/pssr.202200336

[37] Isaac Zarazua, Guifang Han, Pablo P. Boix, Subodh Mhaisalkar, Francisco Fabregat-Santiago, Ivan Mora-Seró, Juan Bisquert, Germà Garcia-Belmonte, Surface Recombination and Collection Efficiency in Perovskite Solar Cells from Impedance Analysis, *J. Phys. Chem. Lett.* **2016**, 7, 5105, https://doi.org/10.1021/acs.jpclett.6b02193

[38] Osbel Almora, Germà Garcia-Belmonte, Light Capacitances in Silicon and Perovskite Solar Cells, *Solar Energy* **2019**, 189, 103, https://doi.org/10.1016/j.solener.2019.07.048

[39] Osbel Almora, Yicheng Zhao, Xiaoyan Du, Thomas Heumueller, Gebhard J. Matt, Germà Garcia-Belmonte, Christoph J. Brabec, Light Intensity Modulated Impedance Spectroscopy (LIMIS) in All-Solid-State Solar Cells at Open-Circuit, *Nano Energy* **2020**, 75, 104982, https://doi.org/10.1016/j.nanoen.2020.104982

[40] Elnaz Ghahremanirad, Osbel Almora, Sunil Suresh, Amandine A. Drew, Towhid H. Chowdhury, Alexander R. Uhl, Beyond Protocols: Understanding the Electrical Behavior of Perovskite Solar Cells by Impedance Spectroscopy, *Adv. Energy Mater.* **2023**, 2204370, https://doi.org/10.1002/aenm.202204370

[41] SETFOS: Simulation Software for Organic and Perovskite Solar Cells and LEDs, https://www.fluxim.com/setfos-intro, accessed: 15.11.2023.

[42] Philip Calado, Ilario Gelmetti, Benjamin Hilton, Mohammed Azzouzi, Jenny Nelson, Piers R. F. Barnes, Driftfusion: an open source code for simulating ordered semiconductor devices with mixed ionic-electronic conducting materials in one dimension, *J. Comput. Electron.* **2022**, 21, 960, https://doi.org/10.1007/s10825-021-01827-z

[43] Rodrigo García-Rodríguez, Antonio J. Riquelme, Matthew Cowley, Karen Valadez-Villalobos, Gerko Oskam, Laurence J. Bennett, Matthew J. Wolf, Lidia Contreras-Bernal, Petra J. Cameron, Alison B. Walker, Juan A. Anta, Inverted Hysteresis in n–i–p and p–i–n Perovskite Solar Cells, *Energy Techn.* **2022**, 10, 2200507, https://doi.org/10.1002/ente.202200507

[44] Caleb C. Boyd, R. Clayton Shallcross, Taylor Moot, Ross Kerner, Luca Bertoluzzi, Arthur Onno, Shalinee Kavadiya, Cullen Chosy, Eli J. Wolf, Jérémie Werner, James A. Raiford, Camila de Paula, Axel F. Palmstrom, Zhengshan J. Yu, Joseph J. Berry, Stacey F. Bent, Zachary C. Holman, Joseph M. Luther, Erin L. Ratcliff, Neal R. Armstrong, Michael D. McGehee, Overcoming Redox Reactions at Perovskite-Nickel Oxide Interfaces to Boost Voltages in Perovskite Solar Cells, *Joule* **2020**, 4, 1759, https://doi.org/10.1016/j.joule.2020.06.004

[45] Osbel Almora, Gebhard J. Matt, Albert These, Andrii Kanak, Ievgen Levchuk, Shreetu Shrestha, Andres Osvet, Christoph J. Brabec, Germà Garcia-Belmonte, Surface versus Bulk Currents and Ionic Space-Charge Effects in $CsPbBr_3$ Single Crystals, *J. Phys. Chem. Lett.* **2022**, 13, 3824, https://doi.org/10.1021/acs.jpclett.2c00804

[46] Wenya Song, Xin Zhang, Stijn Lammar, Weiming Qiu, Yinghuan Kuang, Bart Ruttens, Jan D'Haen, Inge Vaesen, Thierry Conard, Yaser Abdulraheem, Tom Aernouts, Yiqiang Zhan, Jef Poortmans, Critical Role of Perovskite Film Stoichiometry in Determining Solar Cell Operational Stability: a Study on the Effects of Volatile A-Cation Additives, *ACS Appl. Mater. Interfaces* **2022**, 14, 27922, https://doi.org/10.1021/acsami.2c05241

[47] Sandheep Ravishankar, Zhifa Liu, Uwe Rau, Thomas Kirchartz, Multilayer Capacitances: How Selective Contacts Affect Capacitance Measurements of Perovskite Solar Cells, *PRX Energy* **2022**, 1, 013003, https://doi.org/10.1103/PRXEnergy.1.013003

[48] Antonio Riquelme, Laurence J. Bennett, Nicola E. Courtier, Matthew J. Wolf, Lidia Contreras-Bernal, Alison B. Walker, Giles Richardson, Juan A. Anta, Identification of recombination losses and charge collection efficiency in a perovskite solar cell by comparing impedance response to a drift-diffusion model, *Nanoscale* **2020**, 12, 17385, https://doi.org/10.1039/D0NR03058A

*Supporting information:*

# Degradation Analysis of Perovskite Solar Cells via Short-Circuit Impedance Spectroscopy: A case study on NiO$_x$ passivation

Osbel Almora,[1,*] Pilar López-Varo,[2,*] Renán Escalante,[3] John Mohanraj,[4] Luis F Marsal,[1] Selina Olthof,[4] Juan A. Anta[3,*]

[1] Departament d'Enginyeria Electrònica Elèctrica i Automàtica, Universitat Rovira i Virgili, 43007 Tarragona, Spain
[2] Institut Photovoltaïque d'Ile-de-France (IPVF), 91120 Palaiseau, France
[3] Department of Physical, Chemical and Natural Systems, Universidad Pablo de Olavide, Sevilla 41013, Spain
[4] Department of Chemistry, University of Cologne, Greinstrasse 4−6, Cologne 50939, Germany

* osbel.almora@urv.cat, pilar.lopez-varo@ipvf.fr, jaantmon@upo.es

# S1. Introduction

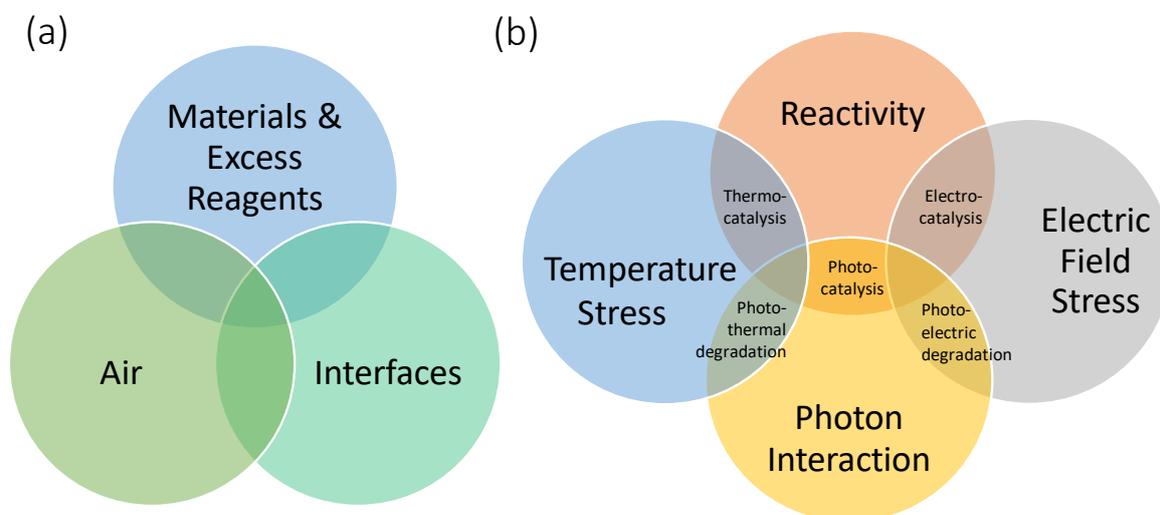

**Figure S1.** Schemed interconnection of different degradation (a) elements and (b) mechanisms.

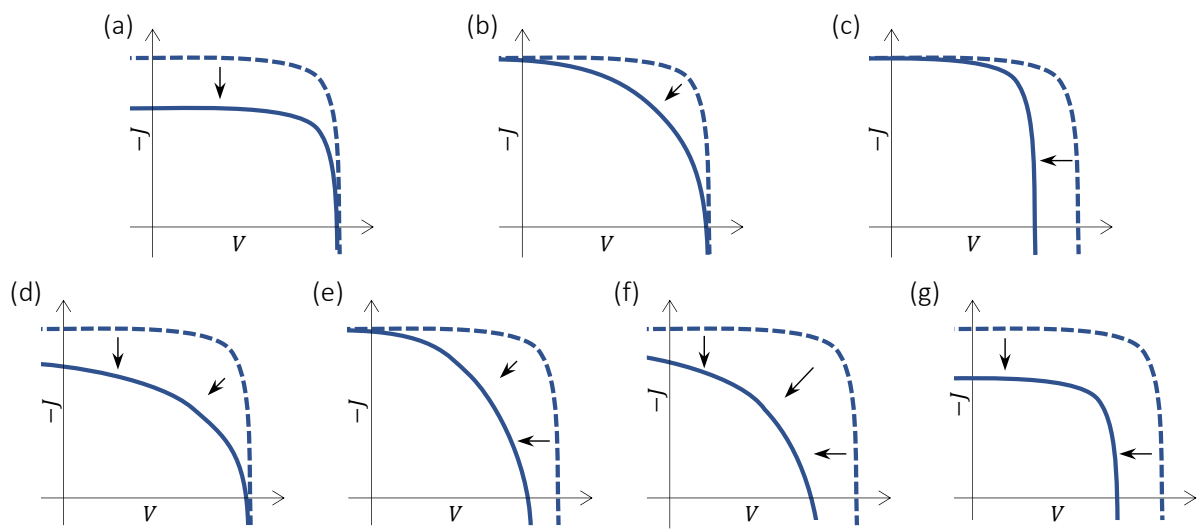

**Figure S2.** Schemed different types of degradation effects on the current-voltage curve: sole or significantly dominant decrease of the (a) short-circuit current density ($J_{sc}$), (b) fill factor ($FF$) and, (c) open-circuit voltage ($V_{oc}$); and combined reduction of (d) $J_{sc} + FF$, (e) $V_{oc} + FF$, (f) $V_{oc} + J_{sc} + FF$, and (f) $V_{oc} + J_{sc}$.

**Table S1.** Dependency of shunt resistance ($R_{sh}$) and/or shor-circuit resistance as a function of illumination intensity ($P_{in}$).

| Absorber material | Dependency | Ref. | Comment |
|---|---|---|---|
| c-Si | $R_{sc} \propto P_{in}^{-1.13}$ | [1] | |
| a-Si:H | $R_{sc} \propto J_{sc}^{-1}$ | [2] | AM1.5G spectrum |
| Si | $R_{sh} = R_0 - R_1 \left(\frac{P_{in}}{P_0}\right)^{1.08}$ | [3] | Polycrystalline silicon; $R_0$, $R_1$ and $P_0$ are independent of $P_{in}$ |
| c-Si | $R_{sh} = \frac{R_0}{1 + \frac{P_0}{P_{in}}}$ | [4] | $R_0$ and $P_0$ are independent of $P_{in}$ |
| CdTe | $R_{sh} \propto P_{in}^{-0.86}$ | [5] | |
| PBDB-T:F-M | $R_p \propto P_{in}^{-1}$ | [6] | |

**Table S2.** Comparative summary between flat-band and short-circuit conditions in perovskite solar cells.

| Properties | Flat-band (bias close to) | Short-circuit |
|---|---|---|
| PCE related parameters | $V_{oc}$, FF, $V_{mpp}$ | $J_{sc}$ |
| Light effect on energy diagram and charge density profile | Low | High |
| Drift versus diffusion contributions | Diffusion>Drift | Drift >>Diffusion |
| Recombination mechanism | Radiative >>Nonradiative | Nonradiative >>Radiative |
| Mobile ion distribution in the perovskite | Bulk | Mostly towards interface |
| Electrical response time | Slower | Faster |
| Current linearity | Lower | Higher |
| Parasitic resistance relation | $R_s$ | $R_{sh}$ |

## S2.  Device fabrication and initial performance characterization.

The description below is adapted from our simultaneous work [ref].

### S2.1.  Material acquisition

For MAPbI$_3$ layer deposition, PbI$_2$ and MAI were purchased from TCI Deutschland GmbH. The dry solvents dimethylformamide ($\geq$ 99.9%), chlorobenzene, ethanol (99.9%), isopropanol (Honeywell 99.9%), and ethanolamine were purchased from Sigma-Aldrich and used as received. Nickel(II) acetate tetrahydrate, HI (57 wt%), Li-TFSI, 1-phenylethylamine, 1,2,-ethanedithiol, 1-iodobutane, C$_{60}$ and bathocuproine were also purchased from Sigma-Aldrich or Merck KGaA. The FTO-coated glass substrates (TEC 10, S2002S1) used for NiO$_x$ deposition as well as solar cells preparation were procured from Ossila BV, The Netherlands.

### S2.2.  Preparation of NiO$_x$ thin films

First, 400 mg of nickel (II) aetate tetrahydrate and 97 µL of ethanolamine were mixed in 3.2 mL of ethanol, resulting in 0.5 M solution. This solution was stirred for at 2 hours (minimum) at room temperature to obtain a transparent dark blue-greenish colored solution. Subsequently, the solution was filtered with a 0.2 µm PTFE filter, and then 100 µL of it was spin-coated on clean FTO (TEC 10) substrates at 3000 rpm for 30 s. Immediately after, the substrates were annealed at 100 °C for 5 minutes inside a N$_2$ glovebox and then at 450 °C for 30 min in a furnace. Once the substrates were cooled to room temperature, they were transferred into a N$_2$ filled glovebox. Prior depositing any further layer on, the substrates were heated to 100 °C for 10 min.

### S2.3.  Preparation of MAPbI$_3$ thin films

MAPbI$_3$ films for degradation studies and solar cells were prepared by following the conventional antisolvent method. For degradation studies, 1 M of MAI and PbI$_2$ were dissolved in dimethylformamide by stirring the solution for at least 2 h at 50 °C inside a N$_2$ filled glovebox, and the solution was filtered with a 0.2 µm PTFE filter. This was used as a stock solution to prepare 0.5 M, 0.375 M, 0.25 and 0.1 M MAPbI$_3$ solutions by diluting with appropriate amount of dimethylformamide. For film preparation, 80 µL of MAPbI$_3$ solution of desired concentration was spin coated on substrates at 3000 rpm for 50 s. During this process, after 8 s, 200 µL of chlorobenzene was dripped continuously onto the substrates. The substrates were then annealed at 80 °C for 1 h on a hot plate inside a glovebox.

### S2.4. Deposition of surface passivation materials on NiO$_x$

In separate vials, 0.1 M solutions of Li-TFSI, 1-phenylethanol amine, 1-iodobutane, HI, MAI and 1,2-ethanedithiol were prepared in isopropanol and dimethylformamide was used as a solvent for 0.1 M solution of PbI$_2$. These solutions were stirred at room temperature and filtered with 0.2 μm PTFE filter. At that time, 80 μL of the filtered solution of desired passivating material was spin coated on FTO/NiO$_x$ substrates at 3000 rpm for 50 s and then annealed at 100 °C for 15 min on a hotplate inside a N$_2$ filled glove box. Importantly, the substrates were then washed with a copious amount of the respective solvent (isopropanol or dimethylformamide) and dried with a N$_2$ gun to remove any unbound material from the NiO$_x$ surface. These substrates were used as such for further analysis or to prepare further samples by depositing MAPbI$_3$ films.

### S2.5. Fabrication of perovskite solar cells

All solar cells were prepared by using etched FTO substrates with a 20-30 nm thick NiO$_x$ hole transport layer prepared from 0.2 M nickel (II) acetate tetrahydrate and ethanolamine mixture in ethanol. On NiO$_x$ layer, passivating materials (0.1 M solution) and MAPbI$_3$ (from 1 M solution, 350-400 nm thick) were deposited by following the procedures described above. The substrates were then moved to an evaporation chamber, where C$_{60}$ electron transport layer (30 nm) and bathocuproine (8 nm) were evaporated. Finally, the devices were completed with Ag cathode (100 nm) deposition. Each substrate contains 7 devices with an active area of 0.0785 cm$^2$, defined by the deposited Ag cathode (see **Figure S3**).

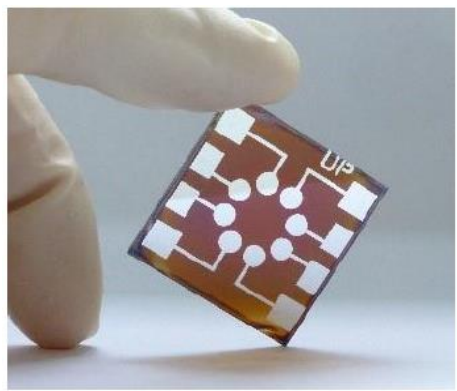

**Figure S3.** Photography of the studied samples with an active area of 0.0785 cm$^2$.

## S2.6. Device performance

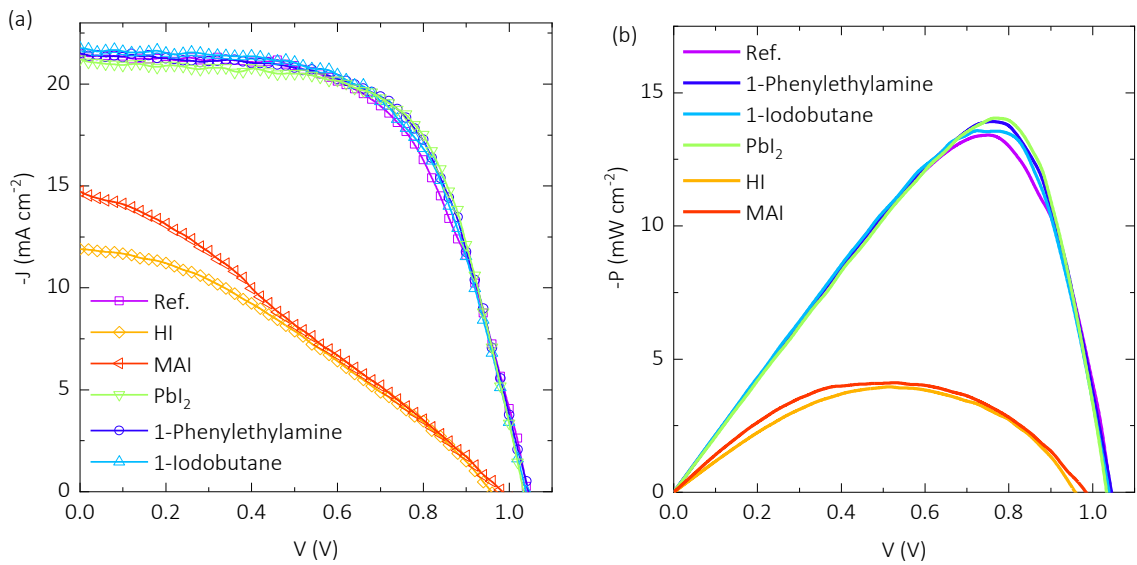

**Figure S4.** Representative as-fabricated current density-voltage ($J-V$) curves under standard 1 sun illumination. In each case, the best two pixels per substrate are shown for two bias scan sweep directions at a scan rate of 70 mV/s.

**Table S3.** Summary of best performance samples in terms of power conversion efficiency from $J-V$ curves under standard 1 sun illumination.

| Sample | Reverse scan | | | | Forward scan | | | |
|---|---|---|---|---|---|---|---|---|
| | $V_{oc}$ (V) | $J_{sc}$ (mA·cm$^{-2}$) | FF (%) | PCE (%) | $V_{oc}$ (V) | $J_{sc}$ (mA·cm-2) | FF (%) | PCE (%) |
| Ref. | 1.04 | 20.0 | 64 | 13.4 | 1.01 | 19.6 | 68 | 13.4 |
| Phenylethylamine | 1.04 | 21.5 | 62 | 13.8 | 1.02 | 20.7 | 68 | 14.4 |
| Iodobutane | 1.04 | 21.8 | 60 | 13.6 | 1.02 | 20.8 | 67 | 14.1 |
| PbI$_2$ | 1.03 | 21.2 | 64 | 14.1 | 1.01 | 20.0 | 73 | 14.7 |
| HI | 0.96 | 11.9 | 35 | 4.0 | 0.96 | 11.8 | 38 | 4.3 |
| MAI | 0.98 | 14.7 | 29 | 4.2 | 0.99 | 15.3 | 34 | 5.2 |

## S3. Impedance spectra in short-circuit at different illumination intensities

### S3.1. Experimental Procedure and Equivalent Circuit Modelling

The impedance spectroscopy was measured with an Autolab PGSTAT302N potentiostat including a FRA32M unit and a kit Autolab Optical Bench from MetroOhm. The samples were illuminated with a white LED at different steady-state illumination intensities, then the short-circuit (SC) condition was applied ($V=0V$), allowing a direct current (DC) density $J_{sc}$. Upon these DC condition the 15 mV perturbation was applied for measuring the impedance spectroscopy (IS) as a function of the $J_{sc}$ for each illumination intensity. Subsequently, the IS spectra were fitted to the equivalent circuit (EC) model of **Figure S5**.

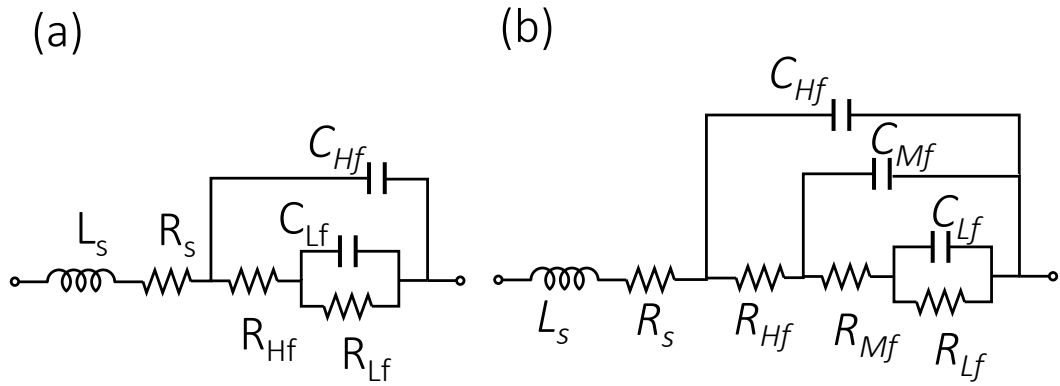

**Figure S5.** Equivalent circuits used for numerical model of impedance spectra. Here $L_s$ and $R_s$ are series inductor and resistor, respectively; $R_{Hf}$, $R_{Mf}$ and $R_{Lf}$ stand for high, medium and low frequency resistors, respectively, and $C_{Hf}$, $C_{Mf}$ and $C_{Lf}$ stand for high, medium and low frequency capacitors, respectively.

The resistance parameters at SC extracted from the EC model were subsequently parameterized as a function of $J_{sc}$, following the empirical equation

$$R_{sc} = R_s + \frac{R_{sh0}}{\left(1 + \frac{J_{sc}}{J_\sigma}\right)} \tag{S1}$$

where $R_s$ is the series resistance; $R_{sh0}$, the equilibrium (dark) shunt resistance; and $J_\sigma$ is the short-circuit current density corresponding to the threshold illumination intensity whose charge carrier concentration at SC increases photoconductivity, thus decreasing the resistance. Notably, $R_{sc} \cong R_{sh}/2$ for $J_{sc} = J_\sigma$.

The capacitance parameters at SC extracted from the EC model were subsequently parameterized as a function of $J_{sc}$, following the empirical equation

$$C_{sc} = C_0 \left(1 + \left(\frac{J_{sc}}{J_\varepsilon}\right)^p\right) \tag{S2}$$

where $C_0$ is the equilibrium (dark) SC capacitance; $p$ is a dimensionless power parameter, and $J_\varepsilon$ is the short-circuit current density corresponding to the threshold illumination intensity whose charge carrier concentration at SC enables a capacitive regime transition from a purely dielectric response (geometric or space-charge capacitance) to a presumably mobile ions-related behavior. Notably, $C_{sc} \cong 2R_{sh}$ for $J_{sc} = J_\varepsilon$ and dielectric contributions which are independent of the light intensity would require significantly large threshold currents ($J_\varepsilon \to \infty \Rightarrow C_{sc} = C_0$)

The characteristic response times for each mechanism can be extracted via the estimation of the corresponding peak maxima across the Bode plots of the imaginary part of the impedance, and/or by considering the resistor-capacitor coupling ($R \cdot C$ product) assumed in the EC model. Either way, the behaviour of the response times follows the empirical equation

$$\tau_{sc} = \frac{\tau_0}{\left(1 + \frac{J_{sc}}{J_\sigma}\right)} \left(1 + \left(\frac{J_{sc}}{J_\varepsilon}\right)^p\right) \tag{S3}$$

where $\tau_0 \approx R_{sh0}C_0$ is the characteristic equilibrium (dark) response time constant. Importantly, Equation (S3) behaves constant for $p = 1$ in the high illumination intensity limit ($J_{sc} \to \infty \Rightarrow \tau_{sc} = \tau_0 J_\sigma/J_\varepsilon$); provided equal threshold currents and/or in the low illumination intensity limit ($J_\sigma = J_\varepsilon \vee J_{sc} \to 0 \Rightarrow \tau_{sc} = \tau_0$).

## S3.2. Experimental impedance spectroscopy data and equivalent circuit fittings for different illumination intensities

### S3.3. NiO$_x$Ref

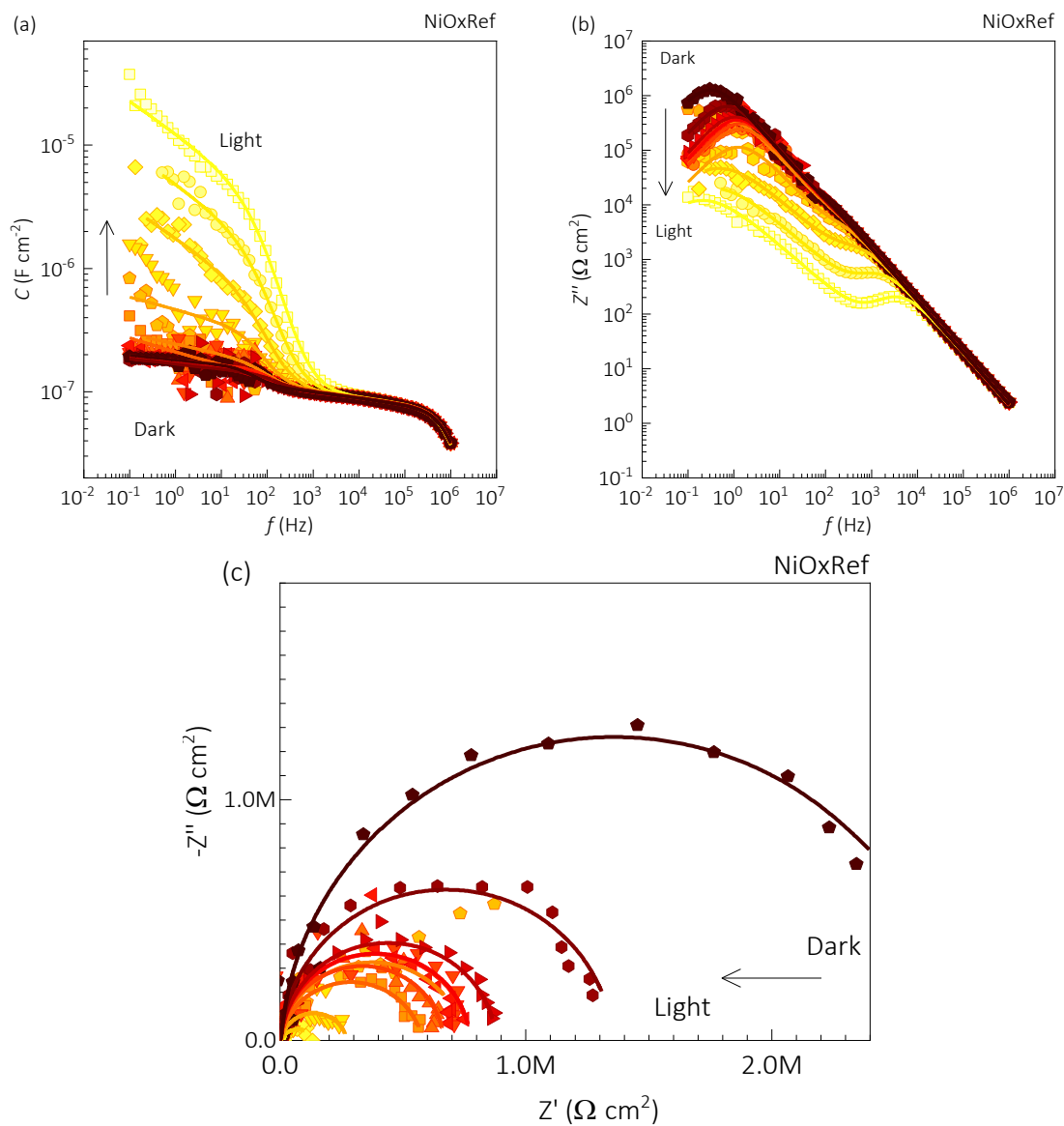

**Figure S6.** Impedance spectra for the **NiOxRef** Sample in SC under different DC illumination intensities in (a) capacitance and (b) imaginary Bode, and (c) impedance Nyquist representations. In each case, the dots indicate experimental data, the lines are the numerical simulations via EC model (see **Figure S5**), and darker and lighter colours indicate low and high illumination intensities (and $J_{sc}$ values), respectively.

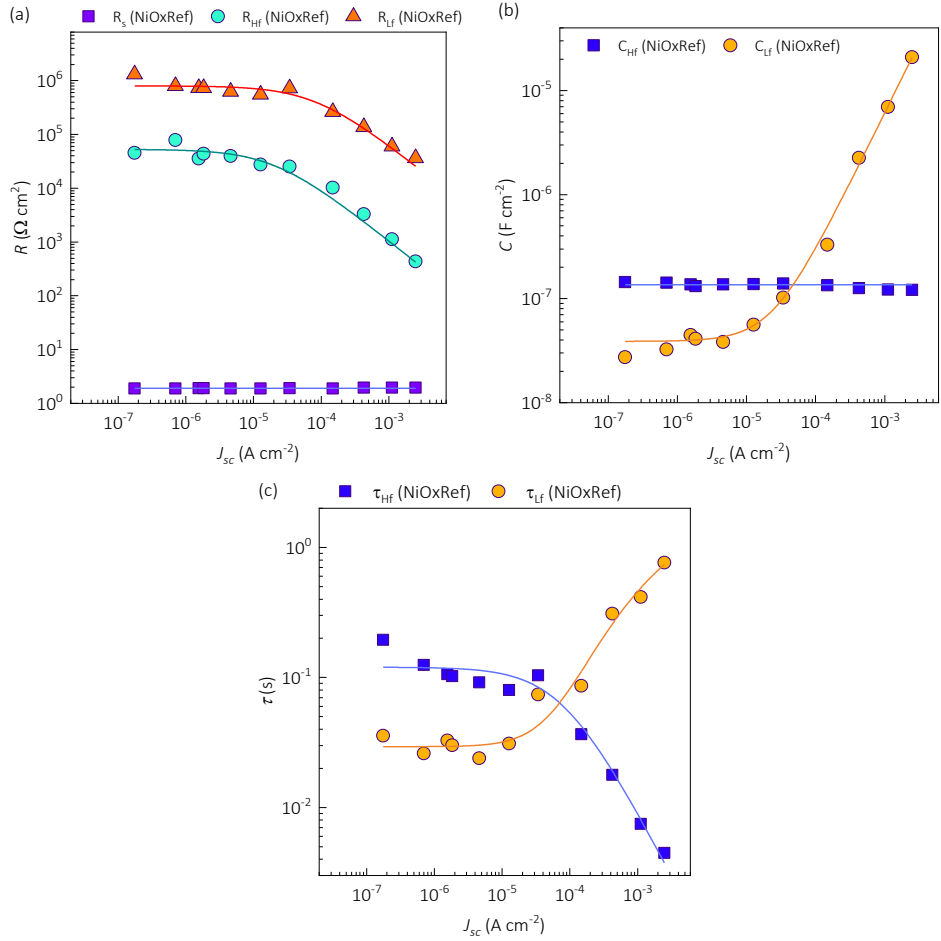

**Figure S7.** Results for (a) resistance, (b) capacitance, and (c) characteristic times extracted from the EC numerical modelling of the impedance spectra in SC of the ***NiO$_x$ Re.f*** sample under different DC illumination intensities. In each case, the dots indicate fitted parameters from the EC model (see **Figure S5**), and the lines are the numerical simulations to the analytical trends of equations (S1), (S2), and (S3) for (a), (b), and (c), respectively.

**Table S4.** Parameterization of photovoltage trend from the EC modelling following Equations (S1), (S2), (S3) and from the IS spectra of the sample **NiO$_x$ Ref.** in SC.

| Parameters | $R_s$ | $R_{Hf}$ | $R_{Lf}$ | $C_{Hf}$ | $C_{Lf}$ | $\tau_{Hf}$ | $\tau_{Lf}$ |
|---|---|---|---|---|---|---|---|
| $R_s$ (Ω·cm²) | 1.90 | — | — | — | — | — | — |
| $R_{sh}$ (Ω·cm²) | — | 58053 | 8E5 | — | — | — | — |
| $J_\sigma$ (A cm⁻²) | — | 2E-5 | 8.2E-5 | — | — | 8e-5 | 3.6E-4 |
| $C_0$ (F·cm⁻²) | — | — | — | 1.38E-7 | 3.93E-8 | — | — |
| $J_\varepsilon$ (A cm⁻²) | — | — | — | — | 2.42E-5 | — | 5E-5 |
| p (a.u.) | — | — | — | — | 1.35 | — | 1.35 |
| $\tau_0$ (s) | — | — | — | — | — | 0.0472 | 0.03 |
| $V_{ph}$ (V) | — | — | — | — | — | — | — |
| $J_V$ (A cm⁻²) | — | — | — | — | — | — | — |

## S3.4. **Amine**

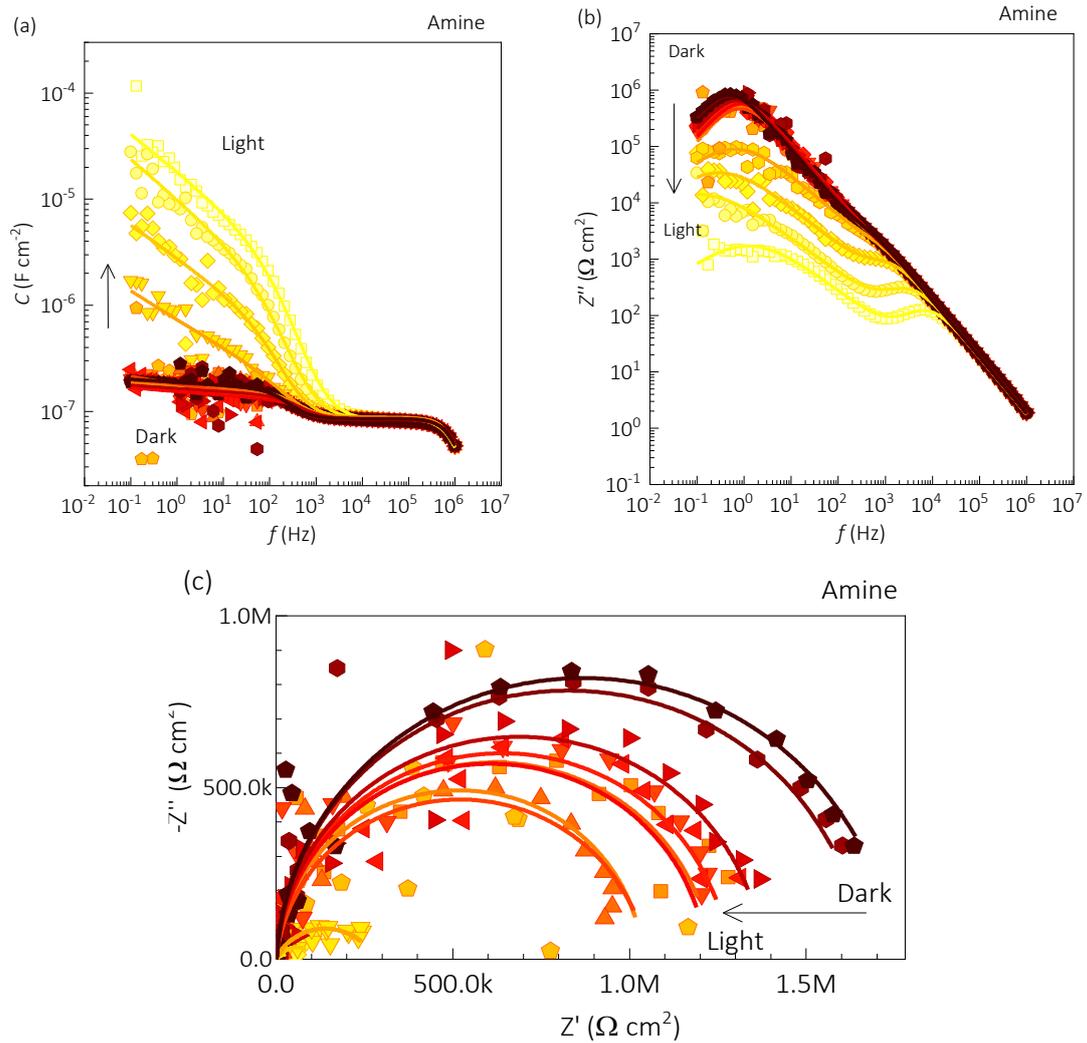

**Figure S8.** Impedance spectra for the **Amine** Sample in SC under different DC illumination intensities in (a) capacitance and (b) imaginary Bode, and (c) impedance Nyquist representations. In each case, the dots indicate experimental data, the lines are the numerical simulations via EC model (see **Figure S5**), and darker and lighter colors indicate low and high illumination intensities (and $J_{sc}$ values), respectively.

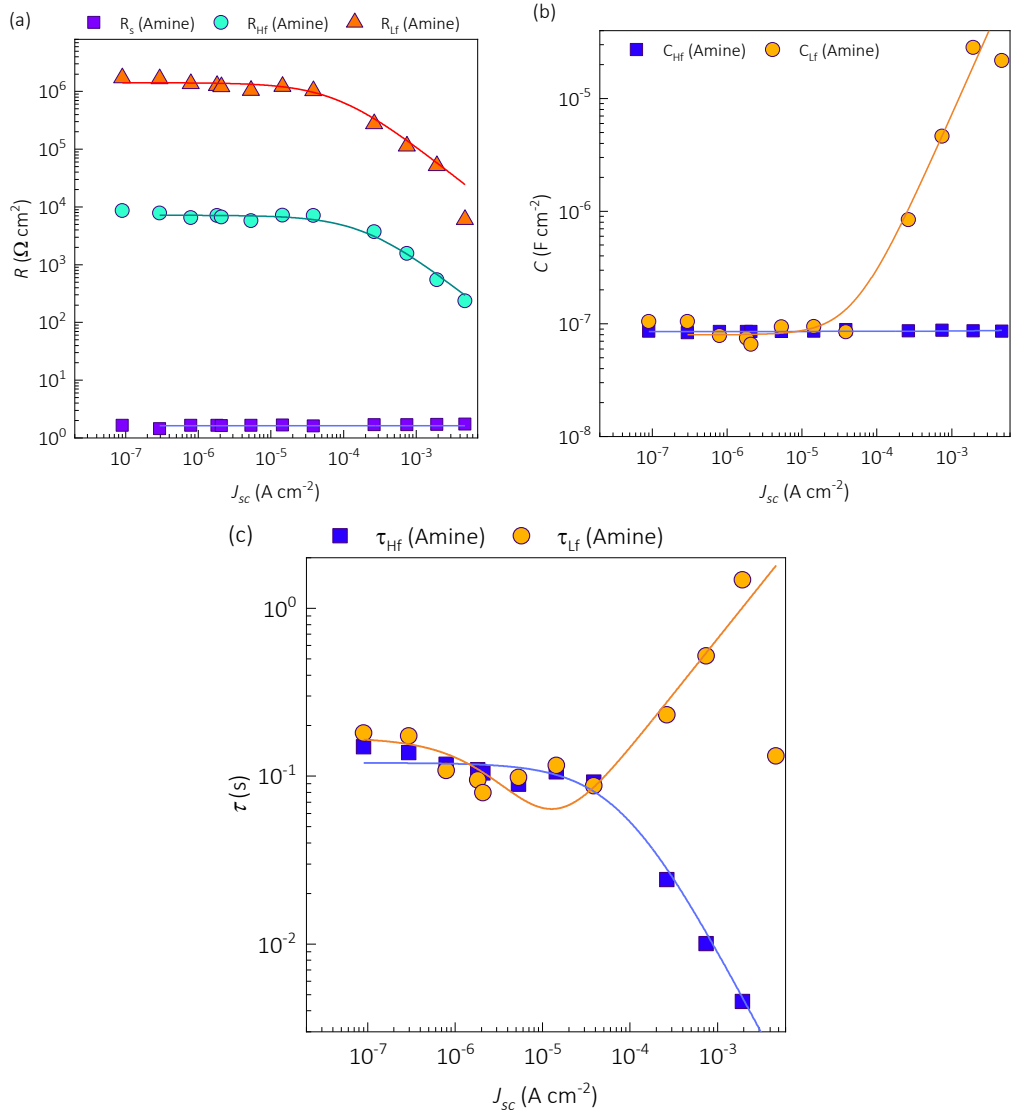

**Figure S9.** Results for (a) resistance, (b) capacitance, and (c) characteristic times extracted from the EC numerical modelling of the impedance spectra of the *Amine* sample in SC under different DC illumination intensities. In each case, the dots indicate fitted parameters from the EC model (see **Figure S5**), and the lines are the numerical simulations to the analytical trends of equations (S1), (S2), and (S3) for (a), (b), and (c), respectively.

**Table S5.** Parameterization of photovoltage trend from the EC modeling following Equations (S1), (S2), (S3) and from the IS spectra of the sample Amine in SC.

| Parameters | $R_s$ | $R_{Hf}$ | $R_{Lf}$ | $C_{Hf}$ | $C_{Lf}$ | $\tau_{Hf}$ | $\tau_{Lf}$ |
|---|---|---|---|---|---|---|---|
| $R_s$ ($\Omega \cdot cm^2$) | 1.61 | — | — | — | — | — | — |
| $R_{sh}$ ($\Omega \cdot cm^2$) | — | 6959 | 1.28E6 | — | — | — | — |
| $J_\sigma$ (A cm$^{-2}$) | — | 2E-4 | 8.2E-5 | — | — | 8.2e-5 | 3E-6 |
| $C_0$ (F·cm$^{-2}$) | — | — | — | 8.7e-8 | 8E-8 | — | — |
| $J_\varepsilon$ (A cm$^{-2}$) | — | — | — | — | 5E-5 | — | 1.3E-5 |
| p (a.u.) | — | — | — | — | 1.5 | — | 1.65 |
| $\tau_0$ (s) | — | — | — | — | — | 0.12 | 0.17 |
| $V_{ph}$ (V) | — | — | — | — | — | — | — |
| $J_V$ (A cm$^{-2}$) | — | — | — | — | — | — | — |

## S3.5. **Iodobutane**

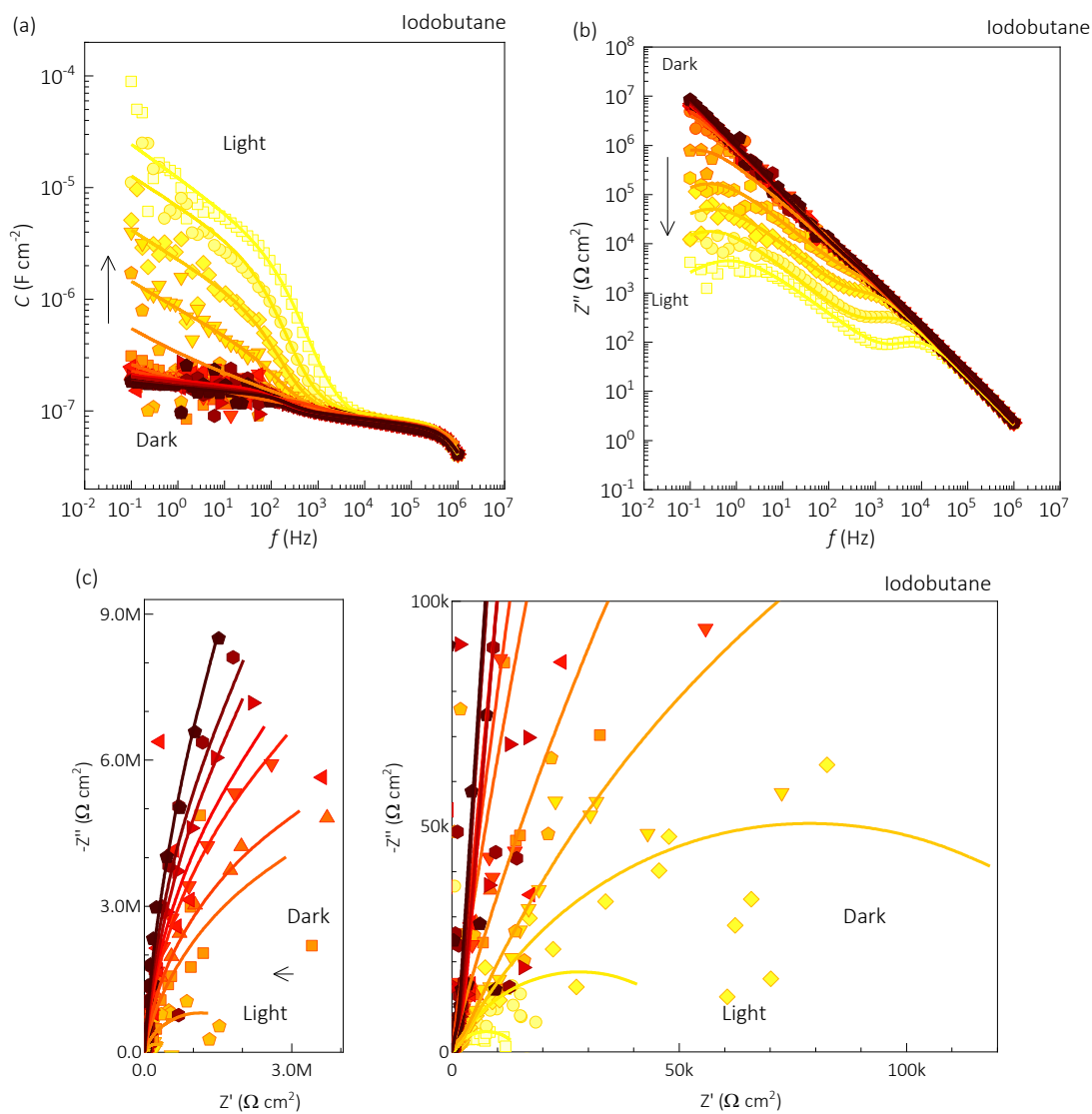

**Figure S10.** Impedance spectra for the **Iodobutane** Sample in SC under different DC illumination intensities in (a) capacitance and (b) imaginary Bode, and (c) impedance Nyquist representations. In each case, the dots indicate experimental data, the lines are the numerical simulations via EC model (see **Figure S5**), and darker and lighter colors indicate low and high illumination intensities (and $J_{sc}$ values), respectively.

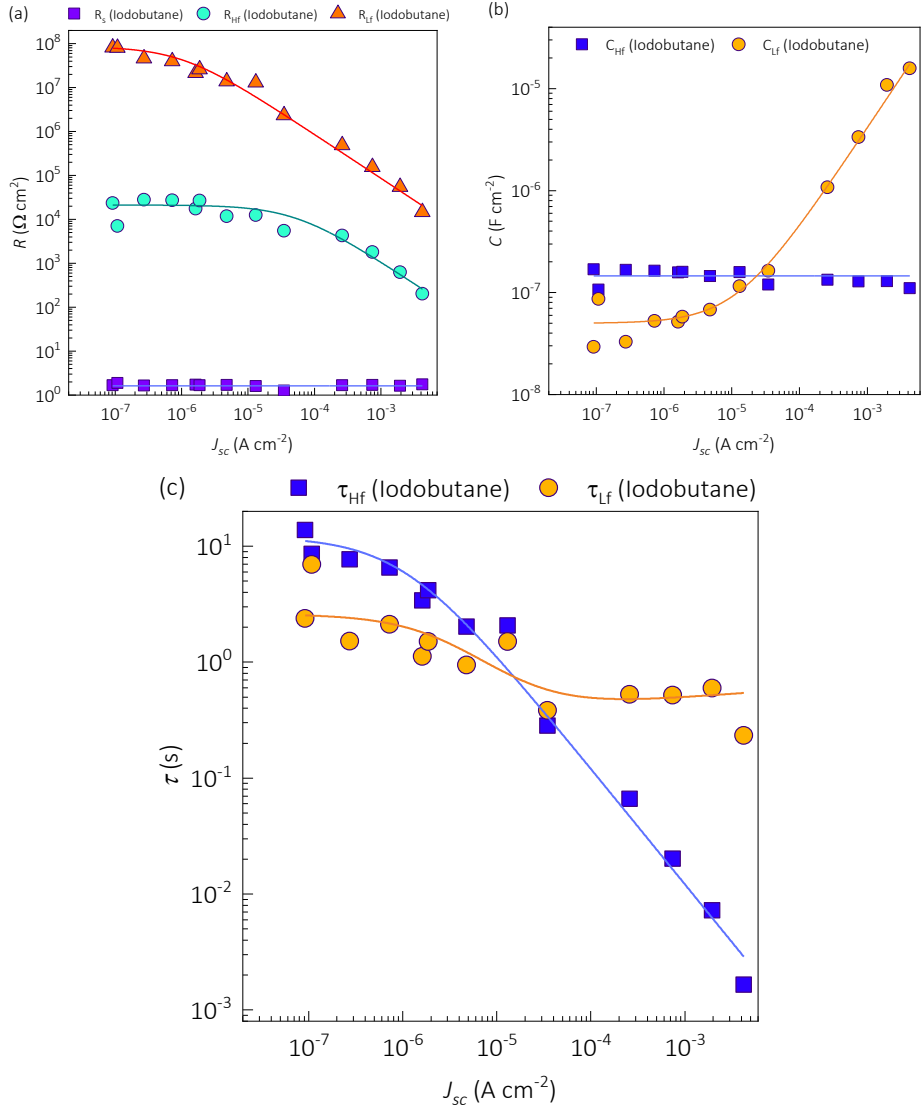

**Figure S11.** Results for (a) resistance, (b) capacitance, and (c) characteristic times extracted from the EC numerical modelling of the impedance spectra of the ***Iodobutane*** sample in SC under different DC illumination intensities. In each case, the dots indicate fitted parameters from the EC model (see **Figure S5**), and the lines are the numerical simulations to the analytical trends of equations (S1), (S2), and (S3) for (a), (b), and (c), respectively.

**Table S6.** Parameterization of photovoltage trend from the EC modelling following Equations (S1), (S2), (S3) and from the IS spectra of the sample ***Iodobutane*** in SC.

| Parameters | $R_s$ | $R_{Hf}$ | $R_{Lf}$ | $C_{Hf}$ | $C_{Lf}$ | $\tau_{Hf}$ | $\tau_{Lf}$ |
|---|---|---|---|---|---|---|---|
| $R_s$ (Ω·cm²) | 1.62 | — | — | — | — | — | — |
| $R_{sh}$ (Ω·cm²) | — | 21227 | 8.62E7 | — | — | — | — |
| $J_\sigma$ (A cm⁻²) | — | 5.2E-5 | 1E-6 | — | — | 1E-6 | 3E-6 |
| $C_0$ (F·cm⁻²) | — | — | — | 1.46E-7 | 5E-8 | — | — |
| $J_\varepsilon$ (A cm⁻²) | — | — | — | — | 1.22E-5 | — | 2E-5 |
| p (a.u.) | — | — | — | — | 1.02 | — | 1.06 |
| $\tau_0$ (s) | — | — | — | — | — | 12.1 | 2.6 |
| $V_{ph}$ (V) | — | — | — | — | — | — | — |
| $J_V$ (A cm⁻²) | — | — | — | — | — | — | — |

S3.6. **HI**

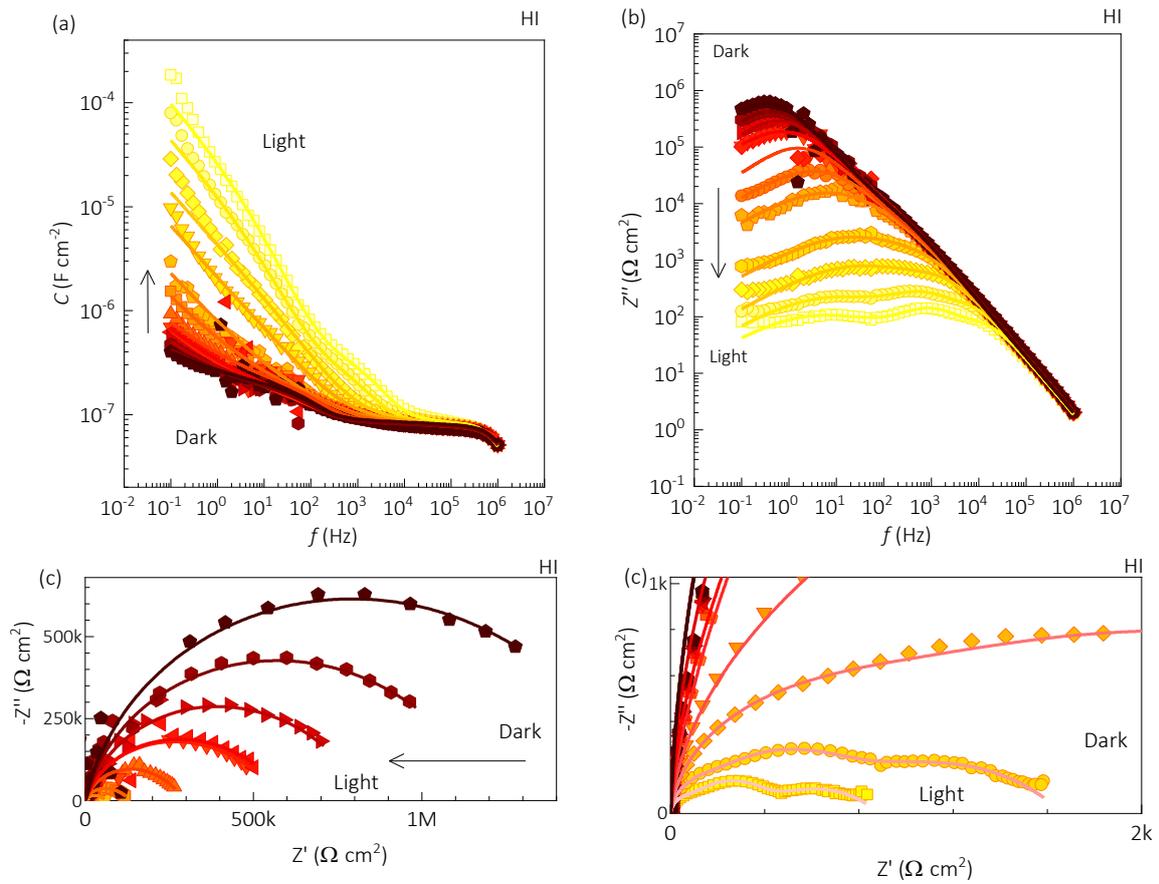

**Figure S12.** Impedance spectra for the *HI* Sample in SC under different DC illumination intensities in (a) capacitance and (b) imaginary Bode, and (c) impedance Nyquist representations. In each case, the dots indicate experimental data, the lines are the numerical simulations via EC model (see **Figure S5**), and darker and lighter colors indicate low and high illumination intensities (and $J_{sc}$ values), respectively.

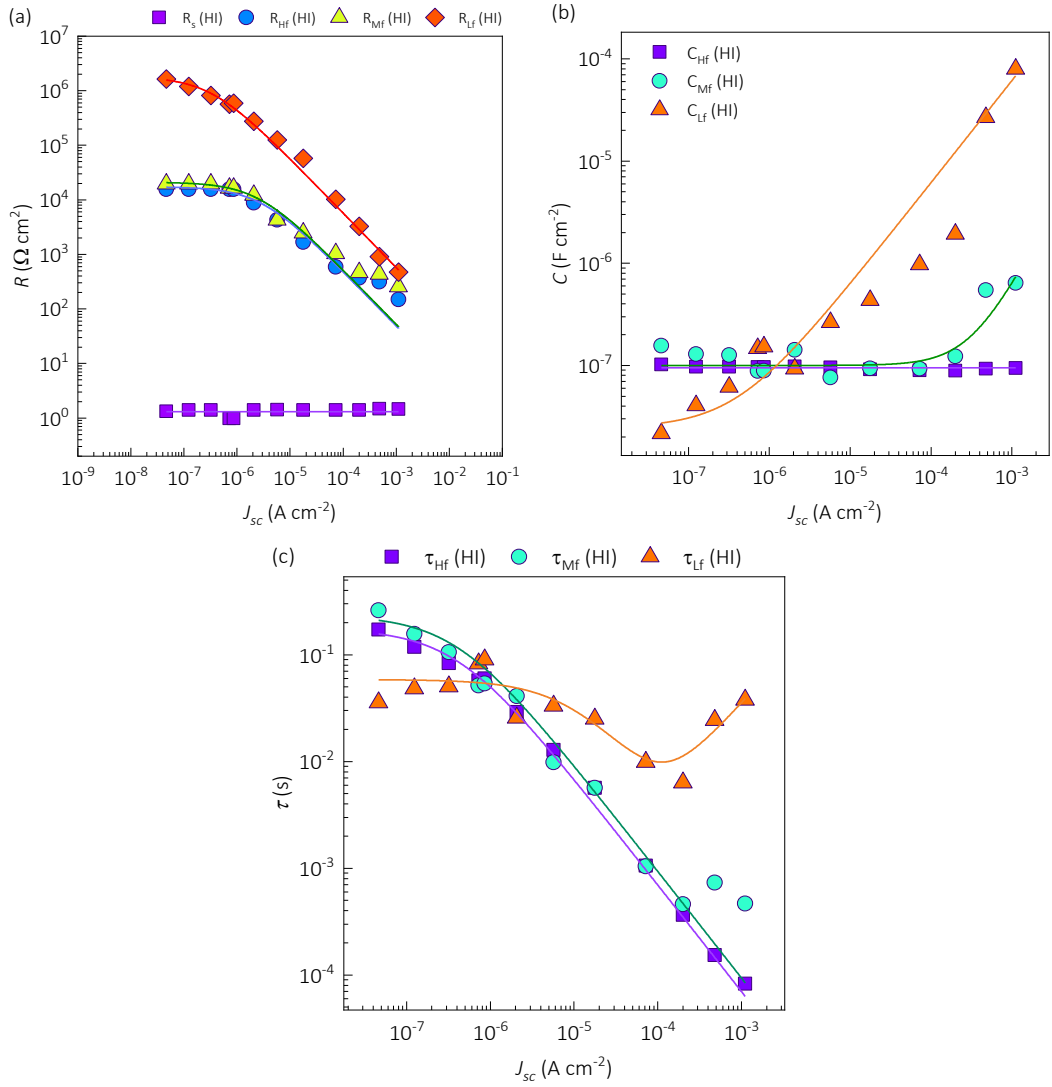

**Figure S13.** Results for (a) resistance, (b) capacitance, and (c) characteristic times extracted from the EC numerical modelling of the impedance spectra of the *HI* sample in SC under different DC illumination intensities. In each case, the dots indicate fitted parameters from the EC model (see **Figure S5**), and the lines are the numerical simulations to the analytical trends of equations (S1), (S2), and (S3) for (a), (b), and (c), respectively.

**Table S7.** Parameterization of photovoltage trend from the EC modelling following Equations (S1), (S2), (S3) and from the IS spectra of the sample *HI* in SC.

| Parameters | $R_s$ | $R_{Hf}$ | $R_{Mf}$ | $R_{Lf}$ | $C_{Hf}$ | $C_{Mf}$ | $C_{Lf}$ | $\tau_{Hf}$ | $\tau_{Mf}$ | $\tau_{Lf}$ |
|---|---|---|---|---|---|---|---|---|---|---|
| $R_s$ ($\Omega\cdot$cm$^2$) | 1.31 | — | — | — | — | — | — | — | — | — |
| $R_{sh}$ ($\Omega\cdot$cm$^2$) | — | 17342 | 21043 | 1.79E6 | — | — | — | — | — | — |
| $J_\sigma$ (A cm$^{-2}$) | — | 2.82E-6 | 2.48E-6 | 3.19E-7 | — | — | — | 4E-7 | 4E-7 | 1.04E-5 |
| $C_0$ (F$\cdot$cm$^{-2}$) | — | — | — | — | 9.49E-8 | 1E-7 | 2.45E-8 | — | — | — |
| $J_\varepsilon$ (A cm$^{-2}$) | — | — | — | — | — | 3.23E-4 | 4E-7 | — | — | 1.13E-4 |
| p (a.u.) | — | — | — | — | — | 1.5 | 1.00 | — | — | 1.86 |
| $\tau_0$ (s) | — | — | — | — | — | — | — | 0.175 | 0.234 | 0.0586 |
| $V_{ph}$ (V) | — | — | — | — | — | — | — | — | — | — |
| $J_V$ (A cm$^{-2}$) | — | — | — | — | — | — | — | — | — | — |

## S3.7. **PbI2**

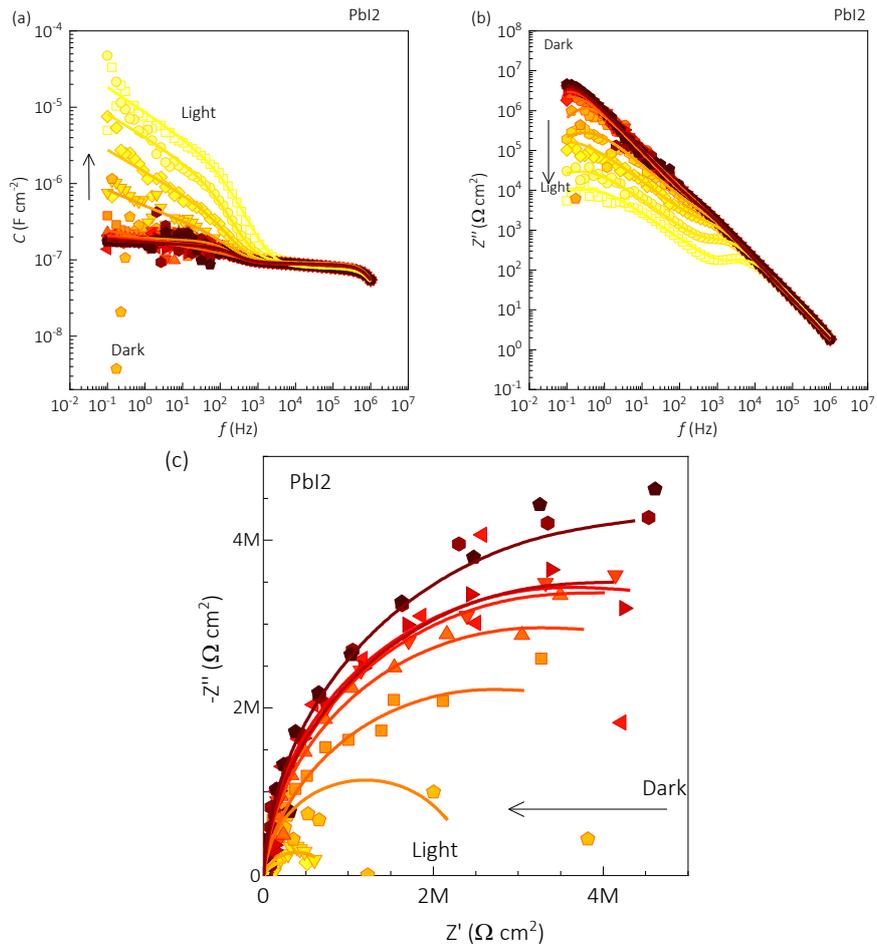

**Figure S14.** Impedance spectra for the *PbI2* Sample in SC under different DC illumination intensities in (a) capacitance and (b) imaginary Bode, and (c) impedance Nyquist representations. In each case, the dots indicate experimental data, the lines are the numerical simulations via EC model (see **Figure S5**), and darker and lighter colors indicate low and high illumination intensities (and $J_{sc}$ values), respectively.

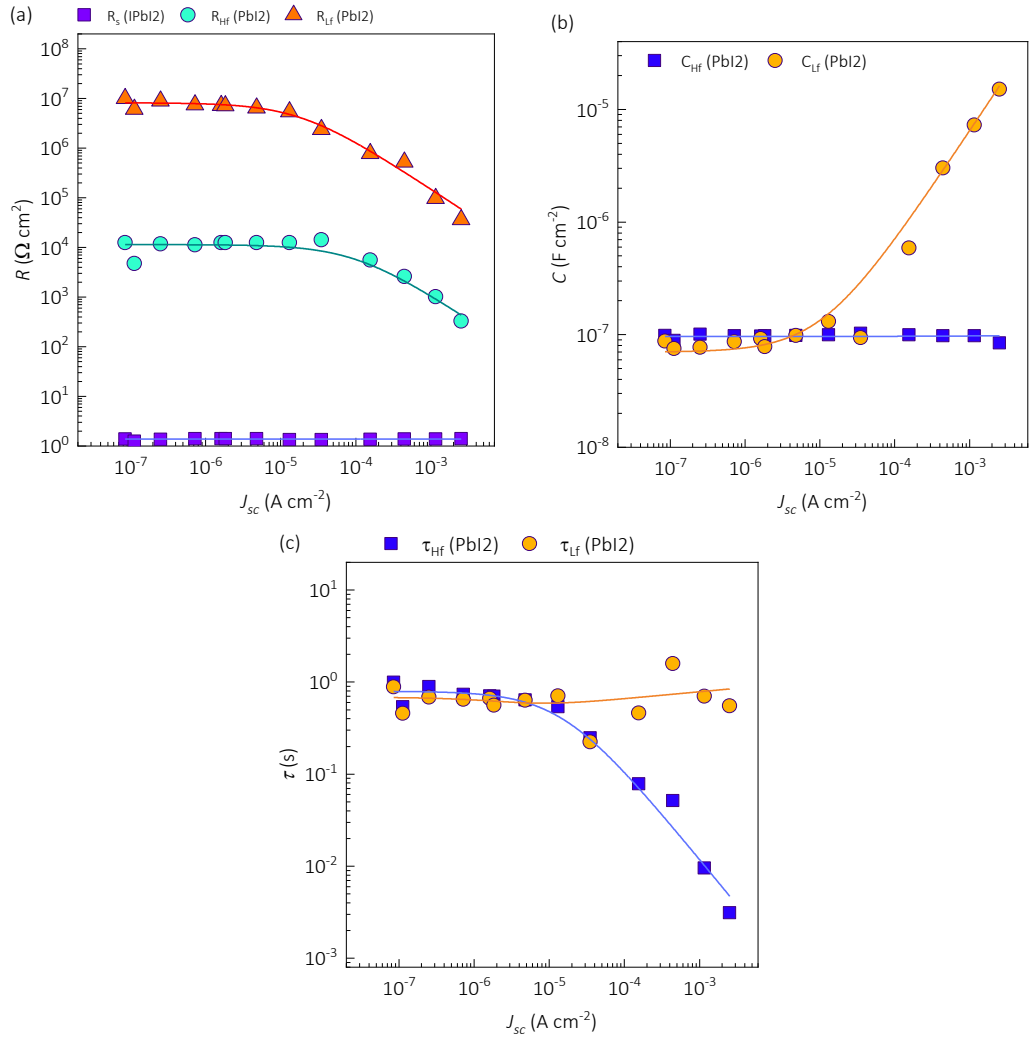

**Figure S15.** Results for (a) resistance, (b) capacitance, (c) characteristic times and (d) internal photovoltages extracted from the EC numerical modelling of the impedance spectra of the *HI* sample in SC under different DC illumination intensities. In each case, the dots indicate fitted parameters from the EC model (see **Figure S5**), and the lines are the numerical simulations to the analytical trends of equations (S1), (S2), and (S3) for (a), (b), and (c), respectively.

**Table S8.** Parameterization of photovoltage trend from the EC modelling following Equations (S1), (S2), (S3) and from the IS spectra of the sample *HI* in SC.

| Parameters | $R_s$ | $R_{Hf}$ | $R_{Mf}$ | $R_{Lf}$ | $C_{Hf}$ | $C_{Lf}$ | $\tau_{Hf}$ | $\tau_{Lf}$ |
|---|---|---|---|---|---|---|---|---|
| $R_s$ ($\Omega\cdot cm^2$) | 1.38 | — | — | — | — | — | — | — |
| $R_{sh}$ ($\Omega\cdot cm^2$) | — | 17342 | 11412 | 8.19E6 | — | — | — | — |
| $J_\sigma$ (A cm$^{-2}$) | — | 2.82E-6 | 1E-4 | 1.84E-5 | — | — | 1.5E-5 | 1.04E-5 |
| $C_0$ (F·cm$^{-2}$) | — | — | — | — | 9.64E-8 | 7E-8 | — | — |
| $J_\varepsilon$ (A cm$^{-2}$) | — | — | — | — | — | 1.09E-5 | — | 1.13E-4 |
| p (a.u.) | — | — | — | — | — | 1.00 | — | 1.86 |
| $\tau_0$ (s) | — | — | — | — | — | — | 1.00 | 0.0586 |
| $V_{ph}$ (V) | — | — | — | — | — | — | — | — |
| $J_V$ (A cm$^{-2}$) | — | — | — | — | — | — | — | — |

## S4. Stability test in short-circuit under 0.2 sun white LED equivalent illumination intensity

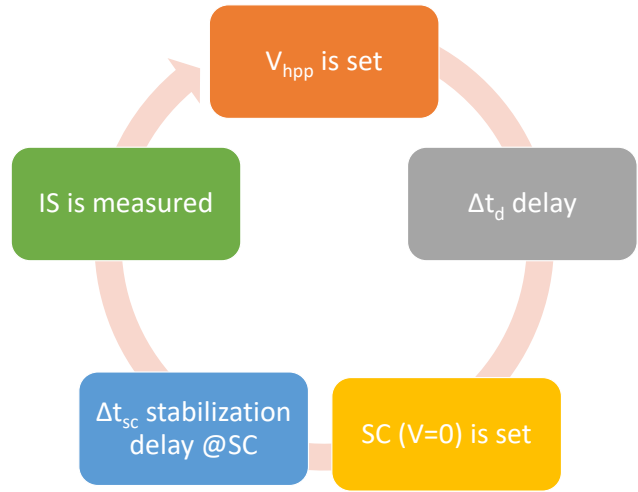

**Figure S16.** Scheme of the degradation test loop. The devices were initially set under illumination to a high-power point ($V_{hpp}$), close to that of the starting maximum-power point (MPP) without correcting hysteresis effect. Next, a delay time ($\Delta t_d$) for degradation was waited while the illumination and the bias $V_{hpp}$ were constant. Subsequently a biasing change from near-to-MPP to short-circuit (V=0V) is set. Another delay time ($\Delta t_{sc}$) is considered allowing the short circuit current to stabilize. Afterwards, the impedance spectrum (IS) is measured at short-circuit (SC). The loop is repeated when a new $V_{hpp}$ is set. For the correction of the $V_{hpp}$ value, an empirical rule was implemented. Provided the initial difference $\Delta V = V_{oc,0} - V_{mpp}$ between the open circuit voltage and the MPP voltage, the open circuit voltage of the n$^{th}$ cycle is measure after the IS and then the n$^{th}$ high-power voltage value is set as: $V_{hpp,n} = V_{oc,n-1} - \Delta V$

## S4.1. High power condition over time

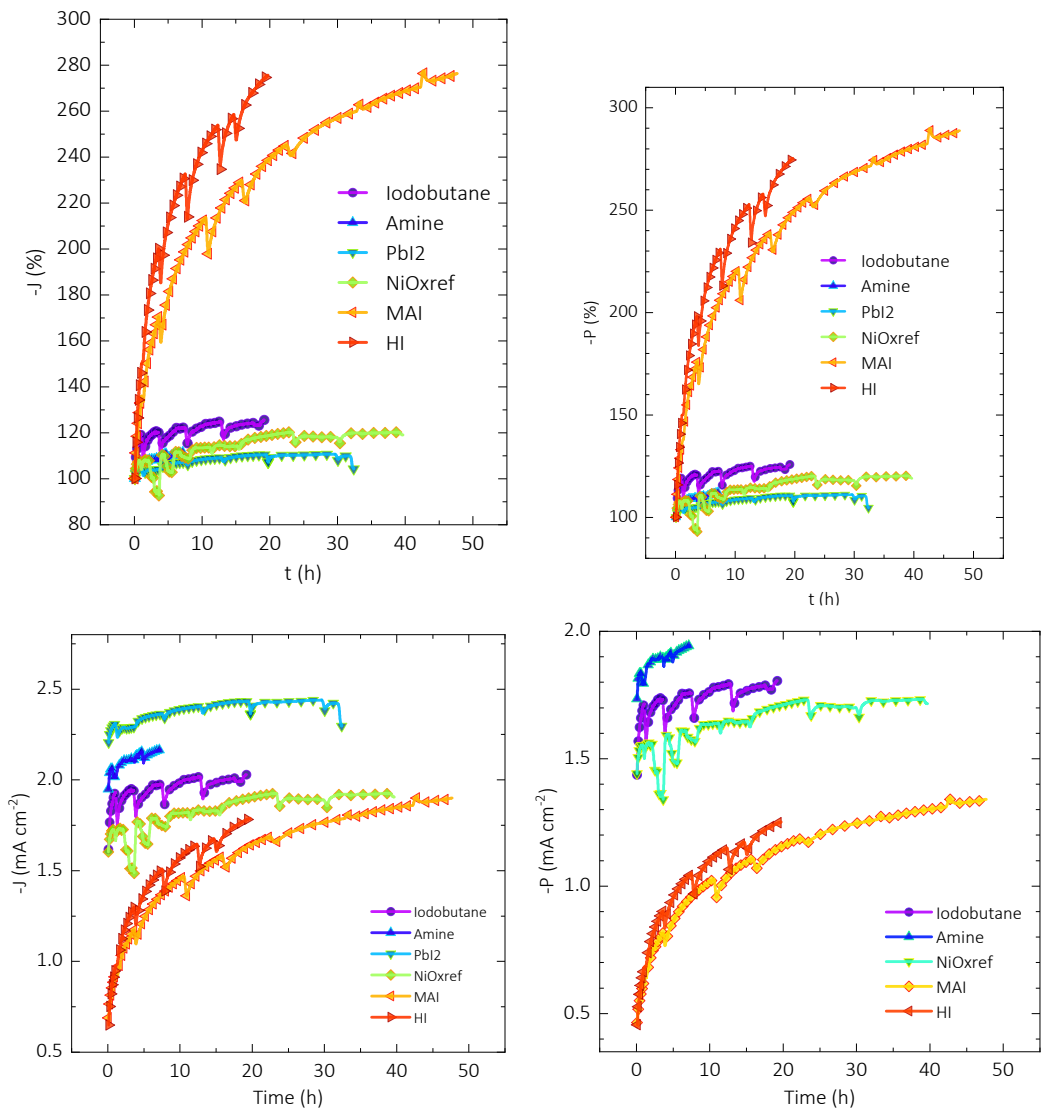

**Figure S17.** Stability test at high power point under 0.2 sun white LED equivalent illumination.

## S4.2. Impedance spectroscopy in short-circuit over time

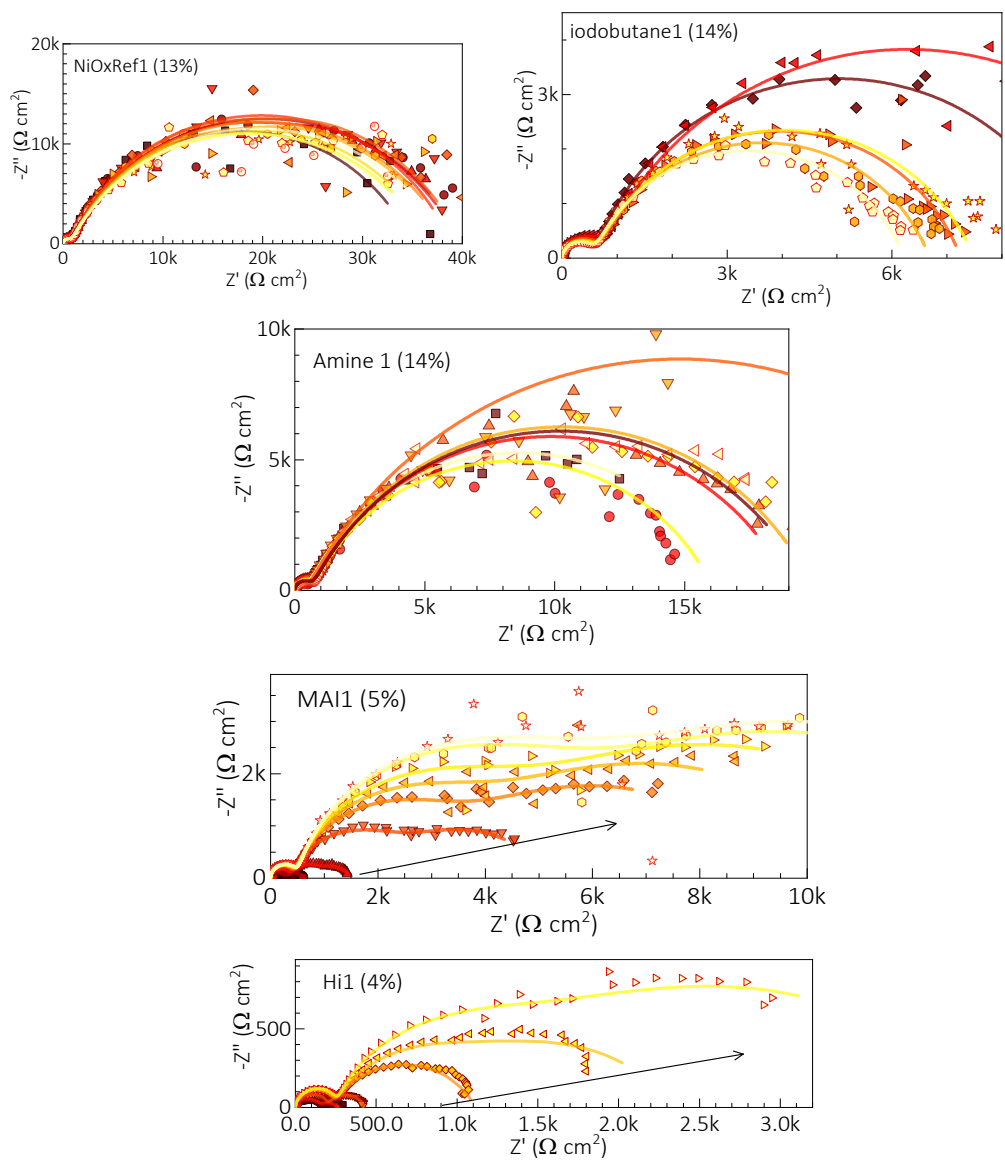

**Figure S18.** Impedance spectra in Nyquist plot representation in short-circuit under 0.2 sun white LED equivalent over time for different sample surface treatment, as indicated. Dots and lines are the experimental data and equivalent circuit model fittings, respectively. The lighter (yellower) the colors the longer the time. The same spectra are shown in **Figure S20** and **Figure S19**

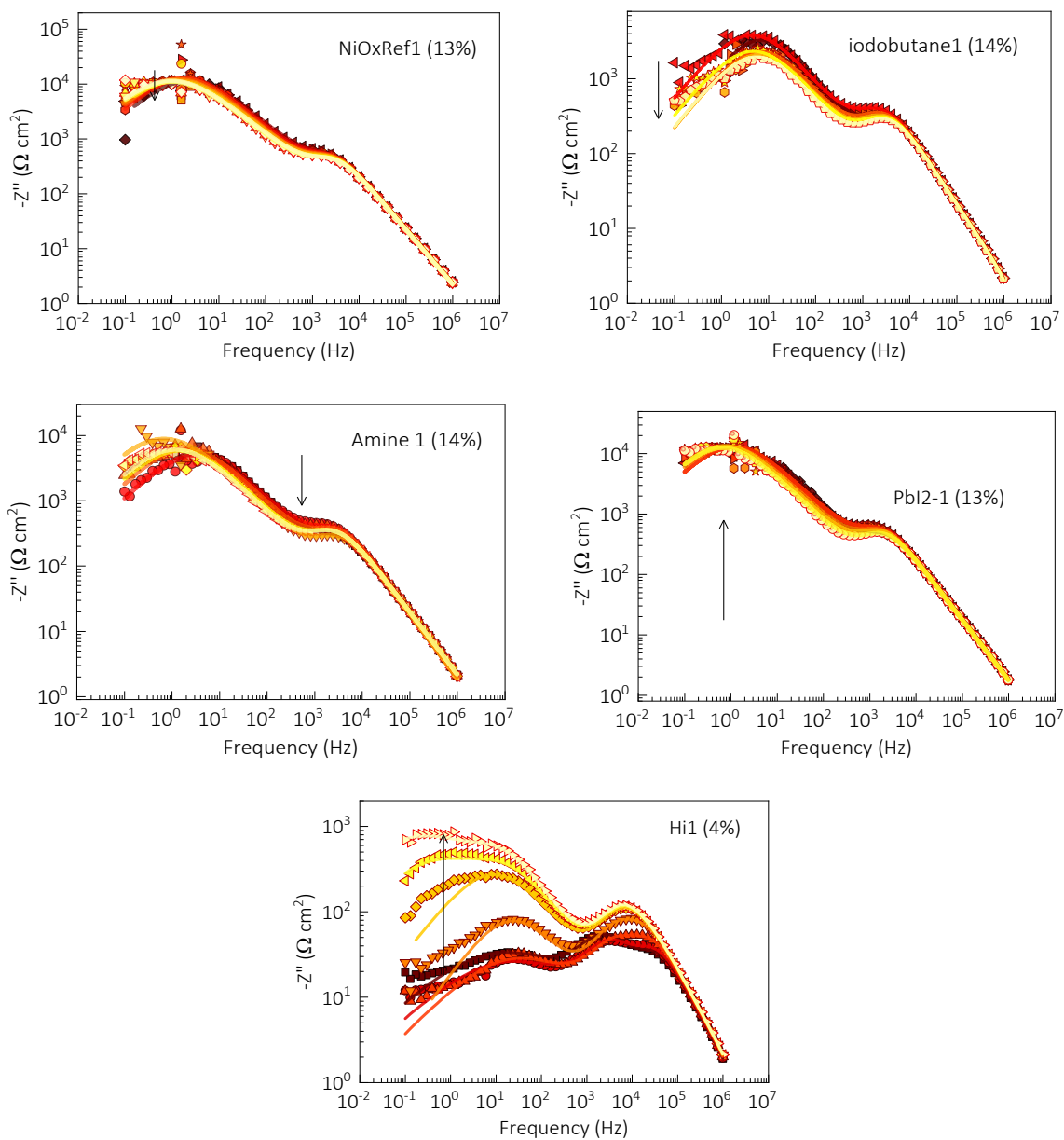

**Figure S19.** Bode plots of imaginary part of impedance spectra over time in short-circuit condition under 0.2 sun white LED equivalent illumination. The dots are the experimental data and lines are the fittings to the equivalent circuit models in **Figure S5**. The same spectra are shown in **Figure S18** and **Figure S20**.

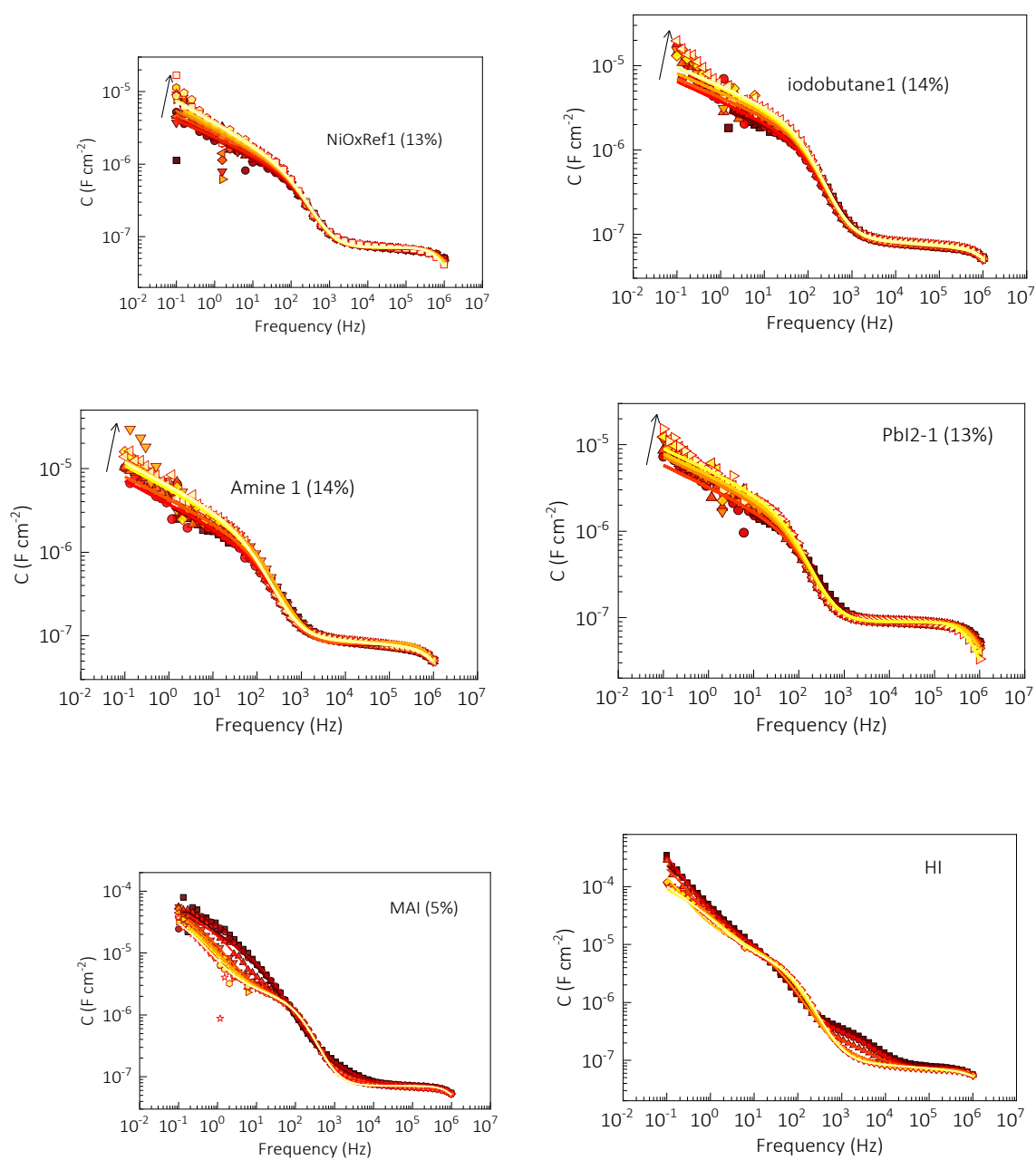

**Figure S20.** Bode plots of capacitance spectra over time in short-circuit condition under 0.2 sun white LED equivalent illumination for different sample surface treatment, as indicated. The dots are the experimental data and lines are the fittings to the equivalent circuit models in **Figure S5**. The same spectra are shown in **Figure S18** and **Figure S19**

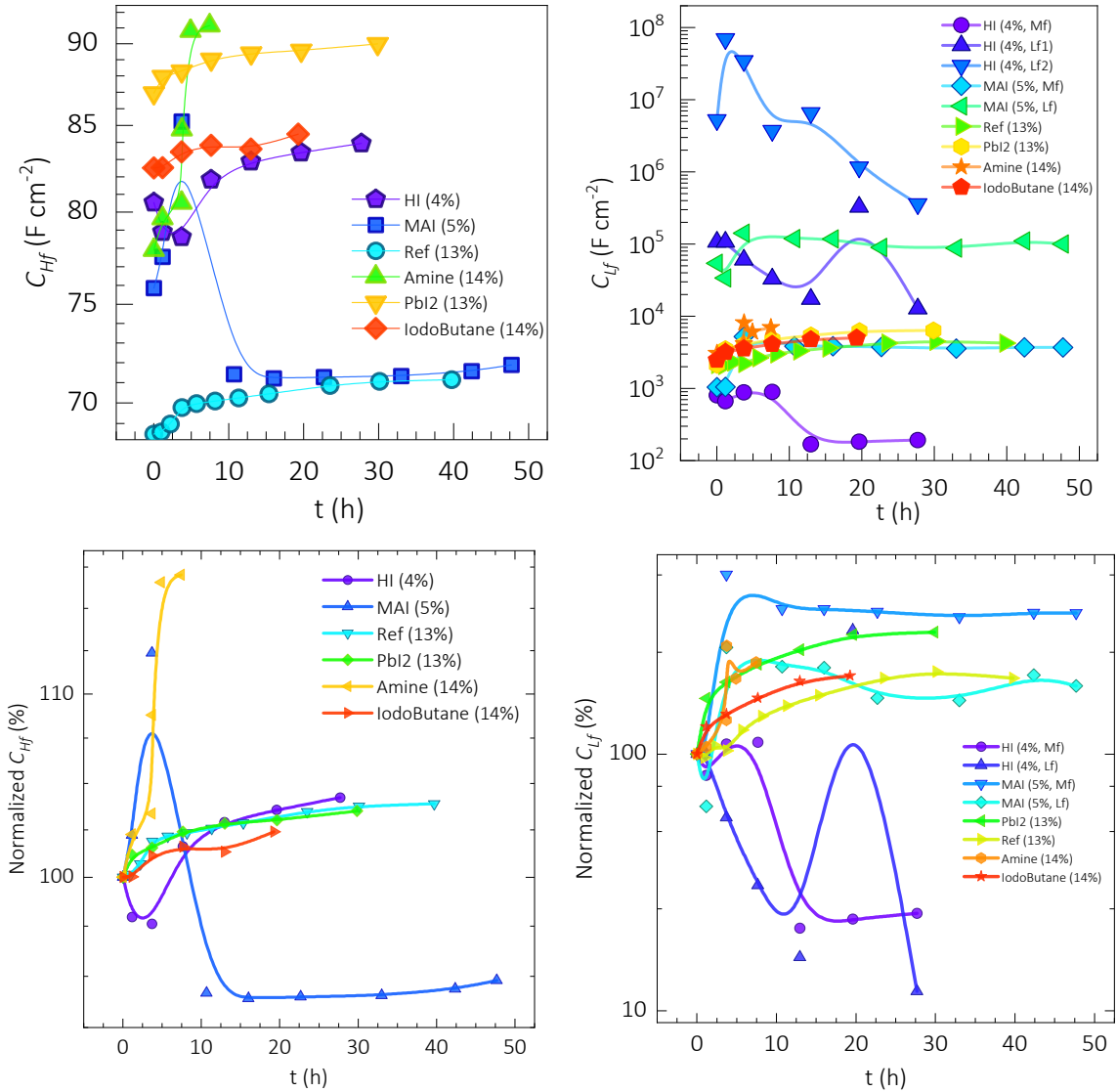

**Figure S21.** Capacitance evolution over time during the stability test in short-circuit condition under 0.2 sun white LED equivalent illumination. The values are the result of numerical fittings of the spectra in **Figure S20** to the equivalent circuit models in **Figure S5**.

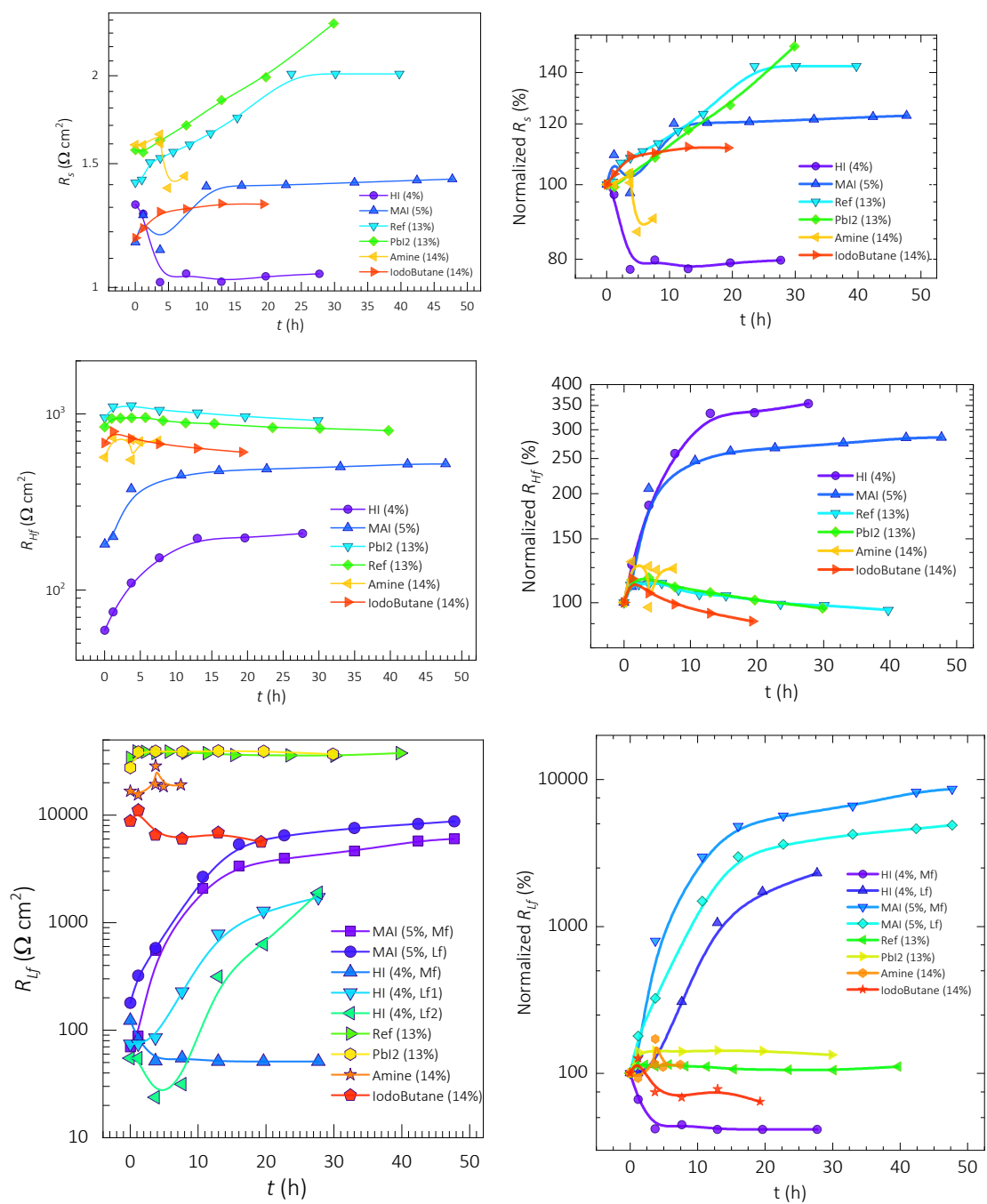

**Figure S22.** Resistance evolution over time during the stability test in short-circuit condition under 0.2 sun white LED equivalent illumination for different samples, as indicated. The values are the result of numerical fittings of the spectra in **Figure S18** to the equivalent circuit models in **Figure S5**.

# S5. SEFTOS-FLUXIM[7] simulations

Table S9. Simulation parameters for Setfos-Fluxim

| Parameter | Value | Unit |
|---|---|---|
| *Active layer (perovskite)* | | |
| Dielectric constant | 24.1 | |
| Thickness | 400 | nm |
| Bimolecular recombination pre-factor | 9.40E-10 | cm$^3$/s |
| Langevin efficiency | 3.25E-04 | |
| Trap density | 1.00E+16 | cm$^3$ |
| Electron/hole pseudo-lifetime | 5.00E-09 | s |
| Electron/ hole capture rates | 2.00E-08 | cm$^3$/s |
| Electron/hole Diffusion coefficient | 5.00E-05 | m$^2$/s |
| Electron/hole mobility | 1.92E+01 | cm$^2$/sV |
| *Hole conductor (NiOx)* | | |
| Dielectric constant | 3 | |
| Thickness | 30 | nm |
| Hole Diffusion coefficient | 2.60E-05 | cm$^2$/s |
| Hole Mobility | 1.00E-03 | cm$^2$/sV |
| Doping density (acceptor) | 1.40E+20 | cm$^{-3}$ |
| *Electron conductor (C$_{60}$)* | | |
| Dielectric constant | 3.9 | |
| Thickness | 30 | nm |
| Electron Diffusion coefficient | 2.31E-05 | cm$^2$/s |
| Electron Mobility | 8.90E-04 | cm$^2$/sV |
| Doping density (donors) | 1.50E+18 | cm$^{-3}$ |
| **External parameters** | | |
| Shunt resistance | 1E6 | Ω cm$^2$ |

S5.1. **Qualitative initial simulation of current-voltage and impedance spectroscopy spectra**

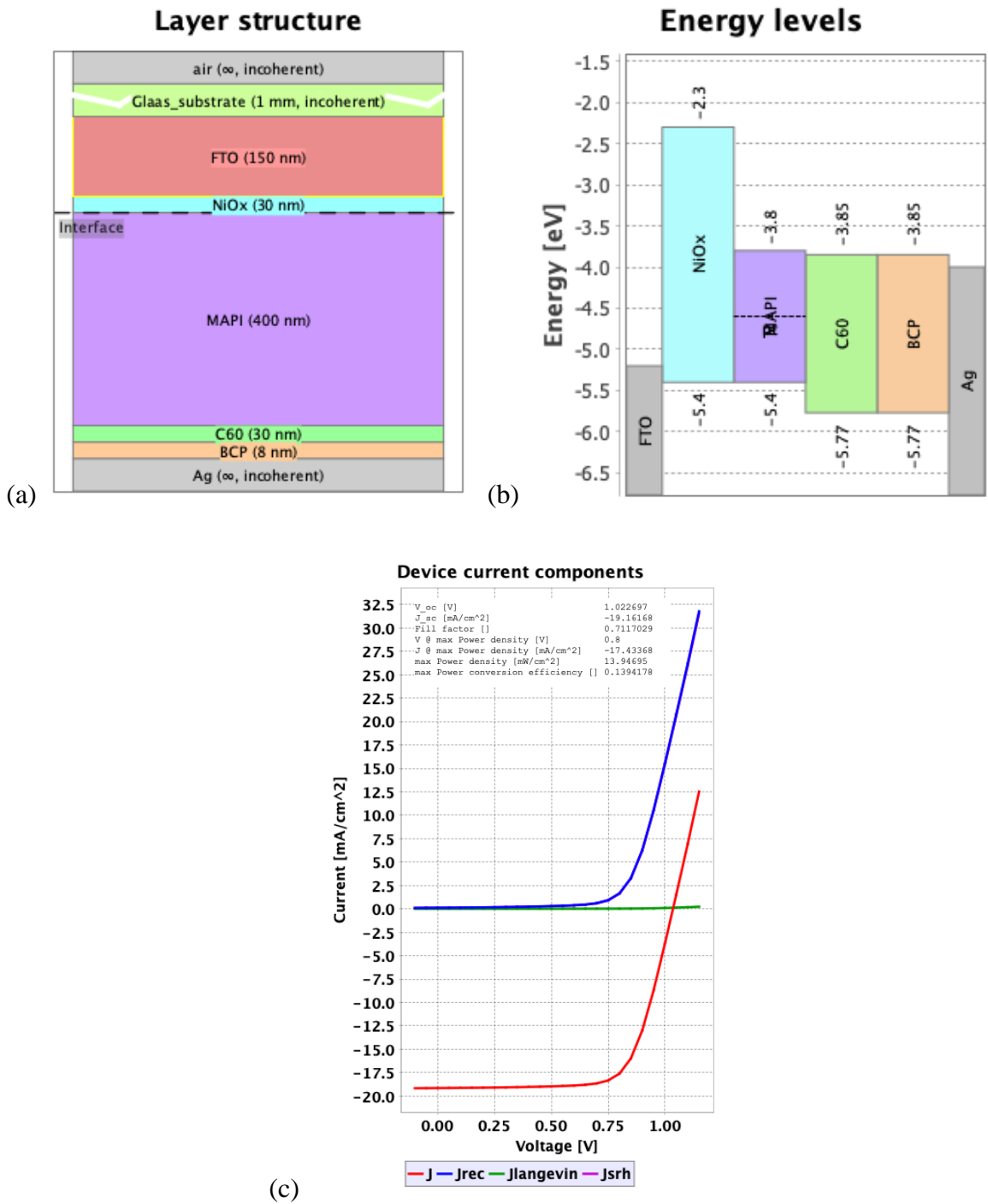

(a)
(b)
(c)

**Figure S23.** Initial settings for Setfos simulator: (a) layer structure, (9) energy diagrams and (c) base simulated current density-voltage curves in dark and under illumination.

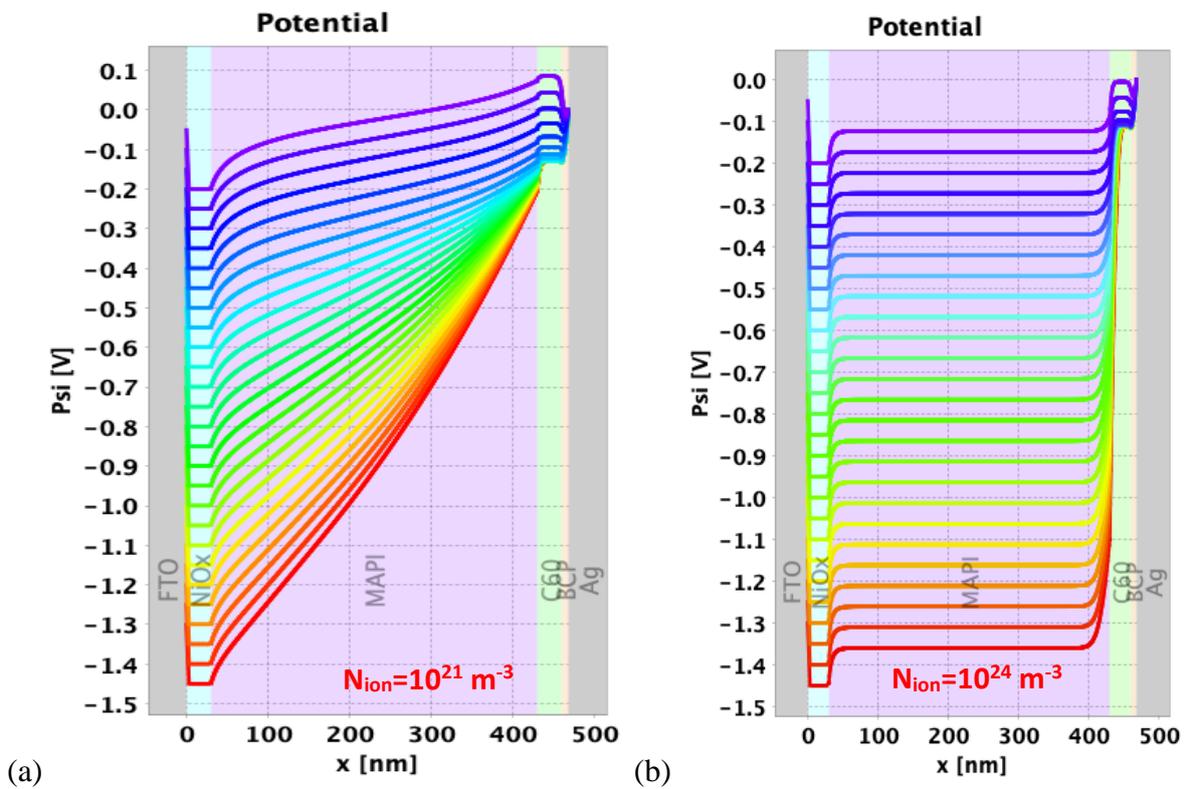

**Figure S24.** Simulated electrical potential (Psi) as a function of the distance (x) between electrodes over the external voltage for mobile ion concentrations of (a) $10^{15}$ cm$^{-3}$ and (b) $10^{18}$ cm$^{-3}$. The simulation parameters are in **Table S9**.

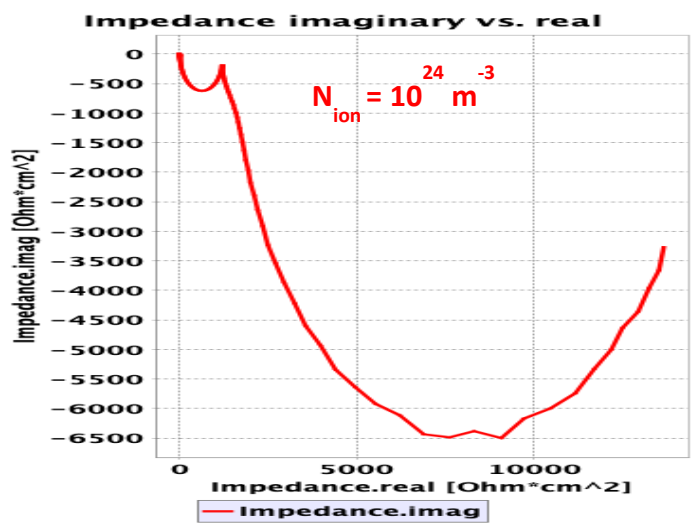

(a)

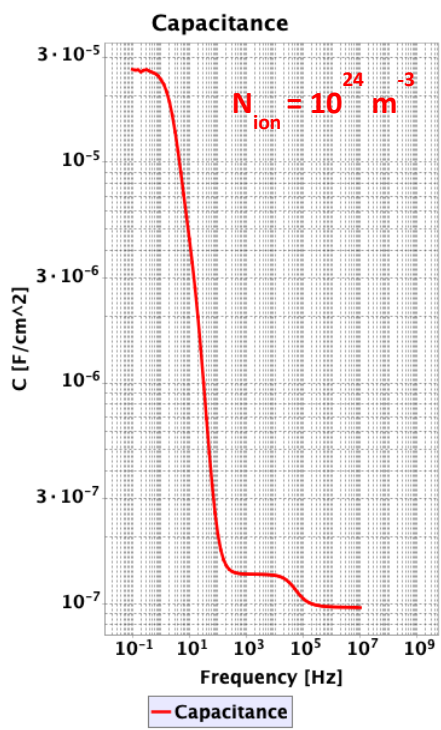

(b)

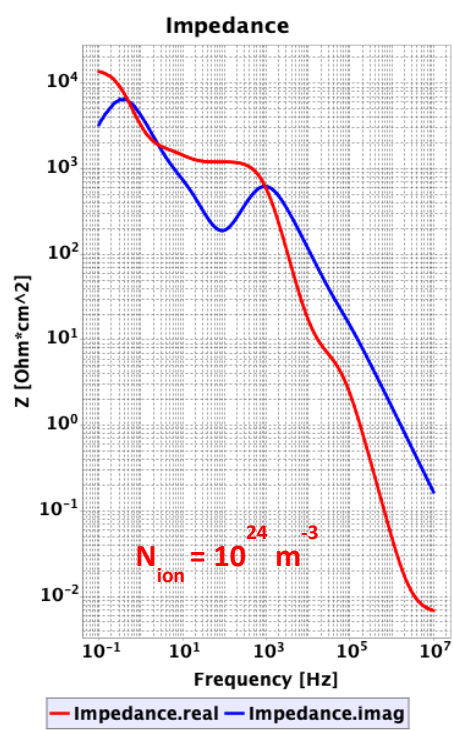

(c)

**Figure S25.** Simulated impedance spectra in short-circuit under 0.2 sun illumination intensity with mobile ion concentrations of $10^{18}$ cm$^{-3}$ in (a) impedance Nyquist plot, (b) capacitance Bode plot and (c) impedance Bode plots. Further simulation parameters are in **Table S9**.

## S5.2. **Incident illumination intensity effects**

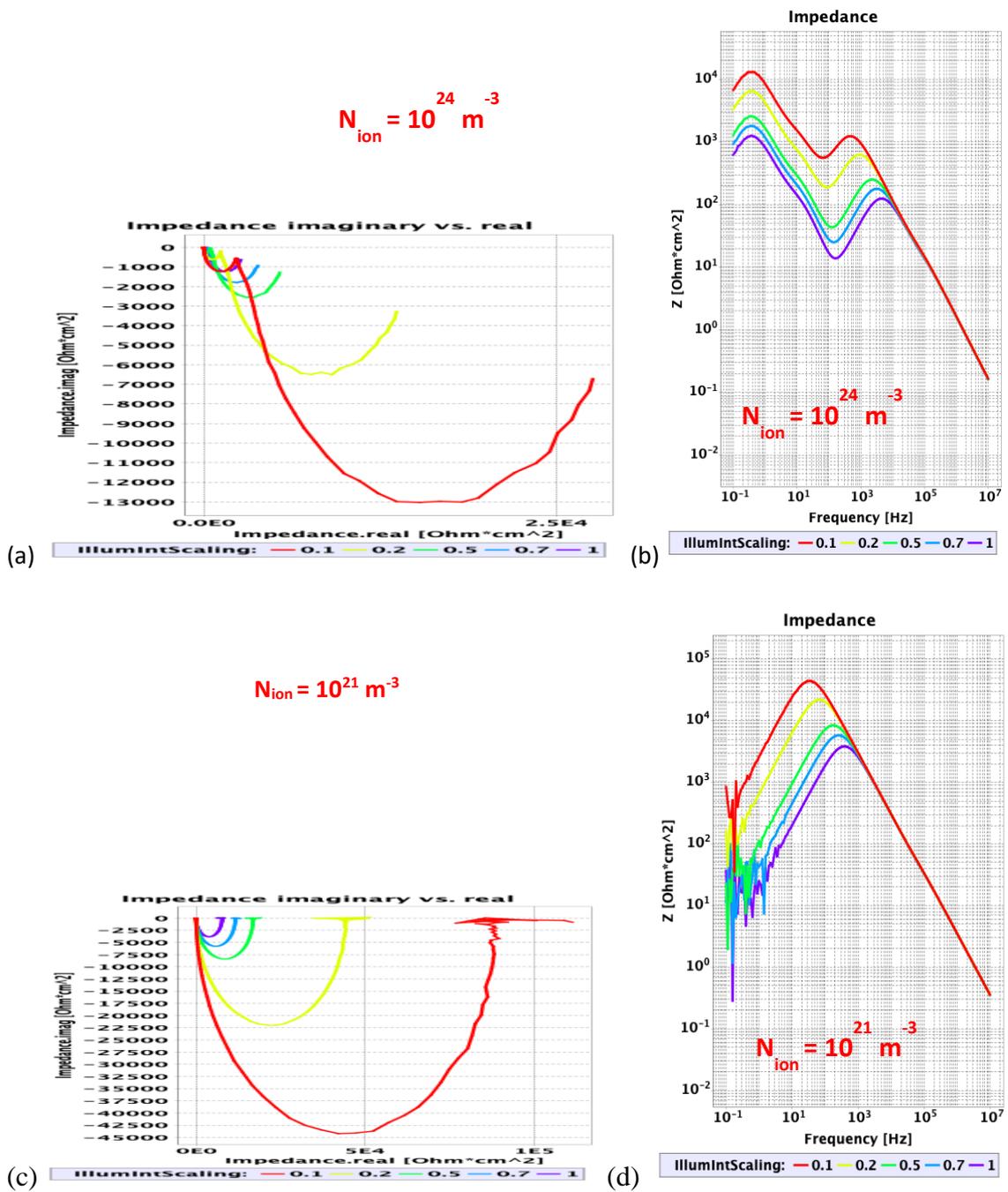

**Figure S26.** Simulated impedance spectra in short-circuit under different illumination intensities (in units of sun) with mobile ion concentrations of (a, b) $10^{18}$ cm$^{-3}$ and (c, d) $10^{15}$ cm$^{-3}$ in (a, c) impedance Nyquist plot, and (b, d) imaginary part of impedance Bode plot. Further simulation parameters are in **Table S9**.

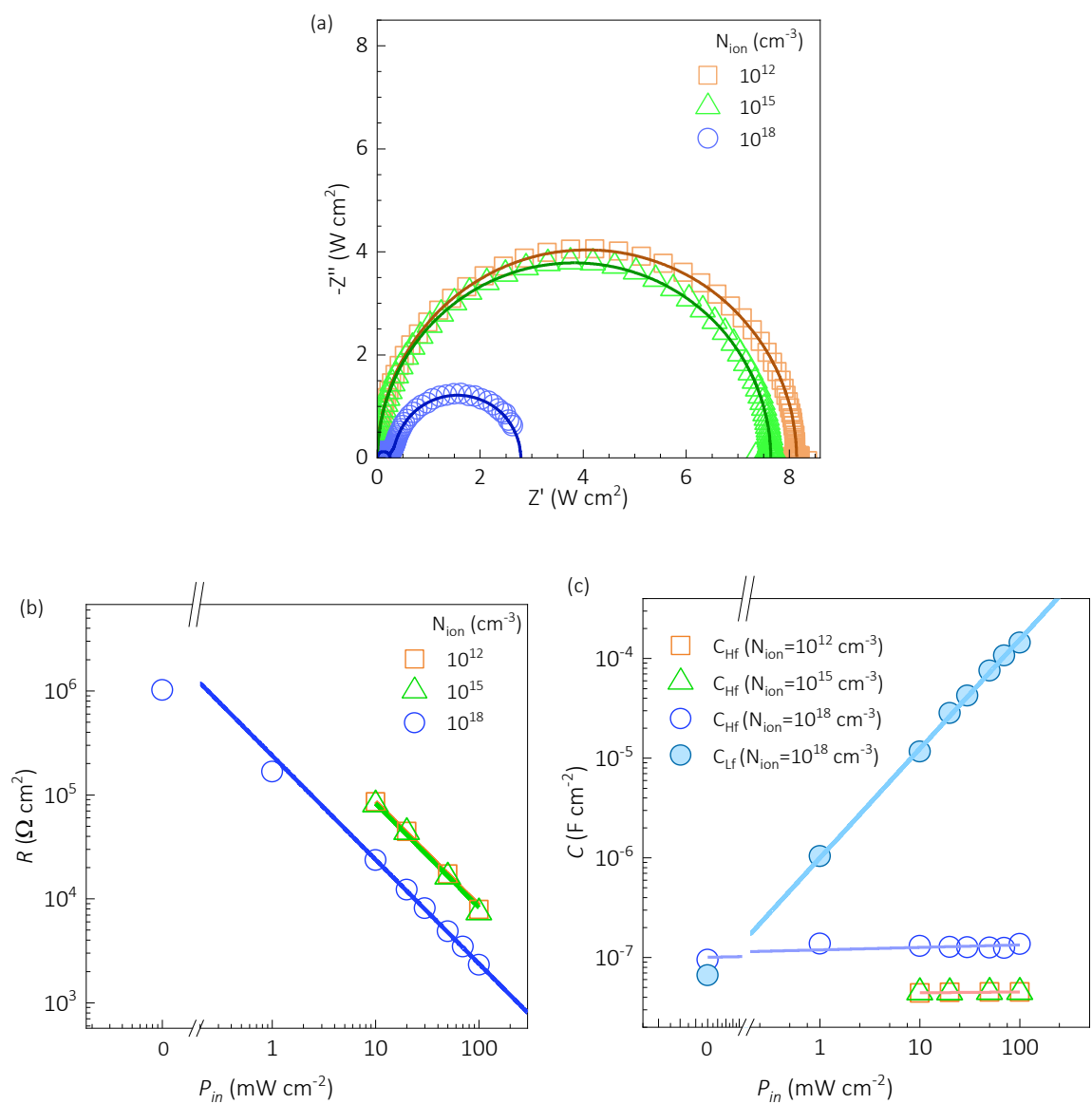

**Figure S27.** Numerical simulation of impedance spectra from the reference sample in short-circuit under different illumination intensities and for different mobile ion concentrations. In (a) there are the Nyquist plots for 1 sun illumination intensity and (b) and (c) show the resultant low frequency resistances and capacitances, respectively, from equivalent circuit modeling. The dots in (a) are Setfos-Fluxim simulations and the lines are Zview equivalent circuit fittings. The solid lines in (b) and (c) are allometric fittings with power -1 and +1, respectively.

## S5.3. Shunt resistance effects

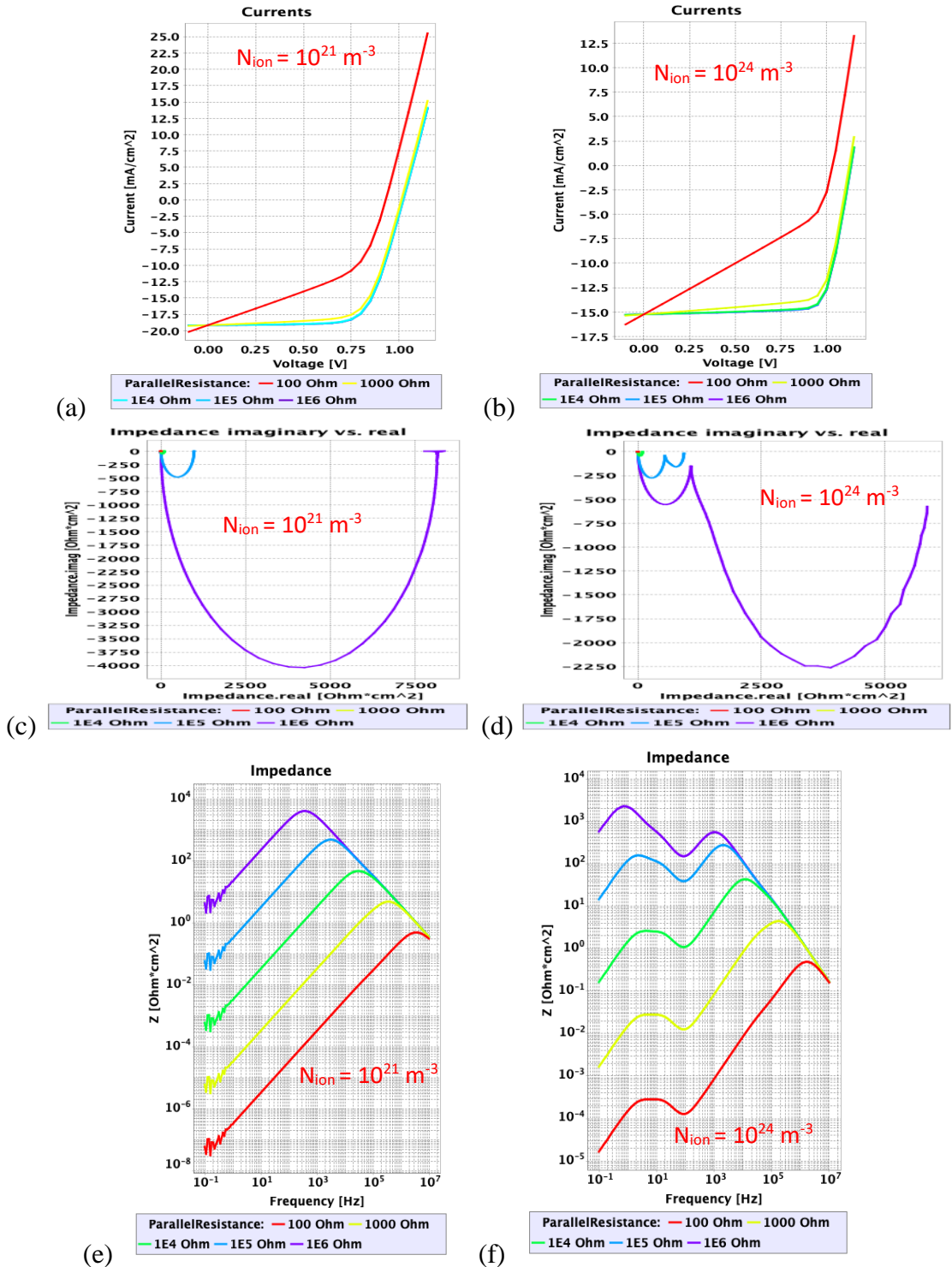

**Figure S28.** Simulated (a, b) current-voltage curves and (c-f) impedance spectra in short-circuit condition under 0.2 sun illumination intensity for different values of shunt resistance, as indicated with mobile ion concentrations of (a, c, e) $10^{15}$ cm$^{-3}$ and (b, d, f) $10^{18}$ cm$^{-3}$ in (c, d) impedance Nyquist plot, and (e, f) impedance imaginary part Bode plots. Further simulation parameters are in **Table S9**.

S5.4. **Ion mobility effects**

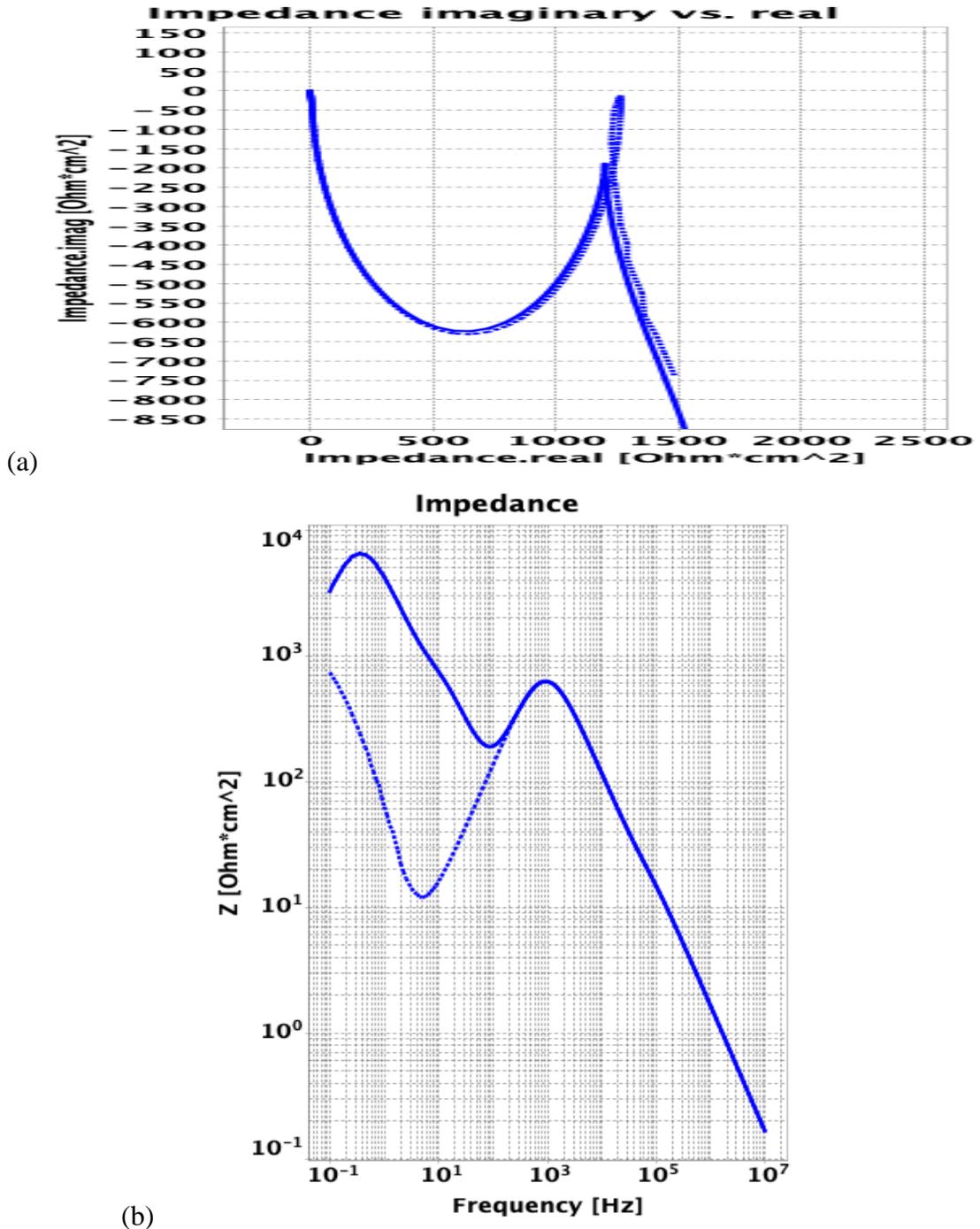

(a)

(b)

**Figure S29.** Simulated impedance spectra in short-circuit under 0.2 sun illumination for a mobile ion concentration of $10^{18}$ cm$^{-3}$ with mobile ion mobility of $10^{-8}$ cm/s (solid line) and $10^{-10}$ cm/s (dashed line). The results are shown in (a) impedance Nyquist plot, and (b) impedance imaginary part Bode plots. The simulation parameters are in **Table S9**.

S5.5.     **Interfacial recombination effect**

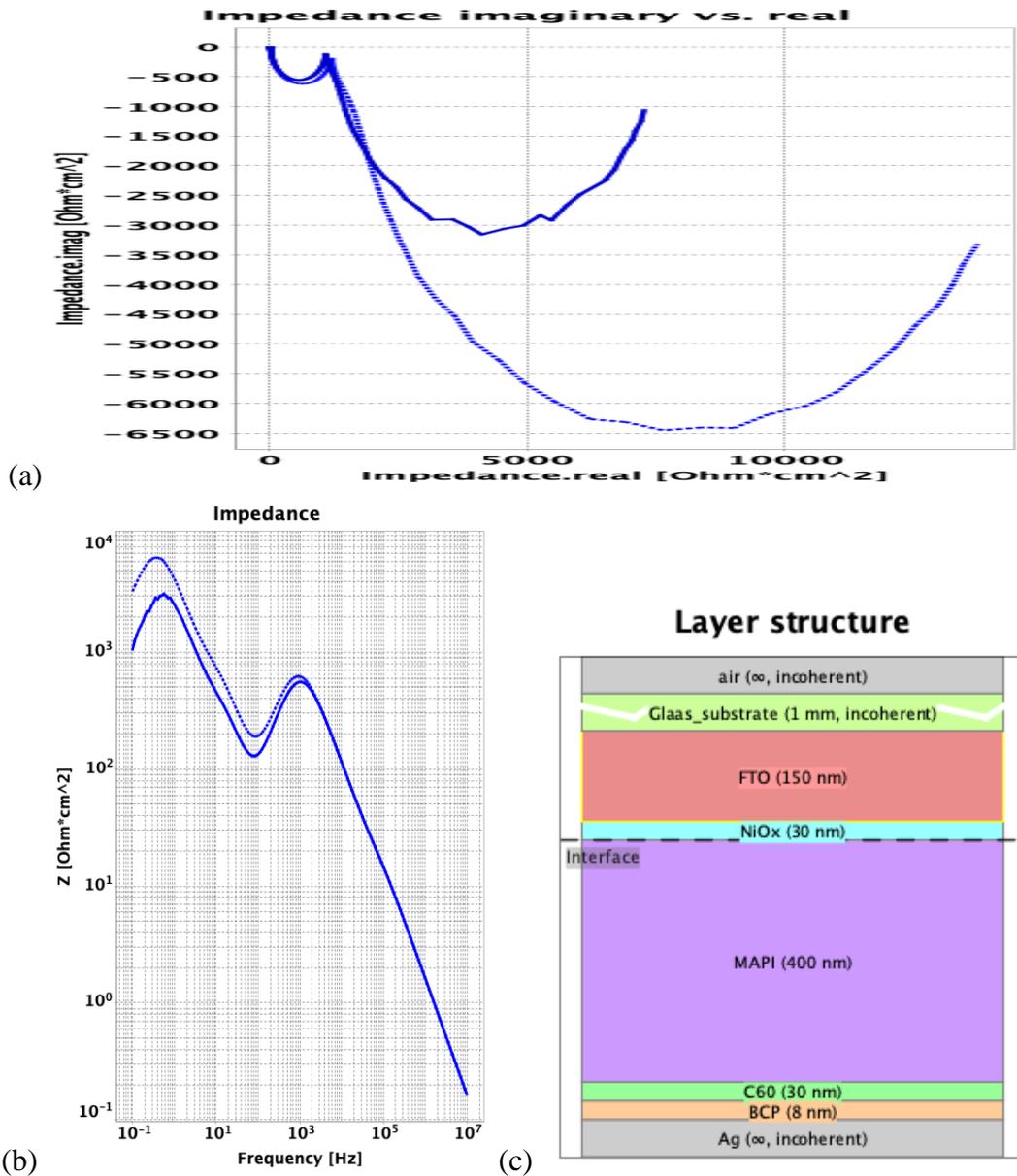

(a)

(b)   (c)

**Figure S30.** Simulated impedance spectra in short-circuit under 0.2 sun illumination for a mobile ion concentration of $10^{18}$ cm$^{-3}$ with an interface (c) recombination velocity of $10^5$ cm/s (solid line) and without interface recombination (dashed line). The results are shown in (a) impedance Nyquist plot, and (b) impedance imaginary part Bode plots. The simulation parameters are in **Table S9**.

## S5.7. HTL (NiOx) doping effect

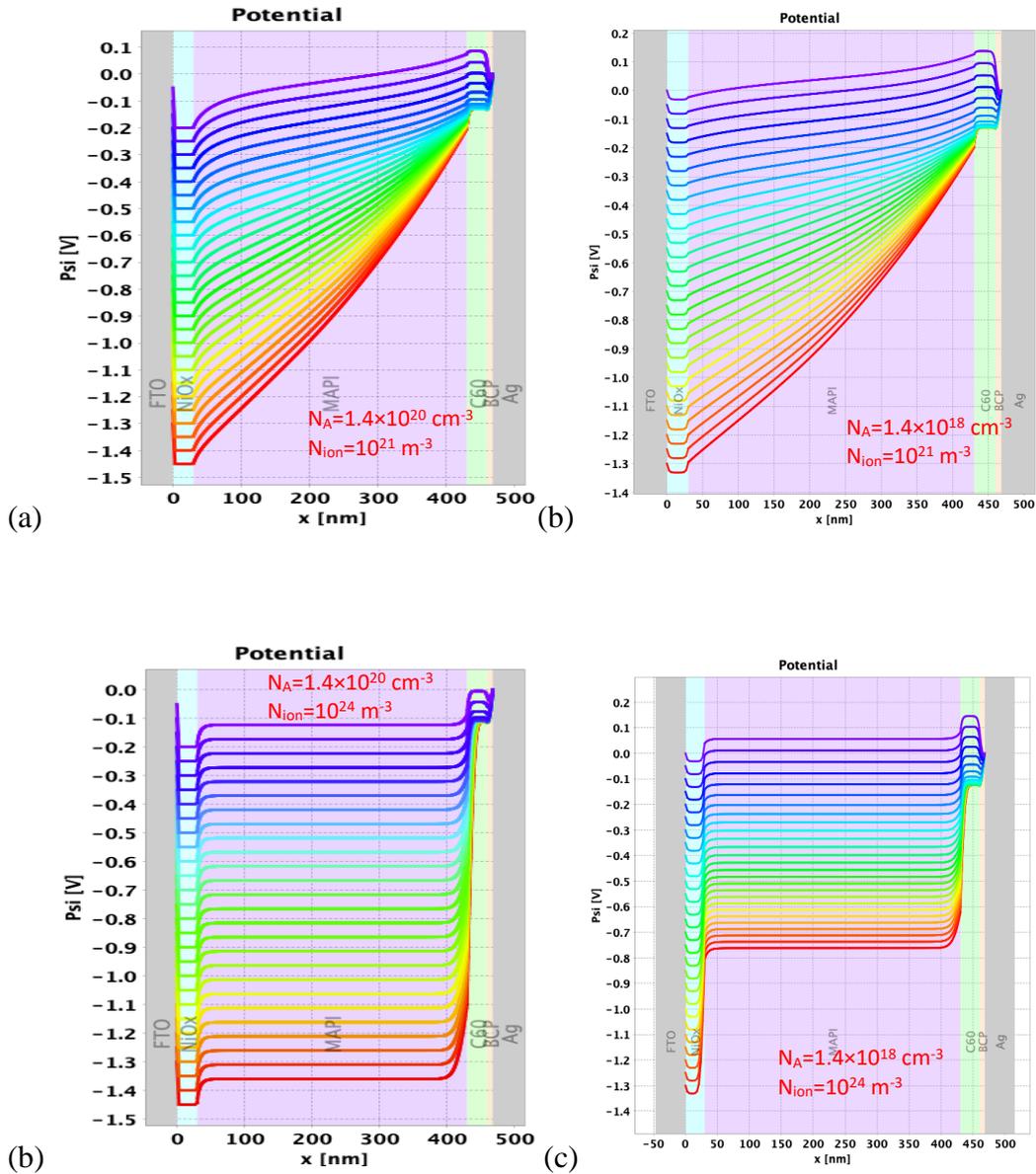

**Figure S31.** Simulated electrostatic potential as a function of external voltage for for different acceptor concentrations at the hole transport layer (HTL), i.e. the NiOx , with (a) $N_A=1.4\times10^{20}$ cm$^{-3}$ and $N_{ion}=10^{15}$ cm$^{-3}$, (b) $N_A=1.4\times10^{18}$ cm$^{-3}$ and $N_{ion}=10^{15}$ cm$^{-3}$, (c) $N_A=1.4\times10^{20}$ cm$^{-3}$ and $N_{ion}=10^{18}$ cm$^{-3}$ and (d) $N_A=1.4\times10^{18}$ cm$^{-3}$ and $N_{ion}=10^{18}$ cm$^{-3}$. Further simulation parameters are in **Table S9**.

# S6. Matlabs's Driftfusion[8] simulations

**Table S10**: Simulation parameters used in this work for the input sheet of Driftfusion.[8]

| Layer type<br>Sack | Contact<br>Anode | Layer<br>NiOx | Interface<br>Interface 1 | Active<br>Perovskite | Interface<br>Interface 2 | Layer<br>$C_{60}$ | Contact<br>Cathode |
|---|---|---|---|---|---|---|---|
| Thickness (cm) | -- | 2E-6 | 1E-6 | 4E-5 | 1E-6 | 3E-6 | -- |
| Electron affinity (eV) | -- | -1.85 | | -3.8 | | -4.1 | -- |
| Ionization potential (eV) | -- | -5.4 | | -5.4 | | -5.8 | -- |
| Work function (eV) | -5.2 | -5.2 | | -4.6 | | -4.25 | -4.2 |
| Trap energy level (eV) | -- | -3.65 | | -4.6 | | -4.95 | -- |
| Effective density of states at the conduction band ($cm^{-3}$) | -- | 1E19 | | 1E19 | | 1E20 | -- |
| Effective density of states at the valence band ($cm^{-3}$) | -- | 1E19 | | 1E19 | | 1E20 | -- |
| Equilibrium concentration of cations ($cm^{-3}$) | -- | 0 | 0 | 1E16 | 0 | 0 | -- |
| Equilibrium concentration of anions ($cm^{-3}$) | -- | 0 | 0 | 1E16 | 0 | 0 | -- |
| Maximum concentration of cations ($cm^{-3}$) | -- | 0 | 0 | 1.21E22 | 0 | 0 | -- |
| Maximum concentration of anions ($cm^{-3}$) | -- | 0 | 0 | 1.21E22 | 0 | 0 | -- |
| Charge carrier mobility of electrons ($cm^2V^{-1}s^{-1}$) | -- | 0.001 | | 2 | | 0.001 | -- |
| Charge carrier mobility of holes ($cm^2V^{-1}s^{-1}$) | -- | 0.001 | | 2 | | 0.001 | -- |
| Charge carrier mobility of cations ($cm^2V^{-1}s^{-1}$) | -- | 0 | 0 | 1E-8 | 0 | 0 | -- |
| Charge carrier mobility of anions ($cm^2V^{-1}s^{-1}$) | -- | 0 | 0 | 1E-10 | 0 | 0 | -- |
| Dielectric constant | -- | 3 | | 64 | | 3 | -- |
| Generation rate ($s^{-1}cm^{-3}$) | -- | 0 | 0 | 2e21 (1 sun) | 0 | 0 | -- |
| Band-to-band radiative recombination rate ($s^{-1}cm^3$) | -- | 6.3E-11 | 0 | 4.8E-11 | 0 | 6.8E-11 | -- |
| Shockley-Read-Hall non-radiative recombination lifetime for electrons (s) | -- | 1E-6 | 1E-10 | 0.8E-7 | 1E-10 | 1E-6 | -- |
| Shockley-Read-Hall non-radiative recombination lifetime for holes (s) | -- | 1E-6 | 1E-10 | 0.8E-7 | 1E-10 | 1E-6 | -- |
| Recombination velocity for electrons ($cm·s^{-1}$) | 1E7 | -- | | -- | | -- | 1E7 |
| Recombination velocity for holes ($cm·s^{-1}$) | 1E7 | -- | | -- | | -- | 1E7 |
| Shunt Resistance ($\Omega cm^2$) | 1E6 | | | | | | |

## S6.1. Qualitative initial simulation of current-voltage and impedance spectroscopy spectra

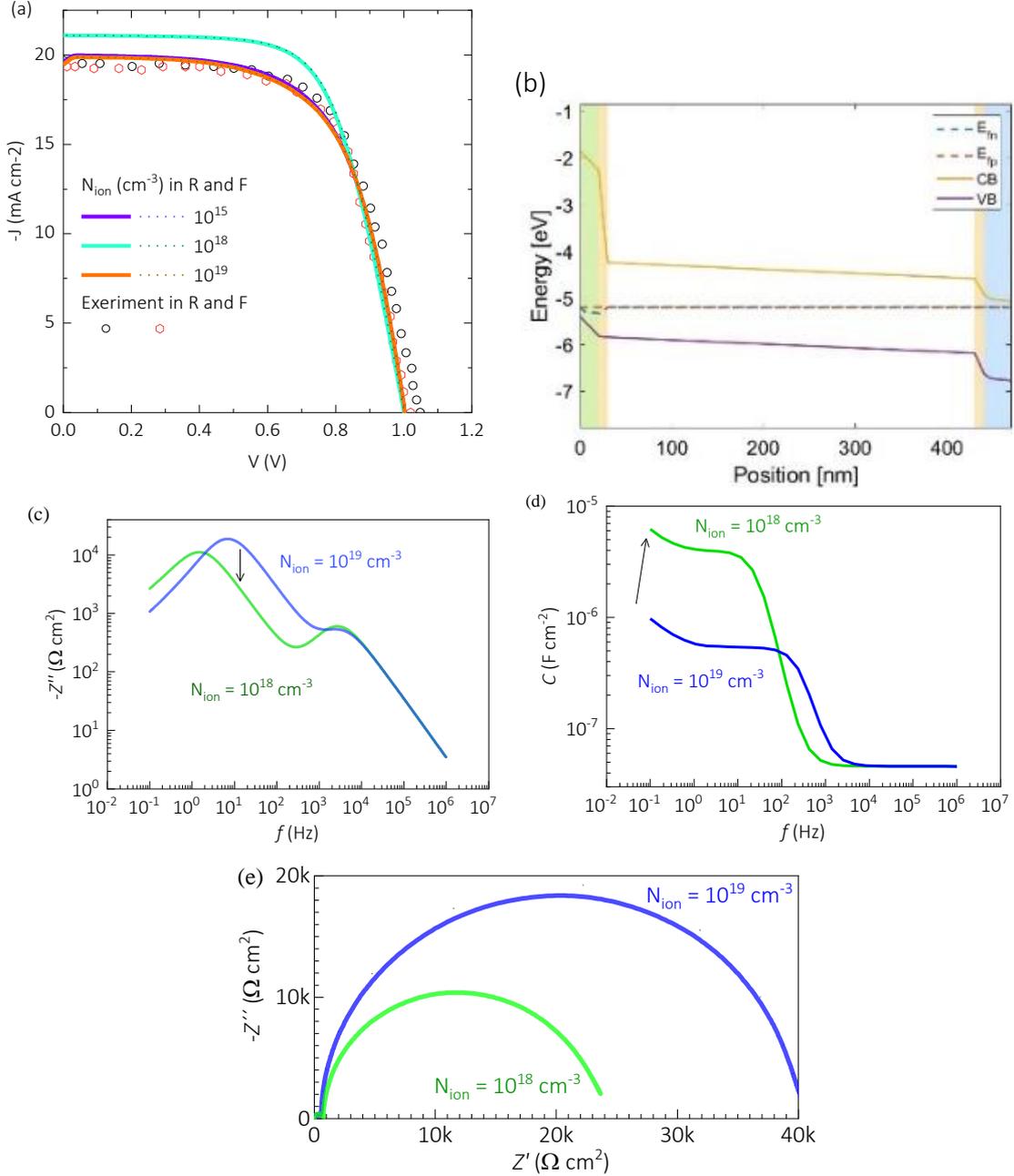

**Figure S32.** (a) The experimental current-voltage curve of the reference structure (dots) and the numerical simulations (lines) considering different ion concentrations and voltage sweep directions. (b) Energy band diagram of the simulated structure: $NiO_x$/perovskite/$C_{60}$ considering $N_{ion}=10^{15}$ cm$^{-3}$, 0.2 sun equivalent illumination intensity, and short-circuit condition. The corresponding simulated impedance spectra are shown the Bode plots of (c) imaginary part of impedance and (d) capacitance and the (e) impedance Nyquist plot. The simulation parameters for the current-voltage curve include $V_{bi}=1$V, $G=33\times10^{20}$ cm$^{-3}$s$^{-1}$, $R_{sh}=1$ MΩ·cm$^2$, in addition to those parameters in **Table S10**.

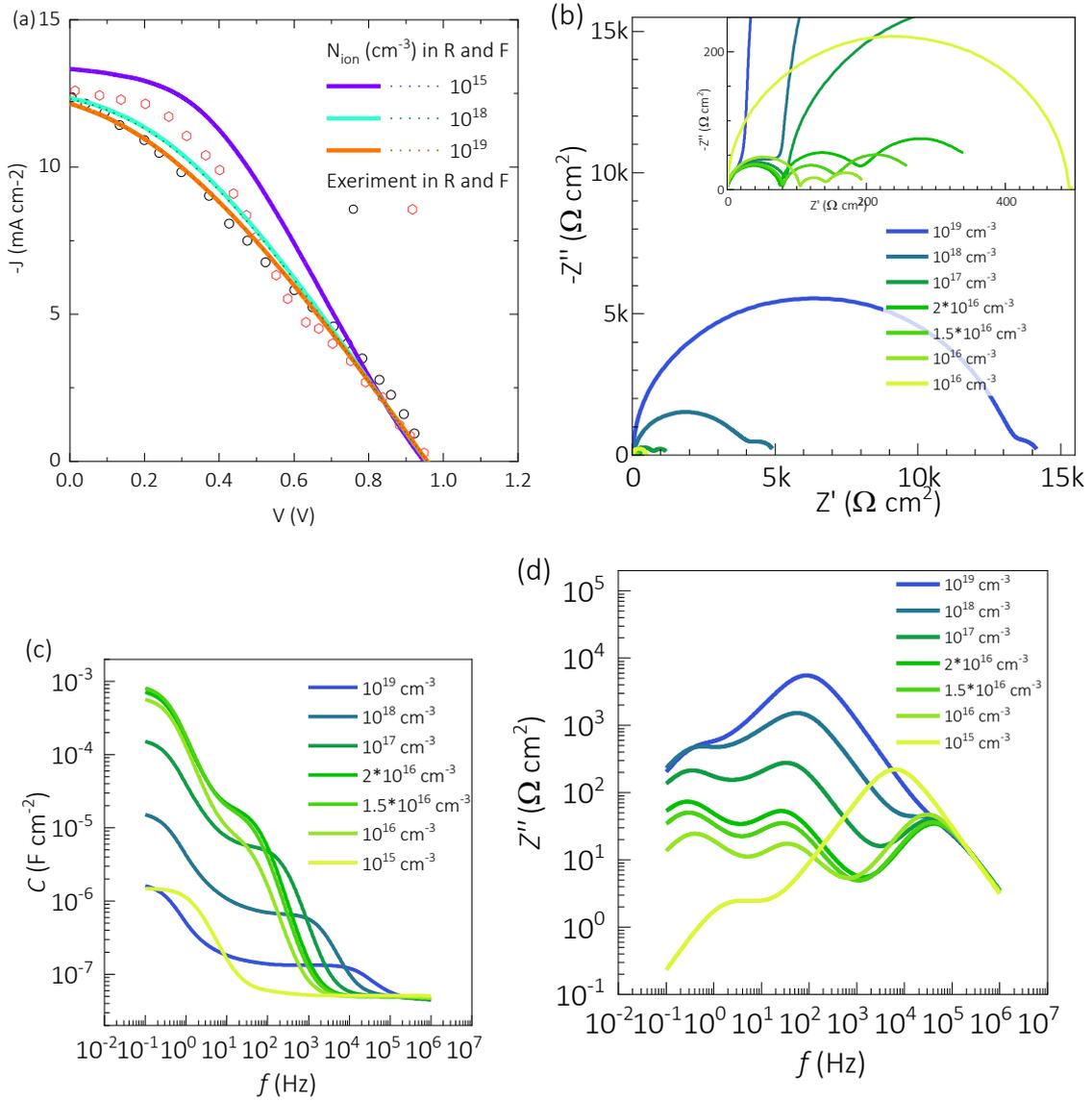

**Figure S33.** Simulations for the the MAI and HI samples. The current-voltage curves in (a) compare the experimental data of (dots) and the numerical simulations (lines) considering different ion concentrations and voltage sweep directions (solid and dashed lines for forward and backward directions). The simulated impedance spectra with the characteristic three arcs are shown in (b) impedance Nyquist plot and the Bode plots of (a) capacitance and (d) imaginary parts of impedance. The simulation parameters for the current-voltage curve include scan rate of 70 mV·s$^{-1}$, $V_{bi}$=1V, $\mu$= 10$^{-4}$ cm$^2$V$^{-1}$s$^{-1}$, $G$=4×10$^{20}$ cm$^{-3}$s$^{-1}$, $R_{sh}$=1 MΩ·cm$^2$, in addition to those parameters in **Table S10**.

## S6.2. Incident illumination intensity effects

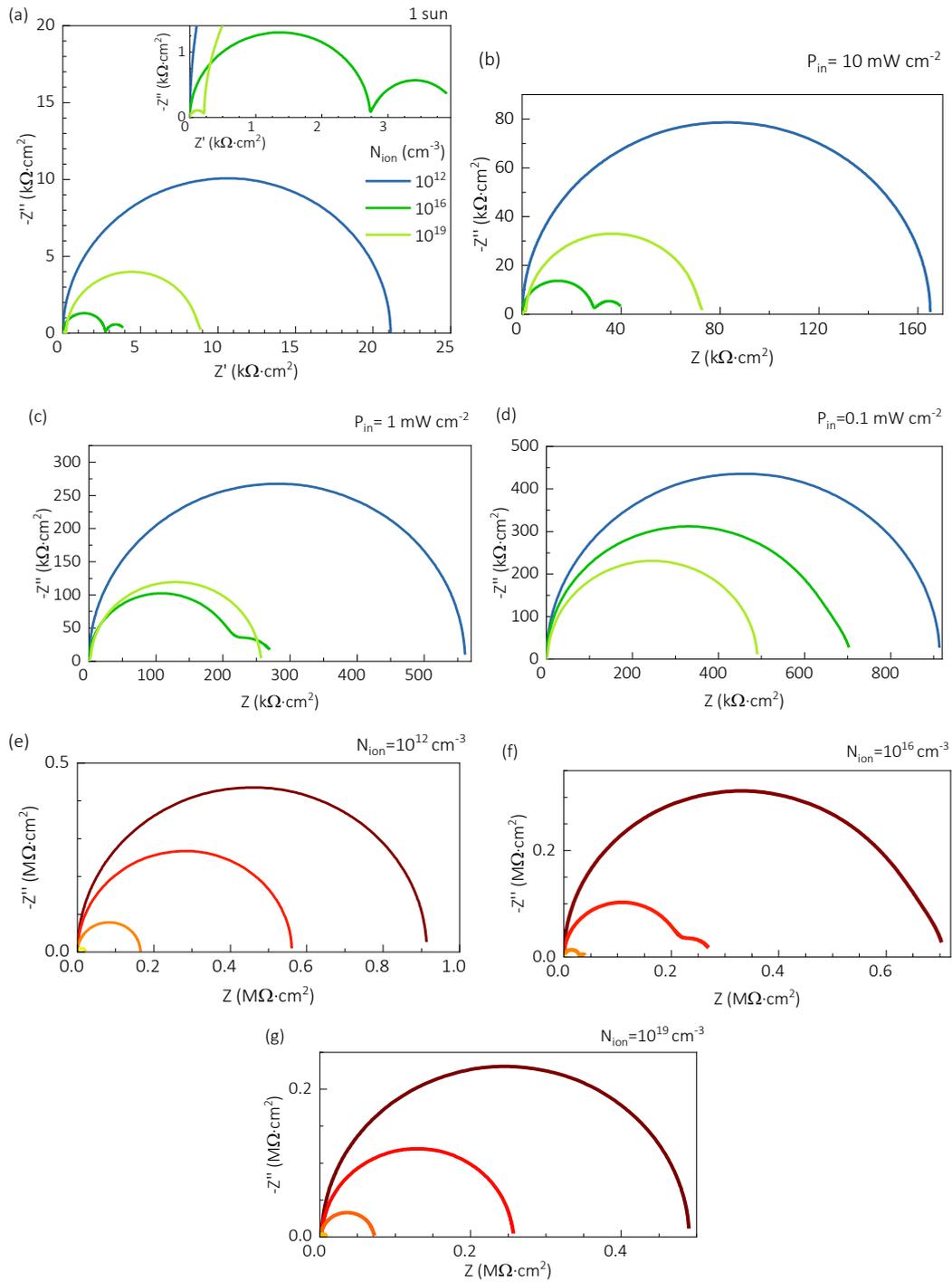

**Figure S34.** Simulated effect of the light intensity on the IS Nyquist plot. The initial generation rate $G = 2 \cdot 10^{21}$ cm$^{-3}$ is reduced by a factor of 1, 0.1, 0.01, 0.001 in (a-d), respectively, for $N_{ion} = 10^{12}$ (blue), $10^{16}$ (green), $10^{19}$ (lime) cm$^{-3}$. The same data is reordered for each ion concentration in (e-g), where darker (brown) line colours indicate lower illumination intensities. The simulated shunt resistance was 1 MΩ.

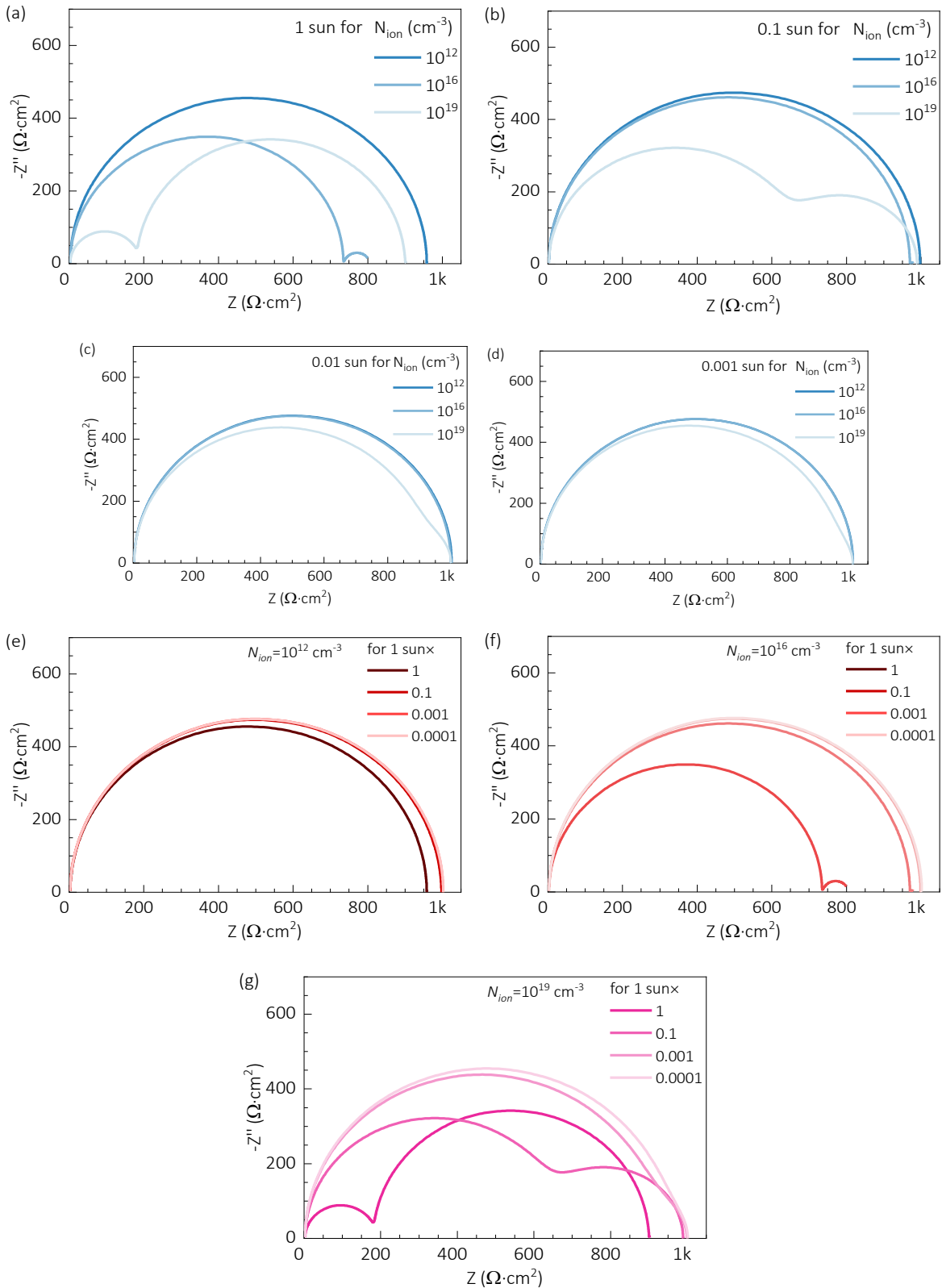

**Figure S35.** Effect of the light intensity (the initial generation rate is reduced by a factor of 1, 0.1, 0.01, 0.001) on the Nyquist plot. At lower light intensity higher is the impact of mobile ions on the total impedance of the system. Simulated ion concentrations are: $N_{ion}= 10^{12}$(blue), $10^{16}$(green), $10^{19}$(red) cm$^{-3}$. The simulation parameters include shunt resistance $R_{sh}=1$ k$\Omega$, besides those in the Table S10.

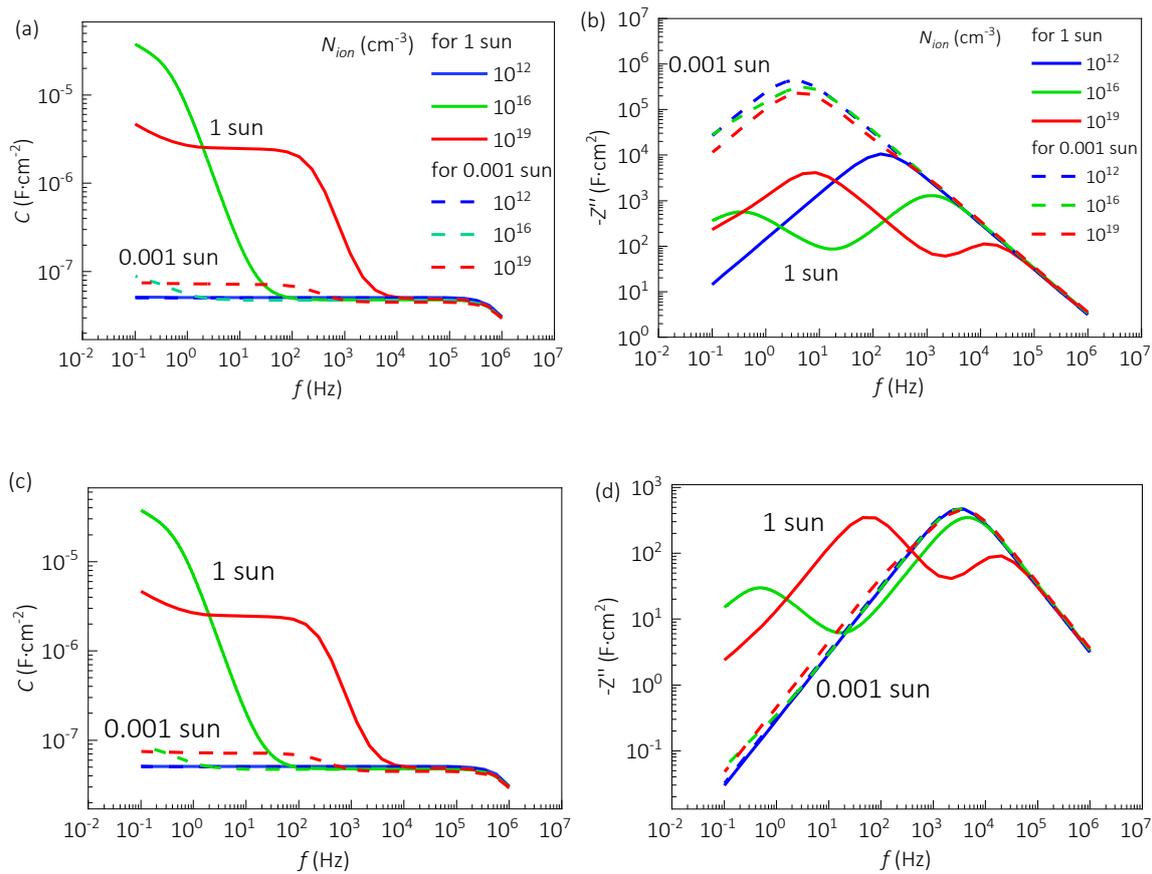

**Figure S36.** Simulated IS spectra with the effect of the light intensity (the initial generation rate is reduced by a factor of 1, 0.001) on the capacitance and imaginary-frequency plot for different values of mobile ion concentration ($N_{ion}$= $10^{12}$(blue), $10^{16}$(green), $10^{19}$(red) cm$^{-3}$) and shunt resistance (1 MΩ and 1 kΩ in top and bottom, respectively). Further simulation parameters are in Table S10.

## S6.3. **Shunt resistance effect**

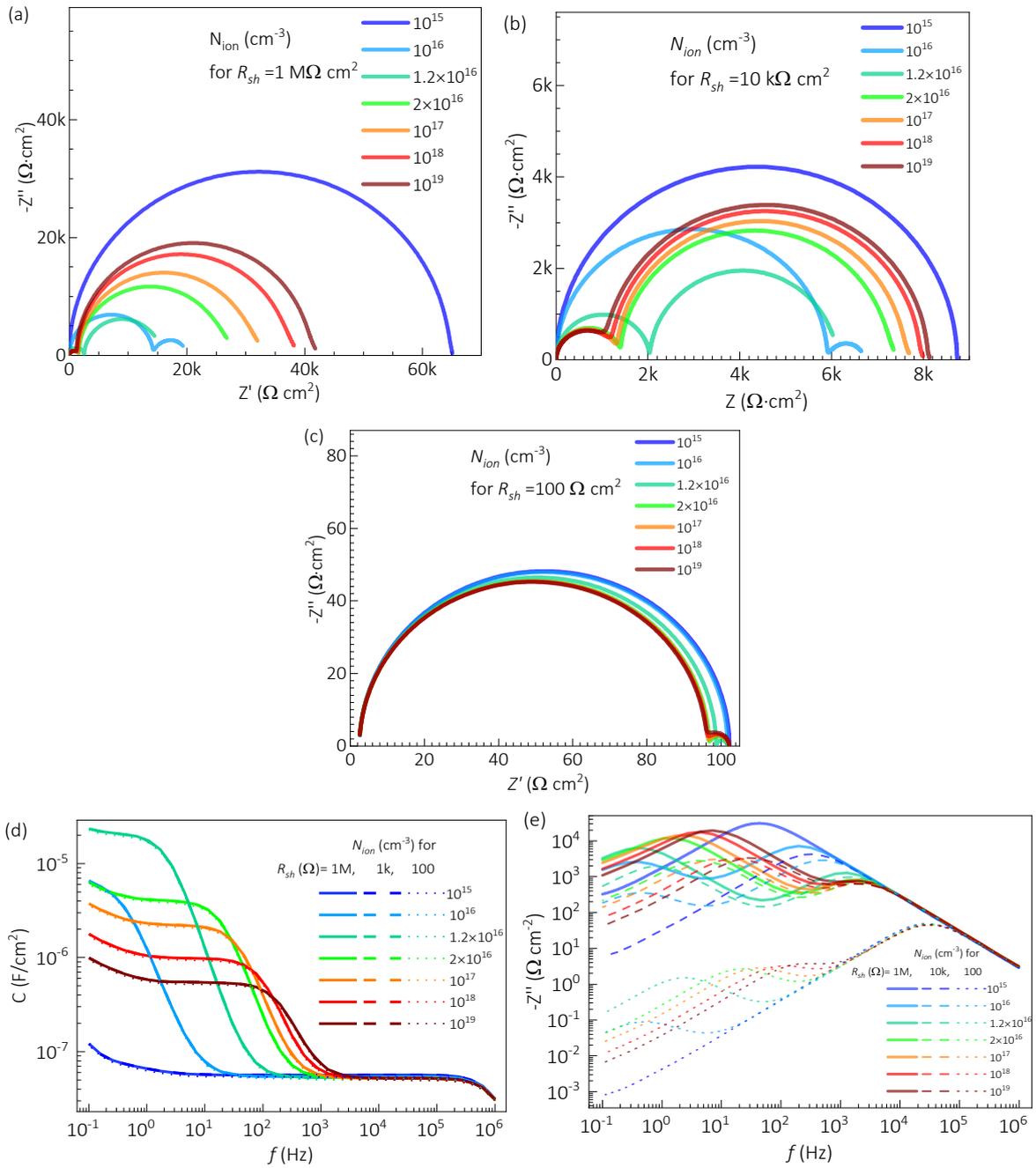

**Figure S37.** Simulated impedance spectra for the reference sample considering different shunt resistance and ion concentrations, as indicated. The physical simulation parameters are in the **Table S10**.

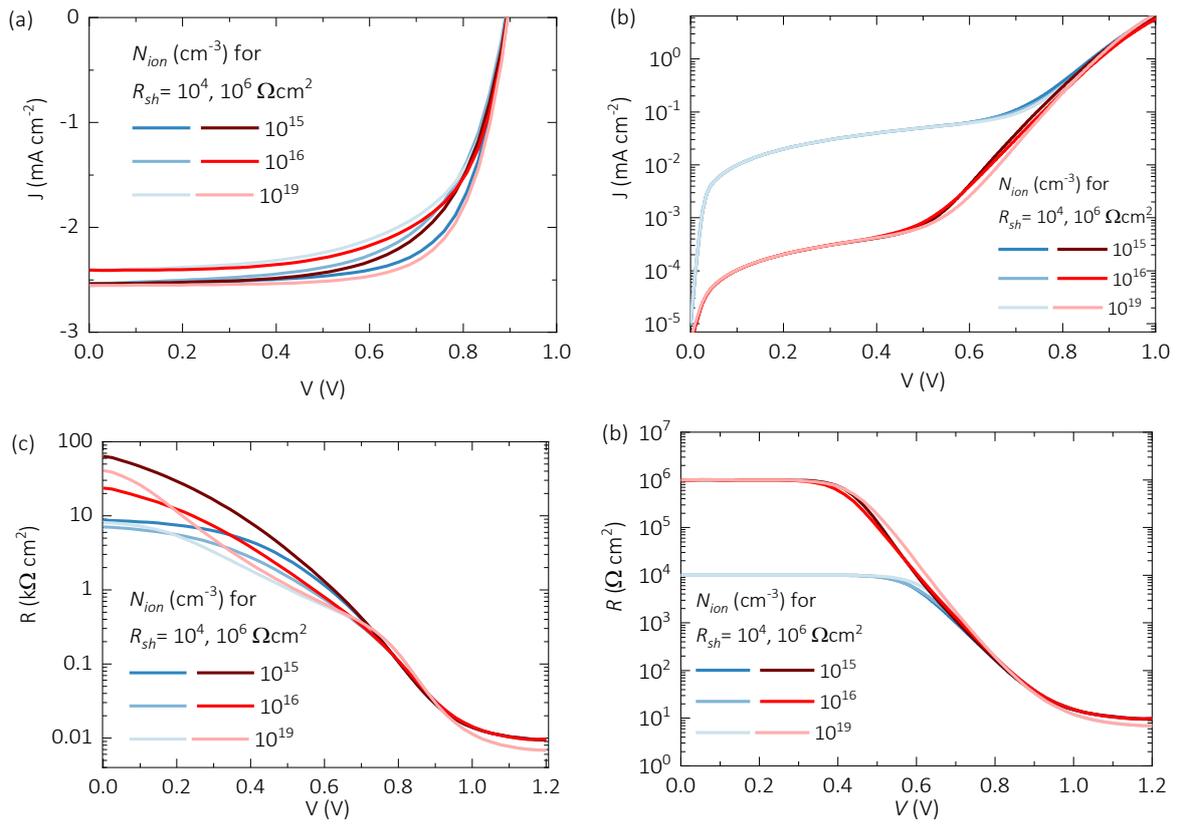

**Figure S38.** Simulated current-voltage curves (a) under 0.2 sun illumination intensity and (b) in dark, with DC resistances (in c, d, respectively) for the reference sample considering different shunt resistance and mobile ion concentrations, as indicated. The physical simulation parameters are in the **Table S10**.

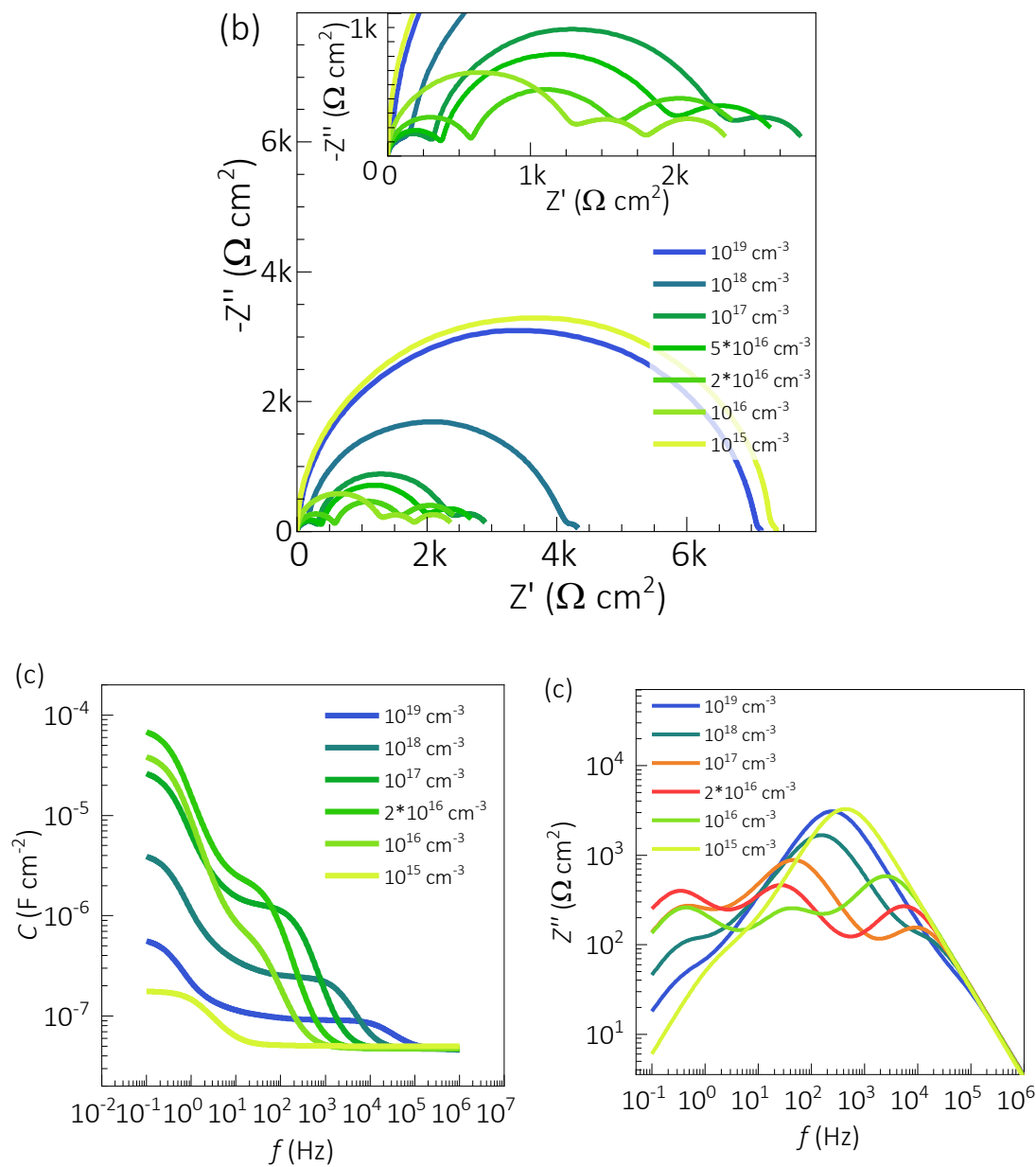

**Figure S39.** Simulated impedance spectra considering different **high shunt resistance** (1 MΩ) ion concentrations in (a) impedance Nyquist representation, (b) capacitance Bode Plot and (c) imaginary part of impedance versus frequency. The physical parameters are in the **Table S10**.

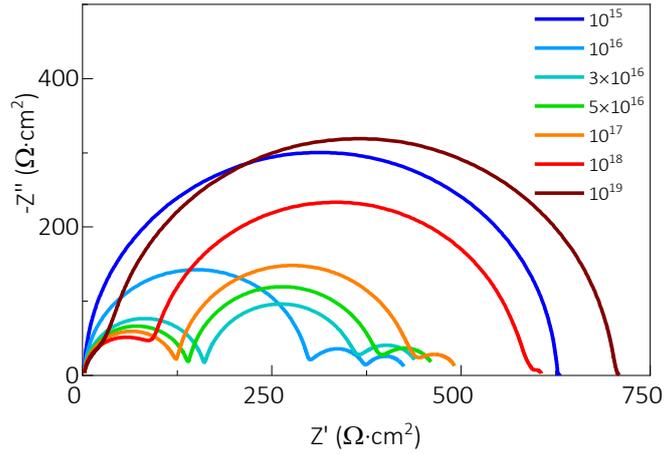

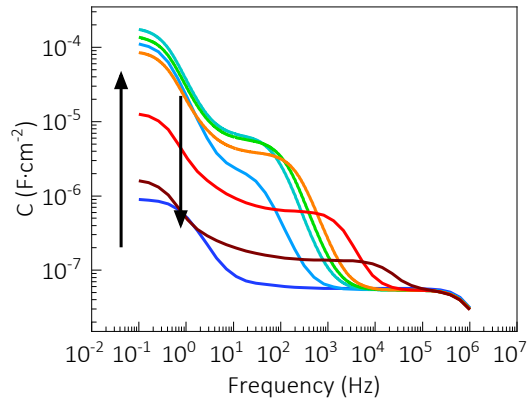

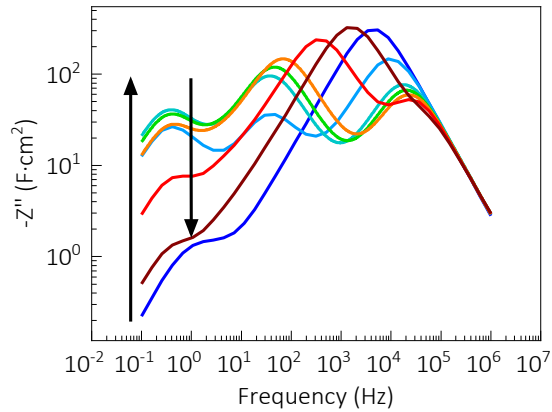

**Figure S40.** Simulated impedance spectra for the HI or MAI samples with combined effect of **low shunt resistance** (1 kΩ cm$^2$), positive/negative ion mobility ($10^{-6}/10^{-8}$ cm$^2$V$^{-1}$s$^{-1}$), low electron/hole mobility (0.2 cm$^2$V$^{-1}$s$^{-1}$) and high light intensity $G_0 = 2\cdot10^{21}$ cm$^{-3}$s$^{-1}$. Simulated ion concentrations are: $N_{ion}$= $10^{15}$, $10^{16}$, $3\times10^{16}$, $5\times10^{16}$, $10^{18}$, $10^{19}$ cm$^{-3}$.

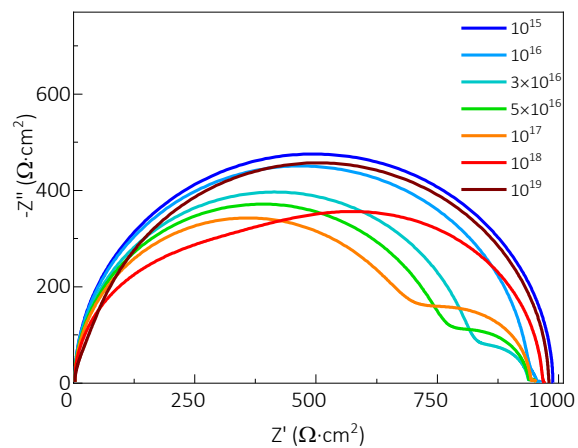

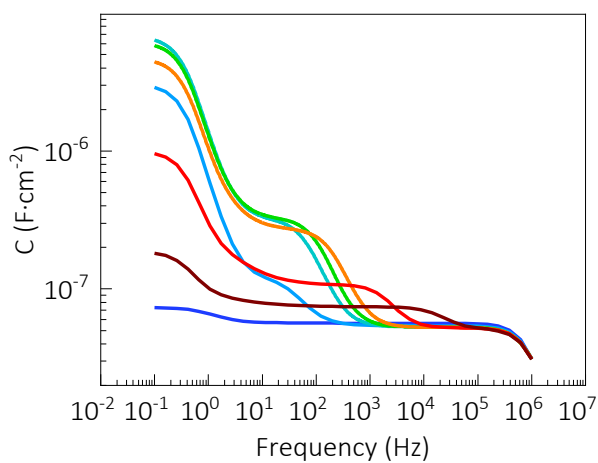

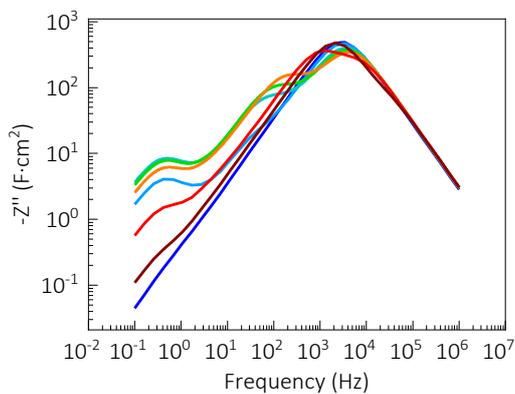

**Figure S41.** Simulated impedance spectra for the HI or MAI samples with lower generation rate than those of **Figure S40**. At low light intensity, or upon reduction of optical absorption, the lower the shunt resistance the more negligible the three-arcs effects within the explored perturbation frequency for high concentration of mobile ions and electron mobility. Low shunt resistance: 1 kΩ cm$^2$, positive/negative ion mobility: $10^{-6}/10^{-8}$ cm$^2$V$^{-1}$s$^{-1}$, electron/hole mobility 2 cm$^2$V$^{-1}$s$^{-1}$ and low light intensity $G_0 = 4\cdot 10^{20}$ cm$^{-3}$s$^{-1}$. Simulated ion concentrations are: $N_{ion}= 10^{15}, 10^{16}, 3\times 10^{16}, 5\times 10^{16}, 10^{18}, 10^{19}$ cm$^{-3}$.

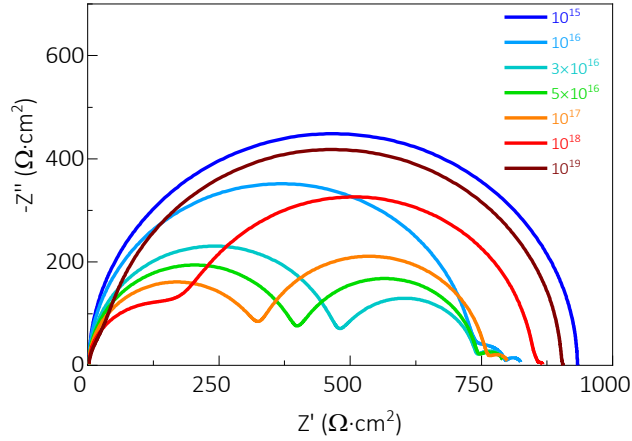

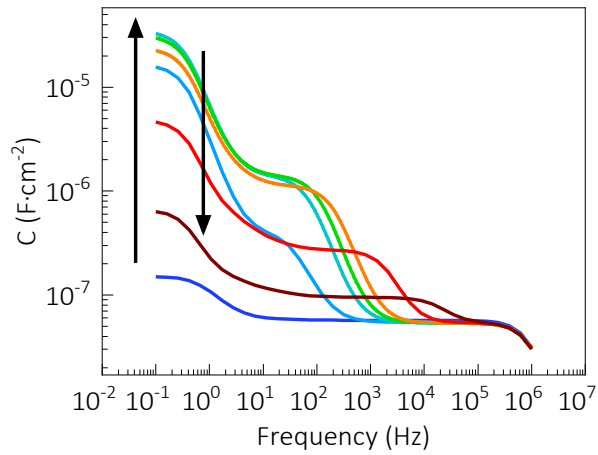

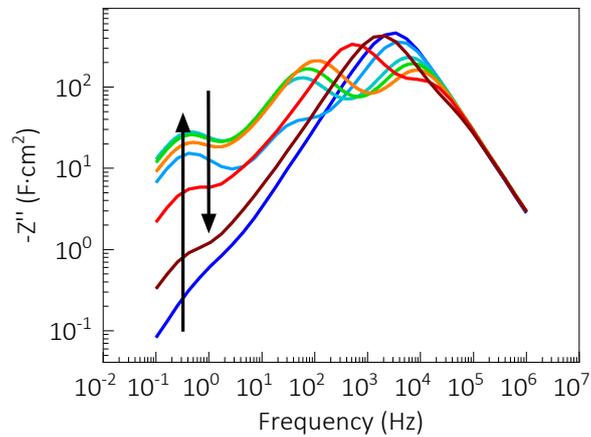

**Figure S42.** Simulated impedance spectra for the HI of MAI sample with combined effect of low shunt resistance (1 k$\Omega$cm$^2$), ion mobility (positive/negative ion mobility: $10^{-6}/10^{-8}$ cm$^2$V$^{-1}$s$^{-1}$), electron/hole mobility (2 cm$^2$V$^{-1}$s$^{-1}$) and low light intensity $G_0 = 2\cdot10^{21}$ cm$^{-3}$s$^{-1}$. Simulated ion concentrations are: $N_{ion}$= $10^{15}$, $10^{16}$, $3\times10^{16}$, $5\times10^{16}$, $10^{18}$, $10^{19}$ cm$^{-3}$.

## S6.4. Mobile ion concentration (in the perovskite) effects

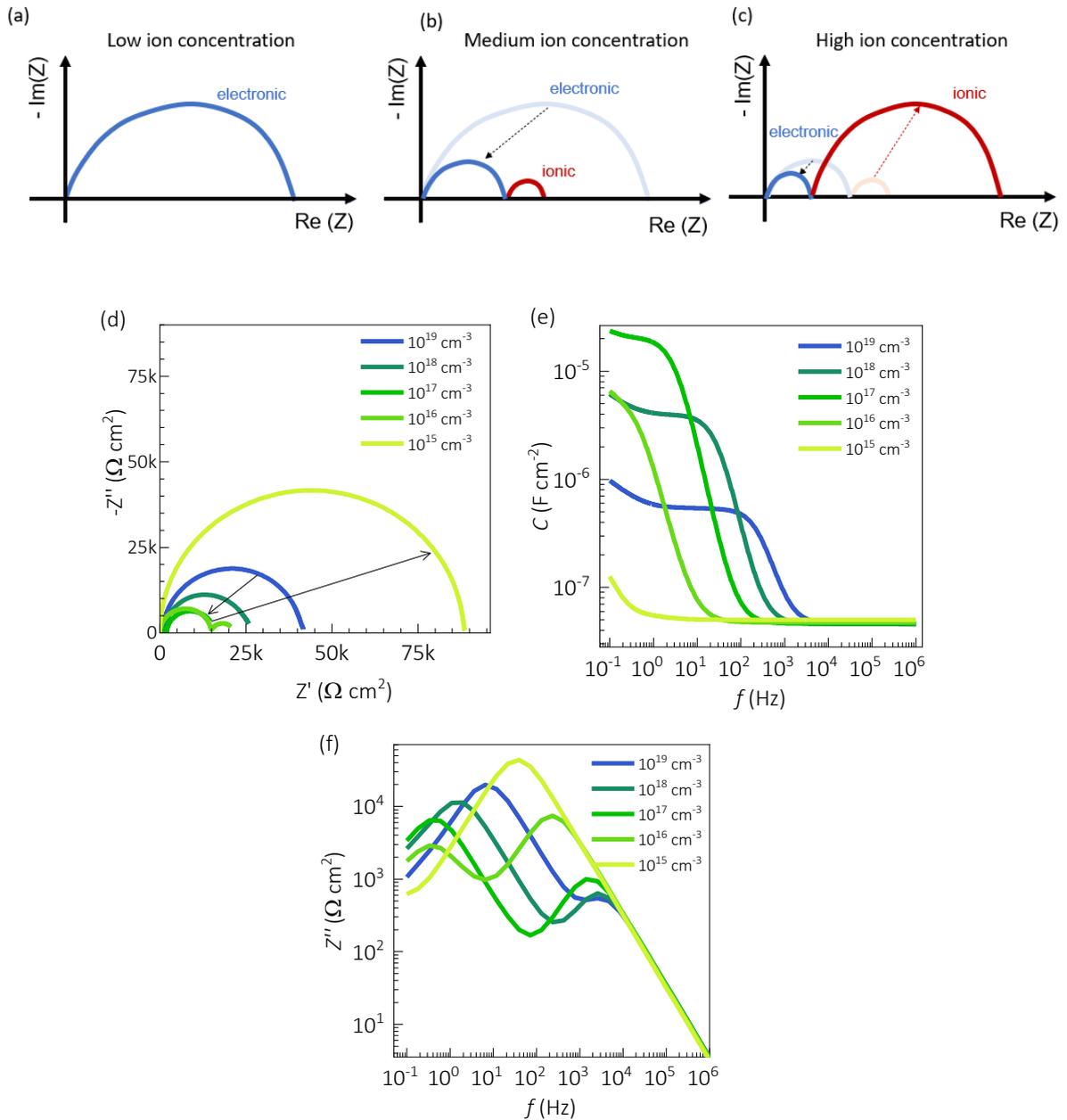

**Figure S43.** Schemed impedance spectra with effects due to different mobile ion concentrations (a-c) and corresponding numerical simulations (d-f). The plots show (d) Nyquist representation, (e) capacitance vs frequency, and (f) imaginary part of impedance vs frequency. Further simulation parameters are in **Table S10**.

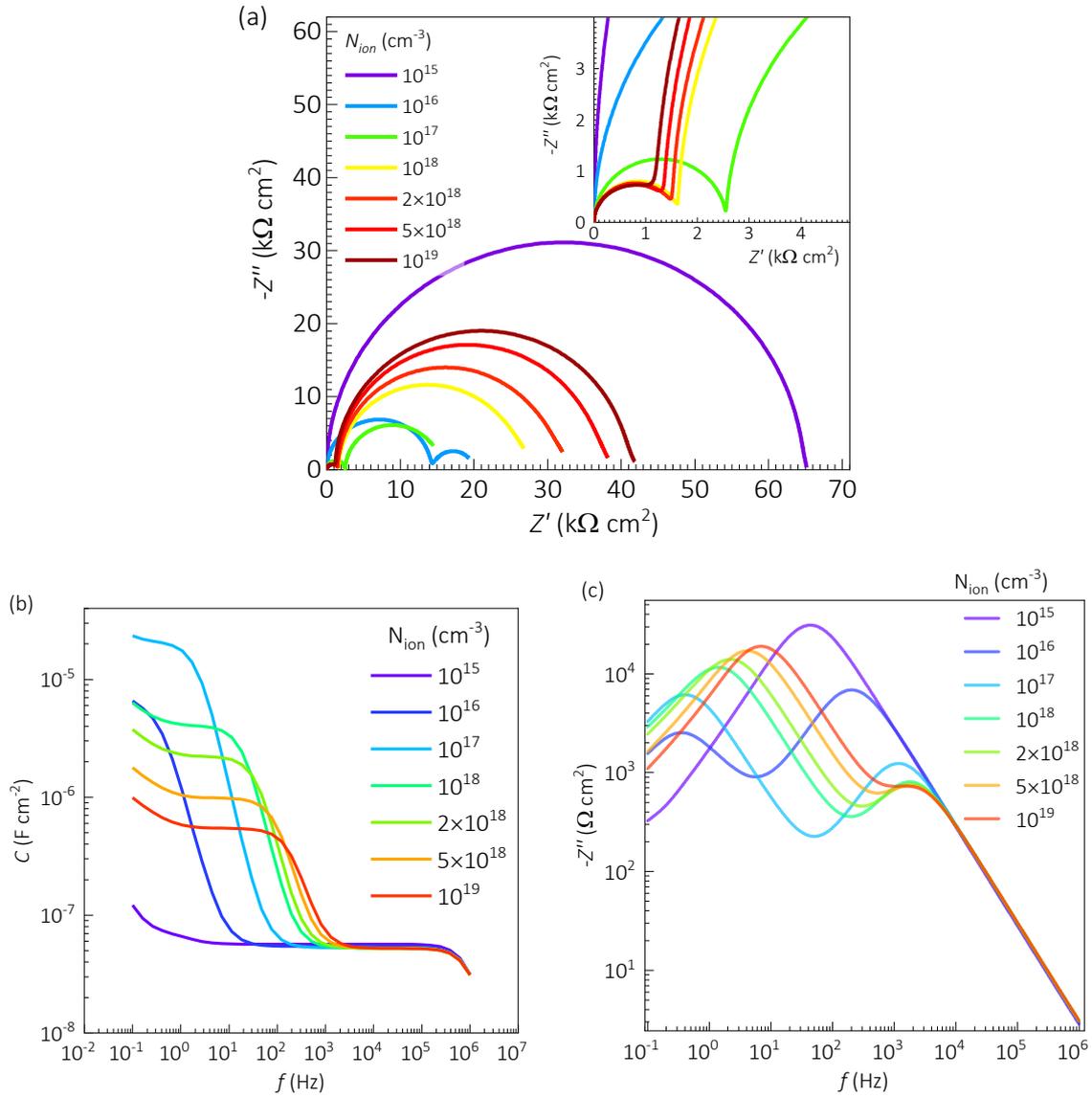

**Figure S44.** Simulated impedance spectra in (a) Nyquist plot and Bode plots of (b) capacitance and (c) imaginary part of impedance for different perovskite mobile ion concentrations, as indicated. The simulation parameters include $R_{sh} = 10^6$ $\Omega \cdot cm^2$; $R_s = 2.5$ $\Omega \cdot cm^2$; $\varepsilon_r = 100$; $G = 4 \times 10^{20}$ $cm^{-3}s^{-1}$, besides those in the **Table S10**.

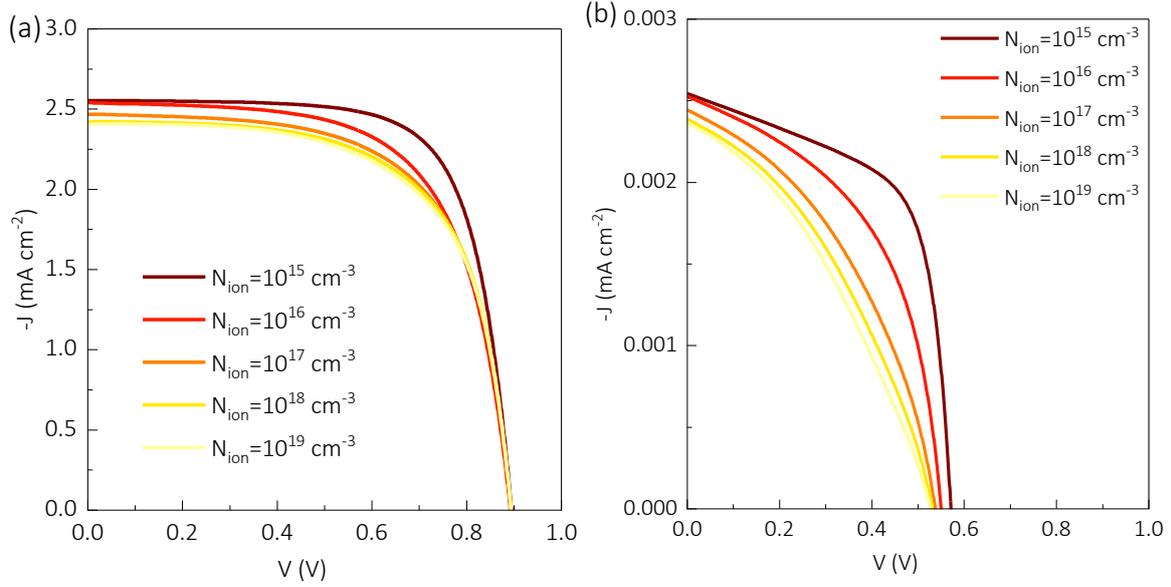

**Figure S45.** Simulated current-voltage curves under 20% of 1 sun illumination intensity for (a) the reference sample and (b) the HI/MAI samples considering different values of mobile ion concentration, as indicated. The simulation parameters include scan rate of 70 mV s$^{-1}$, generation rate of $4\times10^{20}$ cm$^{-3}$s$^{-1}$, in addition to those values in **Table S10**. Notably shunt resistance is $R_{sh}$=1 MΩ·cm$^2$ and 1kΩ·cm$^2$, and electron mobility is $\mu_e$=2 and 0.2 cm$^2$V$^{-1}$s$^{-1}$ in (a) and (b), respectively.

## S6.5. Interface (perovskite/HTL) charge carrier recombination effects

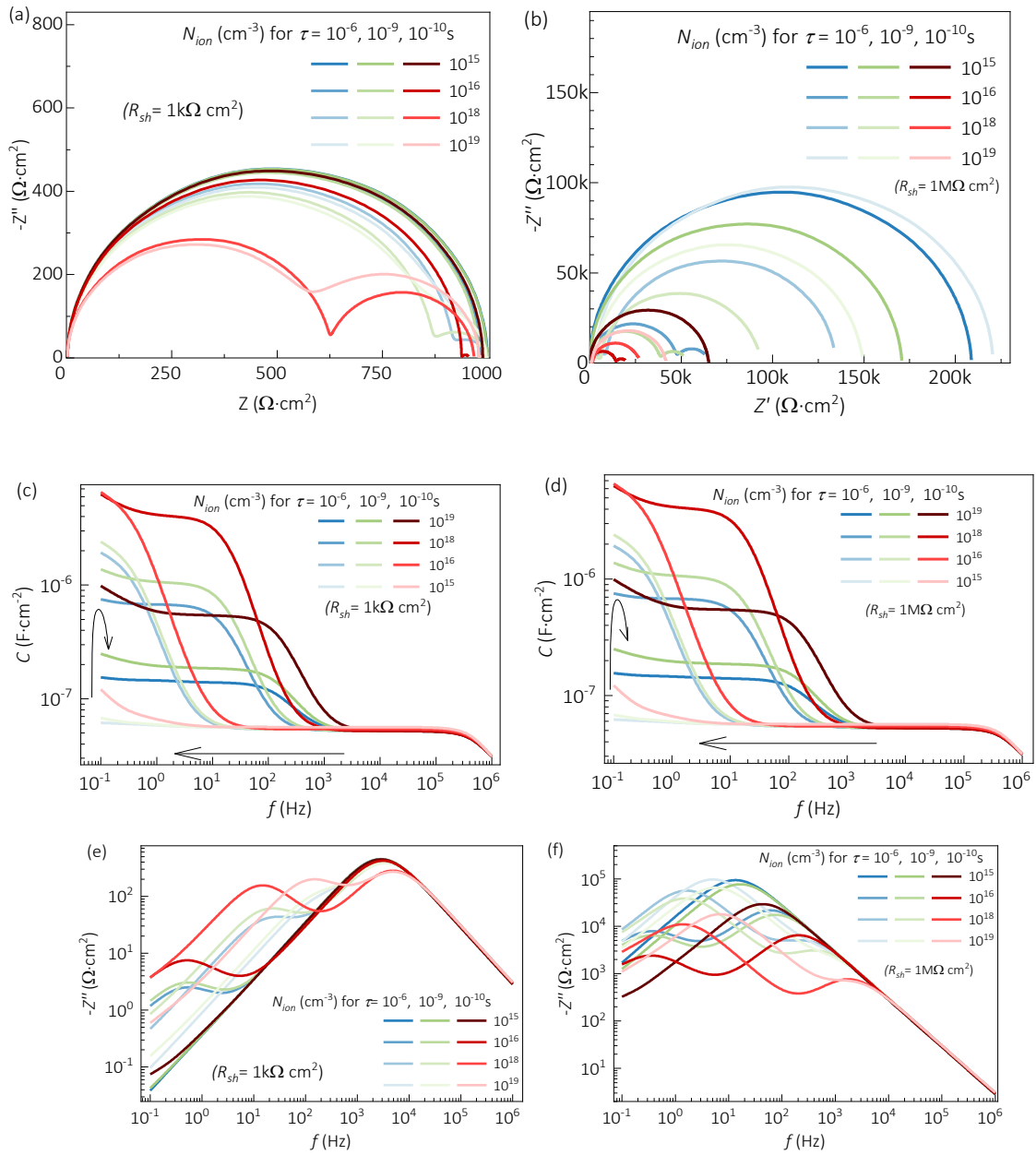

**Figure S46.** Simulated impedance spectra in SC for different concentrations of mobile ions ($N_{ion}$= $10^{15}$, $10^{16}$, $10^{18}$, $10^{19}$ cm$^{-3}$) and trap recombination lifetime values at the interface ($\tau$ = $10^{-6}$, $10^{-9}$, $10^{-10}$ s) considering a shunt resistance of $R_{sh}$ = 1 kΩ and 1 MΩ, for the left and right panels, respectively. Further simulation parameters are in the **Table S10**.

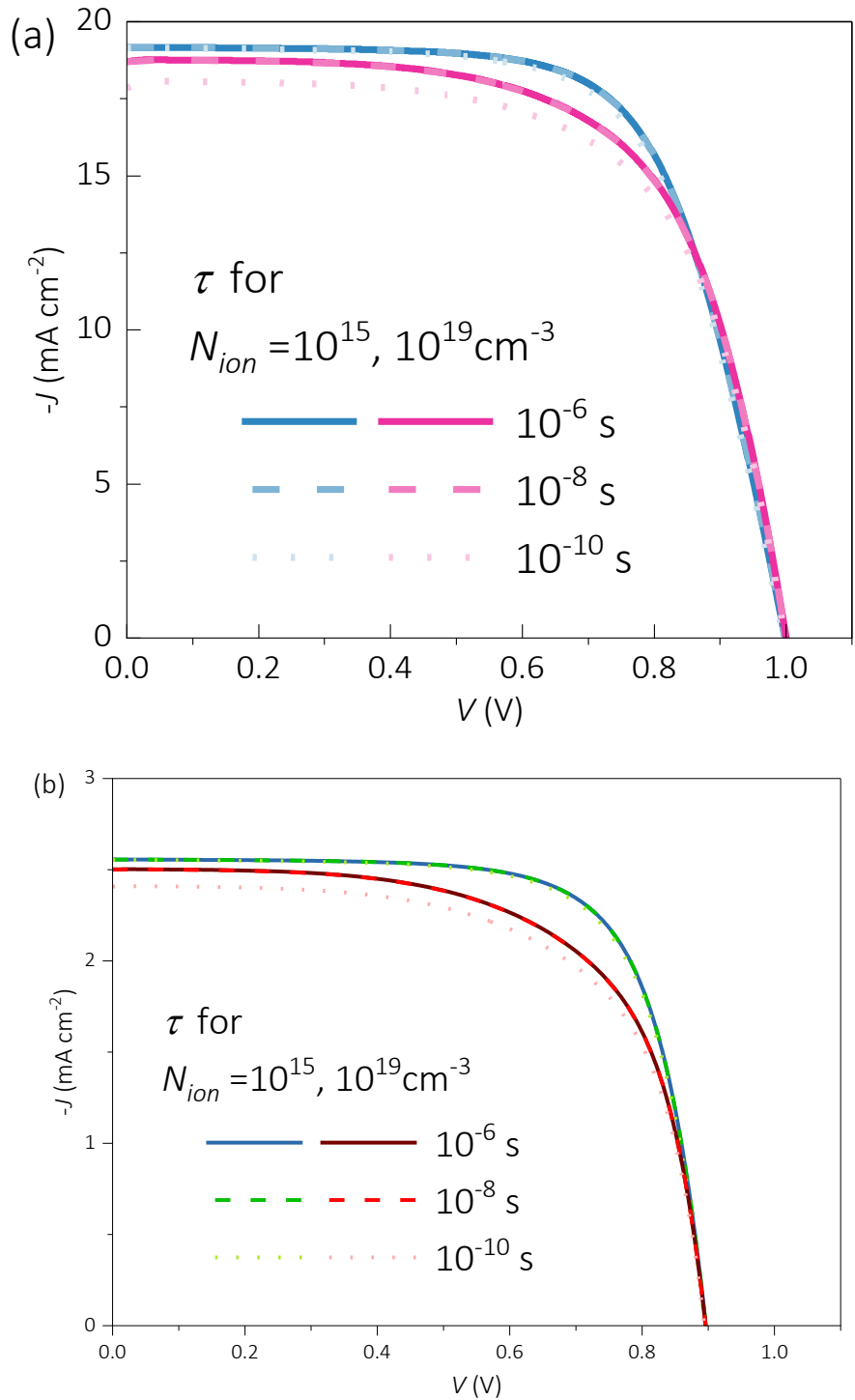

**Figure S47.** Simulated current-voltage curve under (a) 1 and (b) 0.2 sun illumination for the reference sample considering different values of mobile ion concentration and trap recombination lifetime at the interface between the perovskite and the $NiO_x$ transport layer, as indicated. The simulation parameters include scan rate of 70 mV s$^{-1}$, generation rate of $2\times10^{21}$ cm$^{-3}$s$^{-1}$, in addition to those values in **Table S10**.

## S6.6. Bulk (of the perovskite) charge carrier recombination effects.

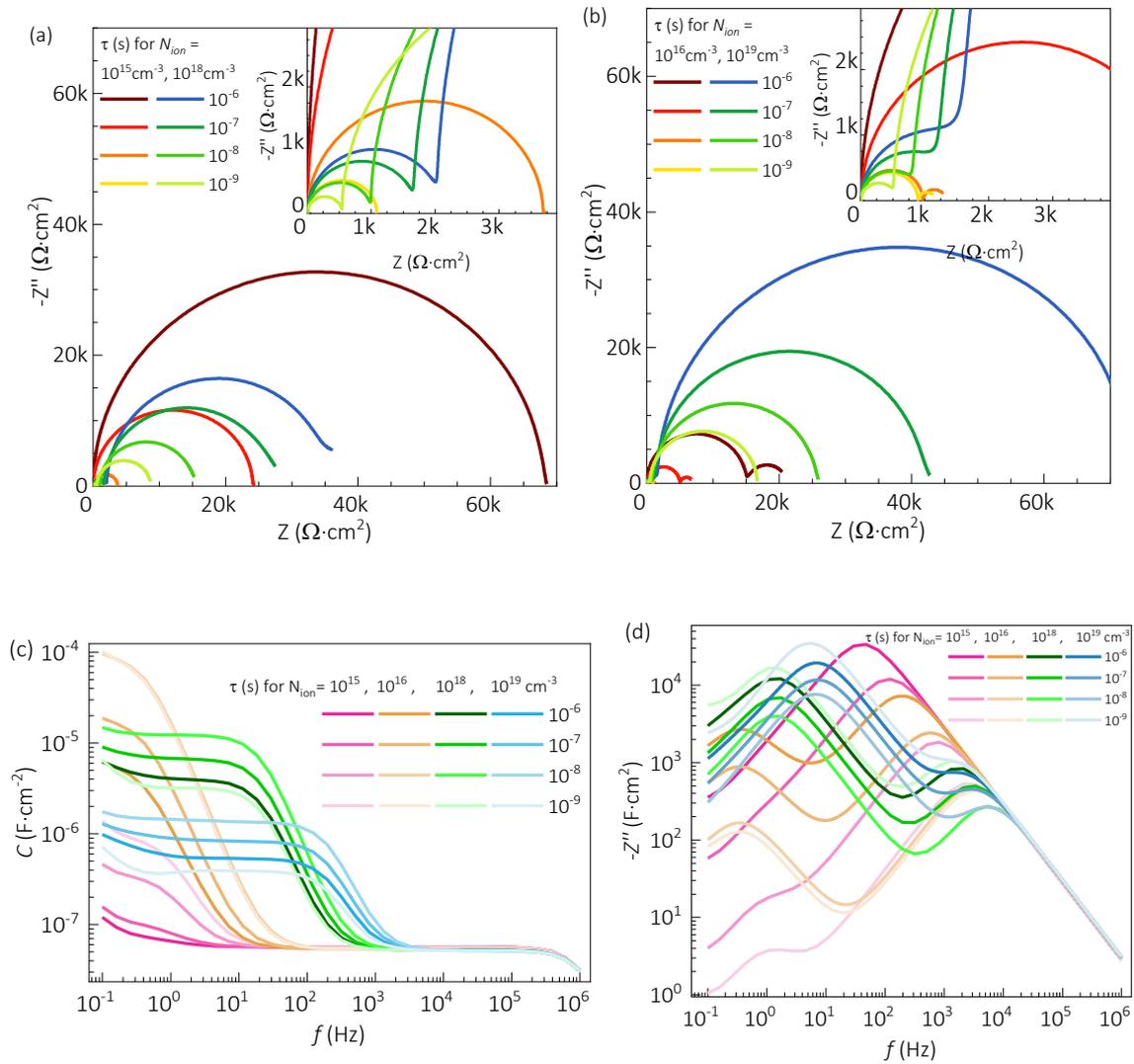

**Figure S48.** Simulated impedance spectra in short-circuit for de reference sample with different concentration of mobile ions and bulk trap recombination lifetime values, as indicated. The physical simulation parameters are in the **Table S10**.

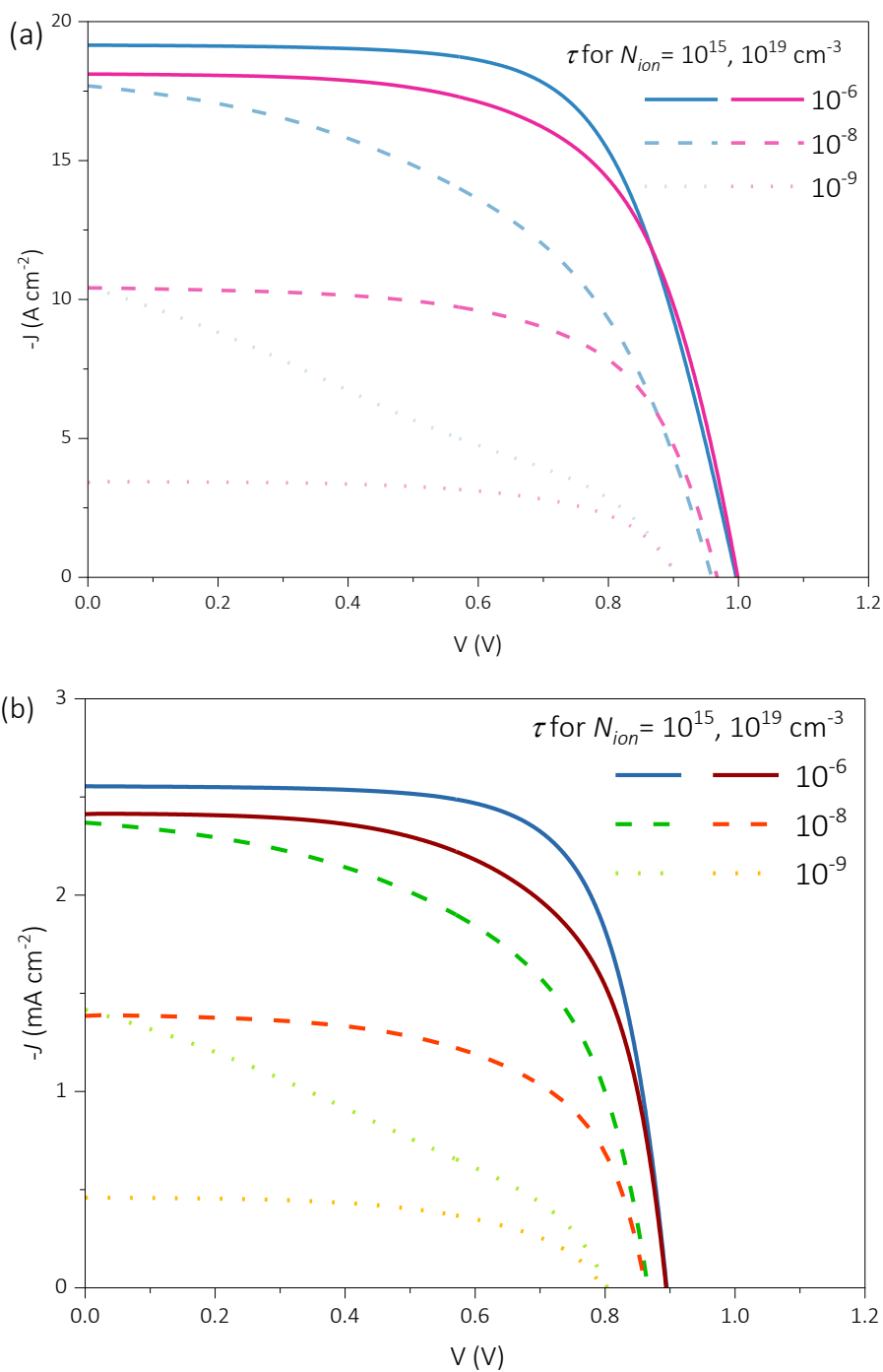

**Figure S49.** Simulated current-voltage curve under (a) 1 and (b) 0.2 sun illumination for the reference sample considering different values of mobile ion concentration and trap recombination lifetime τ within the perovskite bulk, as indicated. The simulation parameters include scan rate of 70 mV s$^{-1}$, generation rate of $4\times10^{20}$ cm$^{-3}$s$^{-1}$, in addition to those values in **Table S10**.

## S6.7. Electron/hole mobility (of the perovskite) effects

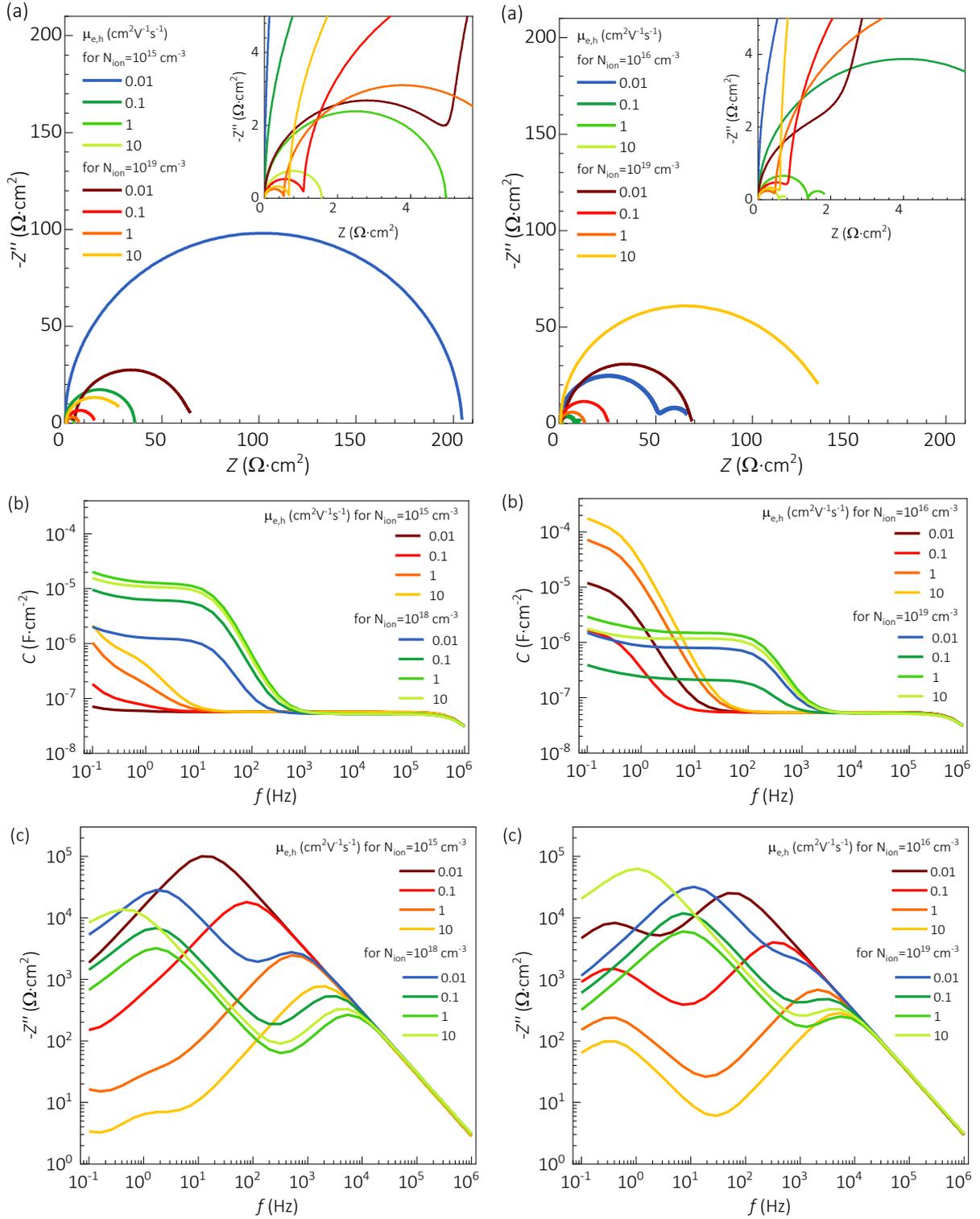

**Figure S50.** Simulated impedance spectra in (a) Nyquist and Bode plots of (b) capacitance and (c) imaginary part of impedance considering different mobile ion concentration and electron/hole mobilities in the perovskite, as indicated. The simulation parameters include $R_{sh}=$ 1 MΩ, in addition to those parameters in the **Table S10**.

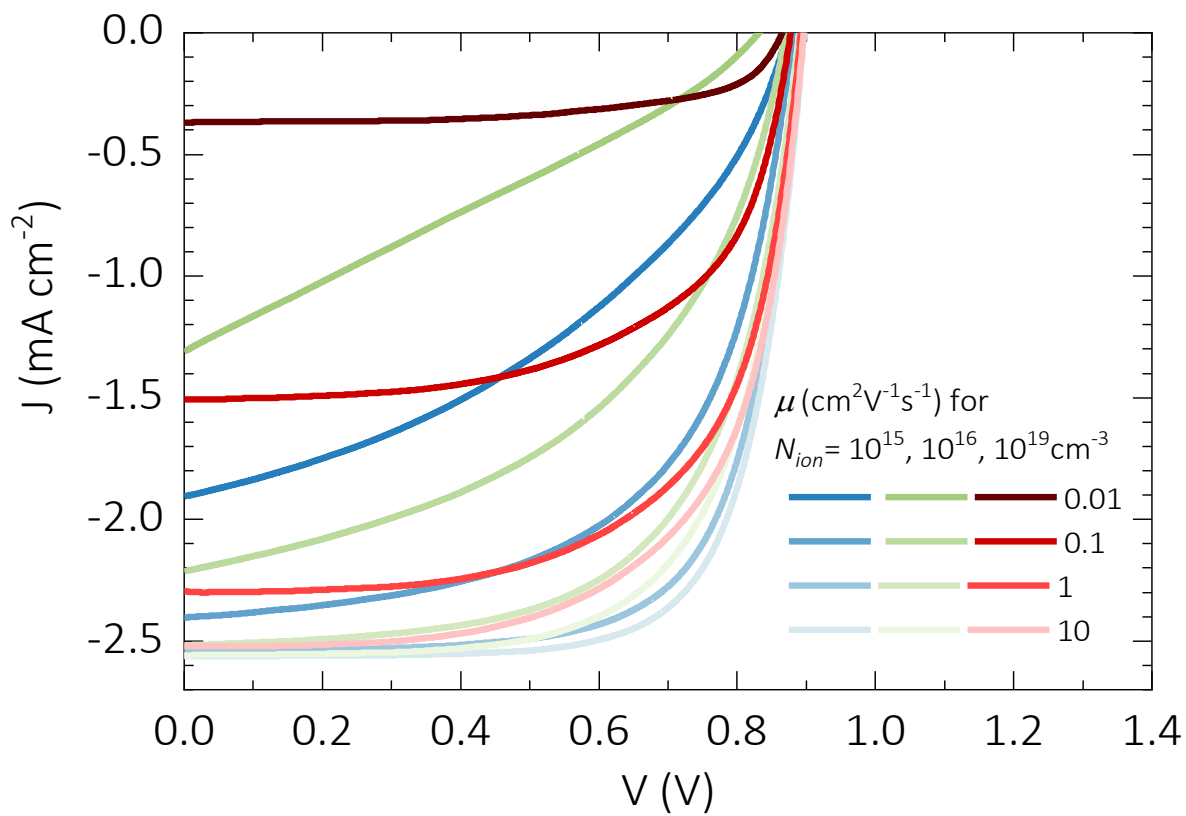

**Figure S51.** Simulated current-voltage curves for the reference sample under 0.2 sun illumination intensity considering different mobile ion concentration and electron and hole mobilities in the perovskite bulk, as indicated. Further simulation parameters can be found in in the **Table S10**.

## S6.8. Hole mobility effects at the NiOₓ hole transport layer

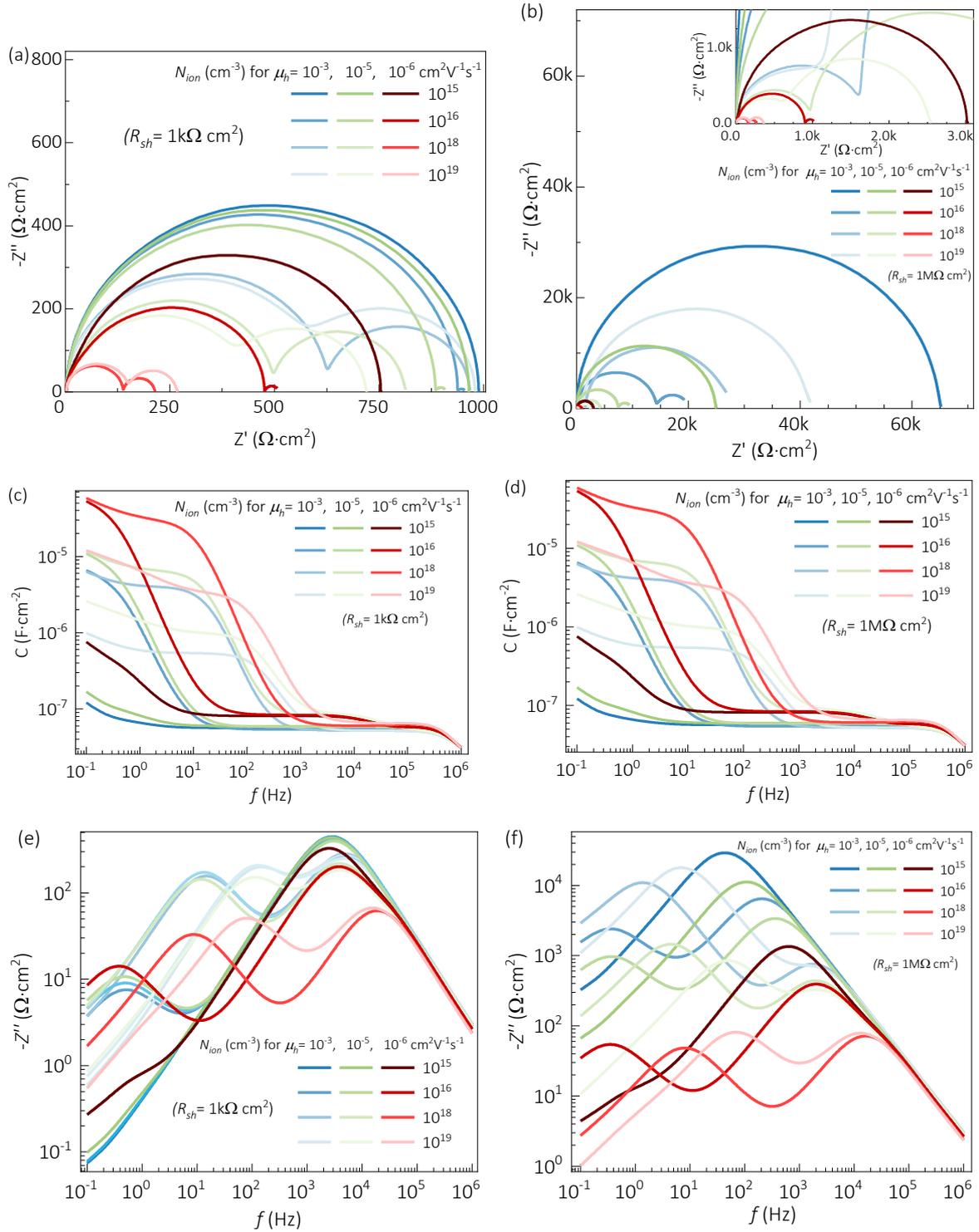

**Figure S52.** Simulated impedance spectra in SC for different concentrations of mobile ions ($N_{ion}$= $10^{15}$, $10^{16}$, $10^{18}$, $10^{19}$ cm$^{-3}$), shunt resistance ($R_{sh}$ = 1 kΩ and 1 MΩ) and hole mobility at the hole transport material ($\mu_h$= $10^{-3}$, $10^{-4}$, $10^{-5}$, $10^{-6}$ cm$^2$/Vs), as indicated. The simulation parameters are in the **Table S10**.

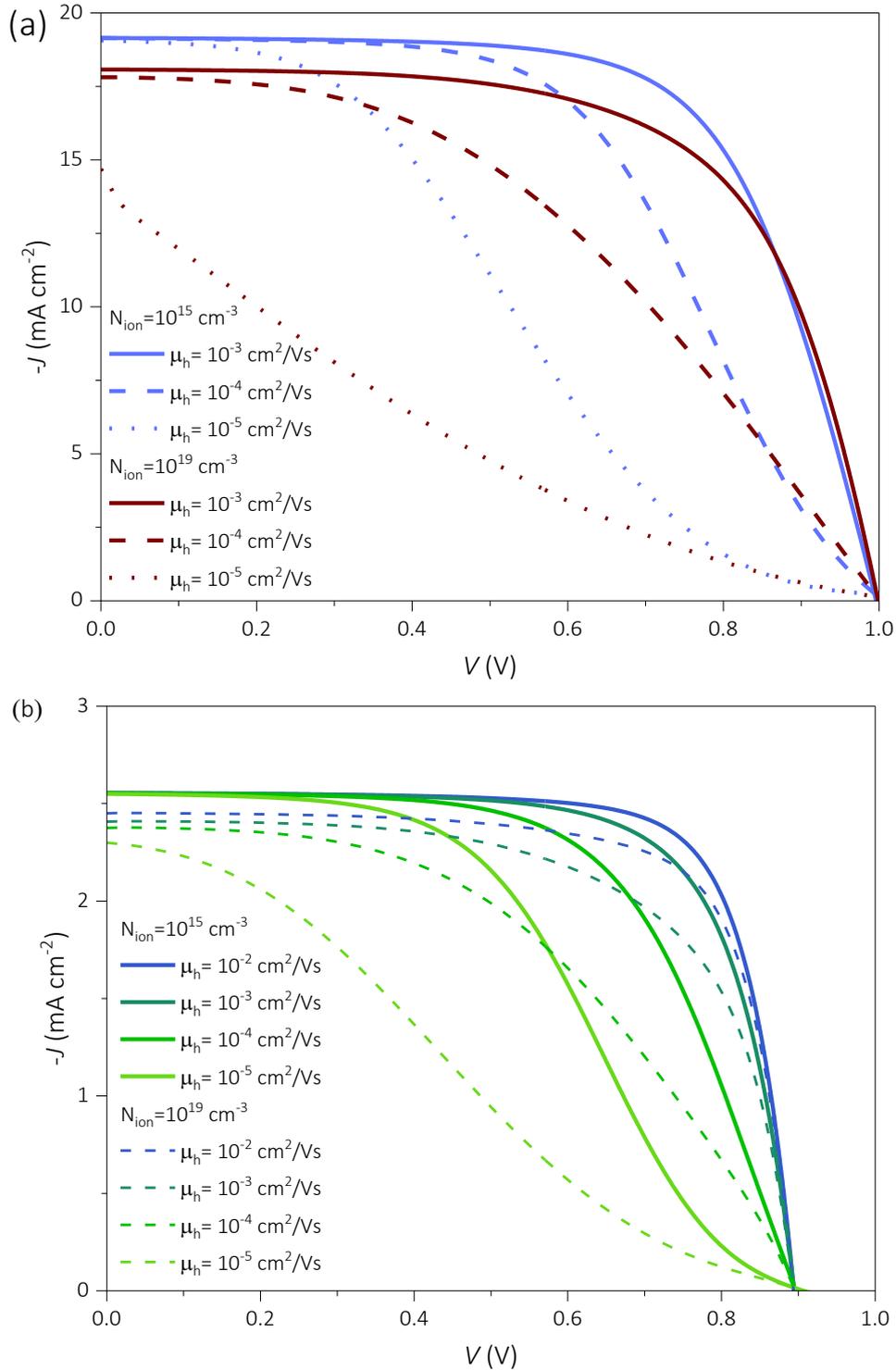

**Figure S53.** Simulated current-voltage curve under (a) 1 and (b) 0.2 sun illumination for the reference sample in SC considering different values of mobile ion concentration and hole mobility at the $NiO_x$ transport layer, as indicated. The simulation parameters include scan rate of 70 mV s$^{-1}$, generation rate of $2\times10^{21}$ cm$^{-3}$s$^{-1}$, shunt resistance of 1MΩcm$^2$, in addition to those values in **Table S10**.

## S6.9. Dielectric permittivity (of the perovskite) effects

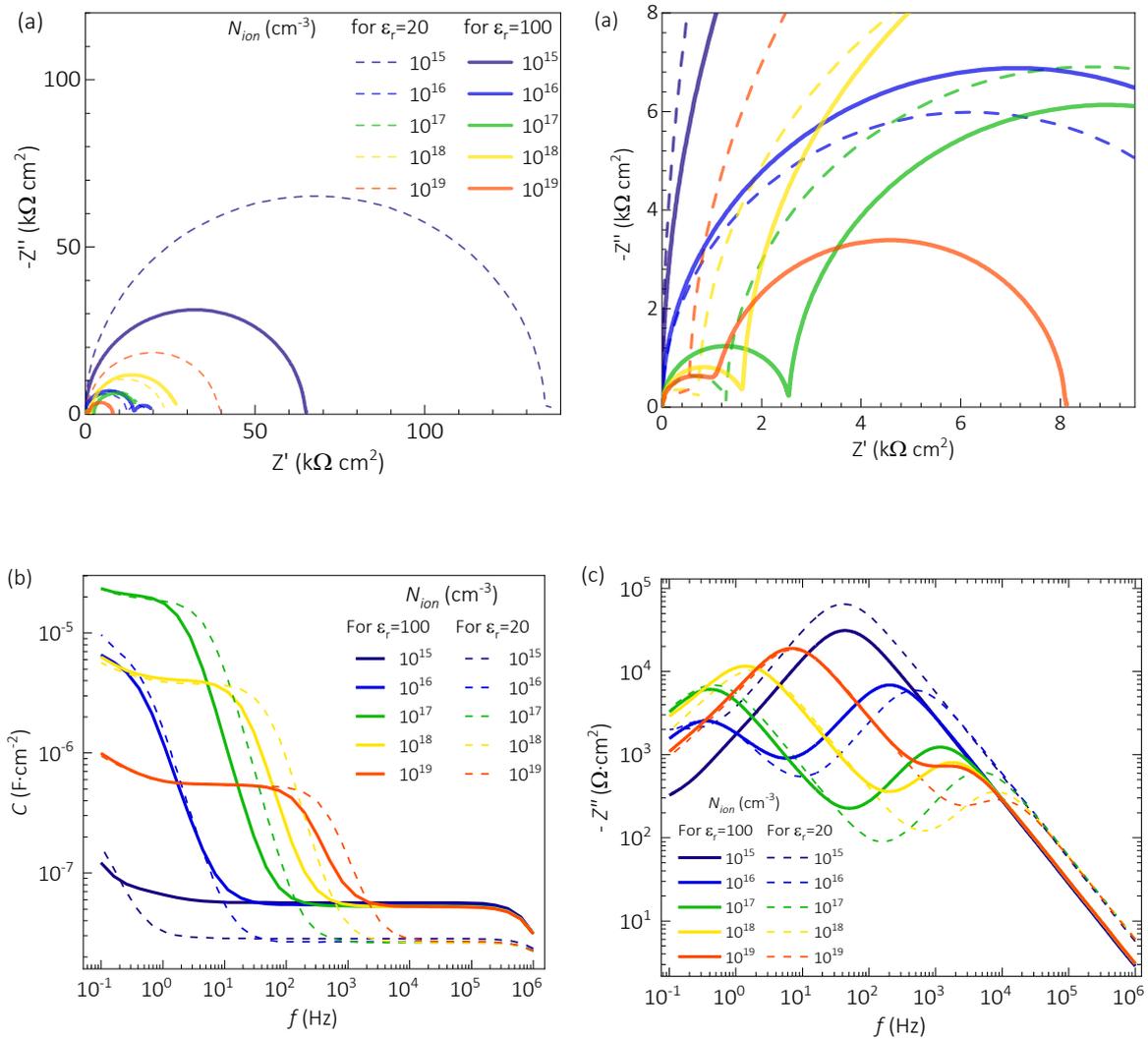

**Figure S54.** Simulated impedance spectra in (a) Nyquist and Bode plots of (b) capacitance and (c) imaginary part or impedance for the reference sample considering different mobile ion concentrations and dielectric permittivity of the perovskite, as indicated. The lower the mobile ion concentration the stronger the influence of the dielectric permittivity. The simulation parameters include $R_{sh}=1$ MΩ cm$^2$, besides those in **Table S10**.

## Acknowledgment

S. Mohamed acknowledges the financial support from Programa Martí i Franquès. M. Ramírez-Como acknowledges the financial support from Diputació de Tarragona under Grant 2021CM14 and 2022PGR-DIPTA-URV04. This work was further supported by the Spanish Ministerio de Ciencia e Innovación (MICINN/FEDER) under Grants PDI2021-128342OB-I00 and RTI2018-094040-B-I00, by the Agency for Management of University and Research Grants (AGAUR) ref. 2017-SGR-1527, and from the Catalan Institution for Research and Advanced Studies (ICREA) under the ICREA Academia Award. O.A. thanks the National Research Agency (Agencia Estatal de Investigación) of Spain for the Juan de la Cierva 2021 grant (FJC2021-046887-I). P.LV. thanks the French Government in the frame of the program of investment for the future (Programme d'Investissement d'Avenir – ANR-IEED-002-01). J.M. and S.O. thank the Ministry of Economic Affairs Innovation, Digitalization and Energy of the State of North Rhine-Westphalia for funding under the grant SCALEUP (SOLAR-ERA.NET Cofund 2, id: 32).

## References

[1] Firoz Khan, Seong-Ho Baek, Jae Hyun Kim, Intensity dependency of photovoltaic cell parameters under high illumination conditions: An analysis, *Appl. Energy* **2014**, 133, 356, https://doi.org/10.1016/j.apenergy.2014.07.107

[2] S. N. Agbo, T. Merdzhanova, U. Rau, O. Astakhov, Illumination intensity and spectrum-dependent performance of thin-film silicon single and multijunction solar cells, *Sol. Energy Mater. Sol. Cells* **2017**, 159, 427, https://doi.org/10.1016/j.solmat.2016.09.039

[3] M. Chegaar, A. Hamzaoui, A. Namoda, P. Petit, M. Aillerie, A. Herguth, Effect of Illumination Intensity on Solar Cells Parameters, *Energy Procedia* **2013**, 36, 722, https://doi.org/10.1016/j.egypro.2013.07.084

[4] Firoz Khan, S. N. Singh, M. Husain, Effect of illumination intensity on cell parameters of a silicon solar cell, *Sol. Energy Mater. Sol. Cells* **2010**, 94, 1473, https://doi.org/10.1016/j.solmat.2010.03.018

[5] Kai Shen, Qiang Li, Dezhao Wang, Ruilong Yang, Yi Deng, Ming-Jer Jeng, Deliang Wang, CdTe solar cell performance under low-intensity light irradiance, *Sol. Energy Mater. Sol. Cells* **2016**, 144, 472, https://doi.org/10.1016/j.solmat.2015.09.043

[6] Dana Lübke, Paula Hartnagel, Markus Hülsbeck, Thomas Kirchartz, Understanding the Thickness and Light-Intensity Dependent Performance of Green-Solvent Processed Organic Solar Cells, *ACS Materials Au* **2023**, 3, 215, https://doi.org/10.1021/acsmaterialsau.2c00070

[7]     SETFOS: Simulation Software for Organic and Perovskite Solar Cells and LEDs, https://www.fluxim.com/setfos-intro, accessed: 15.11.2023.

[8]     Philip Calado, Ilario Gelmetti, Benjamin Hilton, Mohammed Azzouzi, Jenny Nelson, Piers R. F. Barnes, Driftfusion: an open source code for simulating ordered semiconductor devices with mixed ionic-electronic conducting materials in one dimension, *J. Comput. Electron.* **2022**, 21, 960, https://doi.org/10.1007/s10825-021-01827-z